\documentclass[draft]{agujournal2019}
\usepackage{url} 
\usepackage{lineno}
\usepackage{soul}

\usepackage{amsmath}
\usepackage{amssymb}
\usepackage{graphicx}
\usepackage{cleveref}
\usepackage{siunitx}

\usepackage[section]{placeins}
\usepackage{float}
\usepackage{multirow}
\usepackage[final]{pdfpages}

\draftfalse

\journalname{Journal of Advances in Modeling Earth Systems (JAMES)}

\begin{document}

%
%


\title{Gaussian Framework and Optimal Projection of Weather Fields for Prediction of Extreme Events}

%
%




\authors{Valeria Mascolo\affil{1,*}, Alessandro Lovo\affil{1,*}, Corentin Herbert\affil{1}, Freddy Bouchet\affil{2}}

\affiliation{1}{ENS de Lyon, CNRS, Laboratoire de Physique, F-69342 Lyon, France}
\affiliation{2}{LMD/IPSL, CNRS, ENS, Université PSL, École Polytechnique, Institut Polytechnique de Paris, Sorbonne Université, Paris, France}
\affiliation{*}{Authors contributed equally to the article.}




\correspondingauthor{Freddy Bouchet}{freddy.bouchet@cnrs.fr}



\begin{keypoints}
    \item This work presents a new simple framework, called the Gaussian approximation, for a-posteriori and a-priori statistics of extreme events.
    \item Our method provides an interpretable probabilistic forecast of extreme heatwaves which is competitive with off-the-shelf neural networks.
    \item The analysis highlights quasi-stationary Rossby waves and low soil moisture as precursors to extreme heatwaves over France.
\end{keypoints}

%
%

%
%


\begin{abstract}


Extreme events are the major weather-related hazard for humanity. It is then of crucial importance to have a good understanding of their statistics and to be able to forecast them. However, lack of sufficient data makes their study particularly challenging.
In this work, we provide a simple framework for studying extreme events that tackles the lack of data issue by using the entire available dataset, rather than focusing on the extremes of the dataset. To do so, we make the assumption that the set of predictors and the observable used to define the extreme event follow a jointly Gaussian distribution. This naturally gives the notion of an optimal projection of the predictors for forecasting the event.
We take as a case study extreme heatwaves over France, and we test our method on an 8000-year-long intermediate complexity climate model time series and on the ERA5 reanalysis dataset.
For a-posteriori statistics, we observe and motivate the fact that composite maps of very extreme events look similar to less extreme ones.
For prediction, we show that our method is competitive with off-the-shelf neural networks on the long dataset and outperforms them on reanalysis.
The optimal projection pattern, which makes our forecast intrinsically interpretable, highlights the importance of soil moisture deficit and quasi-stationary Rossby waves as precursors to extreme heatwaves.

\end{abstract}

\section*{Plain Language Summary}
Extreme weather events such as heatwaves are responsible for large financial and human costs and their impact can only be expected to grow in the future.
Understanding such events and being able to predict them is therefore of major interest, but suffers from a fundamental problem of lack of data.
In this work, we present a new framework which addresses this issue by making simple assumptions on the statistics of weather fields relevant for heatwaves.
We validate our method using a very long climate simulation. We find that it provides good approximations of atmospheric conditions prevailing during heatwaves, and good prediction capabilities.
It even outperforms existing approaches for short datasets, such as those obtained by combining observations and state-of-the-art weather prediction models, which contain much fewer extreme events than climate simulations but represent more accurately the dynamics of the atmosphere.
This approach explains the observed property that more extreme events are simply stronger versions of less extreme ones, and allows to identify the features of atmospheric patterns which are relevant for making predictions.
The formulation of the method is very general, and it could potentially be applied to other types of extreme events.

%
%

%

%
%
%


\section{Introduction}
\label{sec:intro}

Extreme weather and climate events, often exacerbated by climate change, have led to major disasters in our recent history \cite{IPCC-AR5-extremes}. Heatwaves, in particular, are among the deadliest events. Prolonged exposure to abnormal heat for a certain duration has proven to worsen existing illnesses and to have caused excess deaths during the recent events of the Western European heatwave of 2003 and the Russian heatwave of 2010 \cite{fouilletExcessMortalityRelated2006,garcia-herreraReviewEuropeanSummer2010,barriopedroHotSummer20102011}.
Moreover, losses in the agricultural sector with the subsequent endangerment of the food production system, together with the endangerment of entire ecosystems, allow classifying heatwaves as events which have critical impacts on the whole society, according to the Intergovernmental Panel on Climate Change \cite{IPCC-AR6-11}.

The intensification and the proliferation of these extreme events in the current climate call for urgent progress in our understanding of the mechanisms that drive them, and for developing prediction tools to anticipate risks. However, the most extreme events are the rarest. For this reason, those two classical tasks of analysis and prediction of extreme events suffer from large methodological difficulties associated to a lack of both historical and model data \cite{miloshevichProbabilisticForecastsExtreme2023}.
In this paper we propose a new framework of analysis and prediction, which is effective with rather short datasets, and efficient for the rare unobserved events up to some approximation we fully characterize. Here, we test thoroughly this framework for extreme heatwaves, but we surmise that it can be applied to a large set of other extreme events.

\begin{figure}[tbhp]
        \centering
        \includegraphics[width=1.\textwidth]{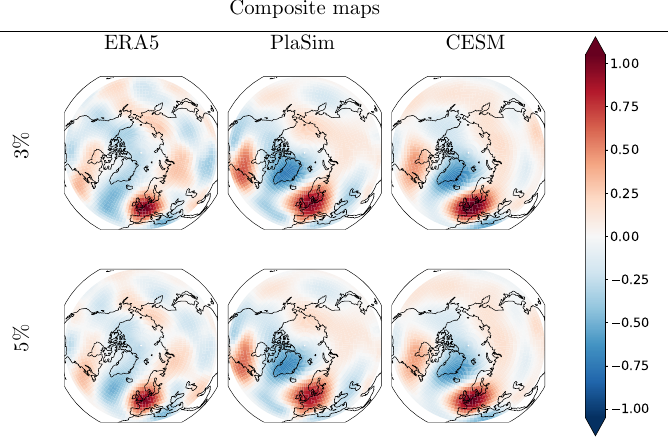}
        \caption{First row: composite maps of \SI{500}{\hecto\pascal} geopotential height anomaly for heatwaves over France, defined as situations with the 3\% most extreme value of two-week-averaged \SI{2}{\meter} temperature anomaly over France. Second row: same but for a 5\% threshold. The maps are normalized pixel-wise by the climatological standard deviation. Composite maps are estimated respectively on ERA5 (daily data from 1940 to 2022), PlaSim (8000 years of simulation) and CESM (1000 years of simulation) datasets. The models reproduce very well ERA5 patterns. Moreover, while the amplitude depends on the threshold defining heatwaves, strikingly the patterns do not. Indeed, we observe in all models and for both thresholds a strong anticyclonic anomaly over western Europe (which is correctly correlated with heatwaves over France). This anticyclonic anomaly is part of a train of a cyclone and an anticyclone which starts over the western part of the United States and continues with a cyclonic anomaly over the North Atlantic Ocean for ERA5, while it is shifted northward over Greenland for both PlaSim and CESM.}
        \label{tab:composite_era_pl_cesm}
\end{figure}

For the task of understanding which weather conditions led to extreme events once they have occurred, composite patterns, i.e. maps of averaged dynamical variables conditioned on the outcome of the extreme event, are the most commonly used statistical diagnostic (see for instance \citeA{grotjahnCompositePredictorMaps2008,sillmannPresentFutureAtmospheric2009,tengProbabilityUSHeat2013,ratnamAnatomyIndianHeatwaves2016,miloshevichRobustIntramodelTeleconnection2023,noyelleInvestigatingTypicalityDynamics2024}).
As visible in \cref{tab:composite_era_pl_cesm} for reanalysis data and two other climate models, the composite patterns associated with very extreme events strikingly resemble those for less extreme ones. This fascinating property has not been addressed in the literature before a recent study \cite{miloshevichRobustIntramodelTeleconnection2023} and has never been explained. Whenever this property holds true, it means that composite maps for rare events can be computed from typical statistics, even if those rare events have not been observed. This is of huge practical interest, and requires understanding. The Gaussian framework we develop in this paper gives a straightforward and enlightening explanation.

For the second task, prediction of future extreme events based on current weather conditions, composite maps are not useful. We clearly demonstrate and explain this in the present paper. The appropriate statistical concept to make predictions is the probability that an extreme event will occur conditioned on the present state of the climate system, the so-called committor function. However, in order to compute this committor function, one actually has to build a forecasting tool able to estimate this probability. Moreover, the committor function is a function of all the variables which characterize the state of the system, called predictors.
For these reasons, it is extremely hard to compute practically and to represent it. Several computations of committor functions have been performed with applications in either geophysical fluid dynamics or in climate sciences~\cite{finkelLearningForecastsRare2021,mironTransitionPathsMarine2021,finkelPathPropertiesAtmospheric2020,lucenteMachineLearningCommittor2019,lucenteCommittorFunctionsClimate2022,lucenteCouplingRareEvent2022}. For climate sciences, methods have been devised using either analogue Markov chains~\cite{lucenteCommittorFunctionsClimate2022}, Galerkin approximations of the Koopman operator \cite{thiedeGalerkinApproximationDynamical2019,strahanLongTimeScalePredictionsShortTrajectory2021}, or neural networks~\cite{lucenteMachineLearningCommittor2019,miloshevichProbabilisticForecastsExtreme2023}. Neural networks seem to be the most efficient and versatile tools. As a matter of fact, there is currently a flourishing literature using neural networks for spatial and temporal predictions of several families of extreme events, such as hurricanes  \cite{NIPS2017_519c8415}, tropical cyclones \cite{giffard-roisinTropicalCycloneTrack2020}, droughts \cite{aganaDeepLearningBased2017, dikshitLongLeadTime2021}, and heatwaves \cite{chattopadhyayAnalogForecastingExtremeCausing2020, jacques-dumasDatadrivenMethodsEstimate2023, miloshevichProbabilisticForecastsExtreme2023}.
However, in \citeA{miloshevichProbabilisticForecastsExtreme2023} the authors clearly demonstrate that machine learning for rare extreme events is most of the time performed in a regime of lack of data and gives suboptimal predictions for typical climate datasets. Moreover, deep learning approaches are, in general, very hard to interpret~\cite{bachPixelWiseExplanationsNonLinear2015,krishnaDisagreementProblemExplainable2022,rudinStopExplainingBlack2019}, and it is extremely difficult to gain some understanding using the forecasting tool.

The main aim of this work is to propose a much simpler alternative method to devise a forecast tool for prediction and to explain the structure of composite maps. This new framework is based on the assumption that the joint probability distribution of the predictors and the extreme event amplitude is Gaussian. Even if this hypothesis is verified only approximately, we show in this paper that the quality of its prediction and its potential for interpretability is extremely high for extreme heatwaves.
We prove that this hypothesis gives a very simple and straightforward explanation of the stability of composite patterns when changing the extreme event amplitude.
For the prediction problem, this Gaussian hypothesis leads to a linear regression problem of the heatwave amplitude on the predictor fields. This is in sharp contrast with regression of fields on scalar values, commonly used in climate sciences. In this case, the predictor is a field in very high dimension, and the predicted value is a scalar.
The key outcome of this procedure is a regression map, which we call the optimal prediction map for the extreme event. This optimal prediction map is a new concept of this study. It is directly interpretable as it gives, at each geographical location, the importance of the predictor field and its sign to determine the heatwave amplitude.
Because of the high dimension of the predictors and because of the not so long dataset length, this regression requires regularization. We analyze thoroughly such optimal prediction maps for extreme heatwaves.

A large part of the work is devoted to the estimate of the accuracy of the results obtained using the Gaussian approximation, compared to the truth. It turns out that this Gaussian approximation is able to give fully interpretable  results which compare very well with the truth.
For instance, it computes composite maps up to errors of the order of 20 to 30\%. Moreover, this Gaussian approximation requires much less data, and it can predict composite maps for unobserved events.
For prediction, it should often be preferred to neural networks for short datasets. For instance, we prove that the Gaussian approximation has a prediction skill close to that of convolutional neural networks on very long datasets and outperforms them on short datasets, like the 80-year long ERA5 reanalysis.

This work is organized as follows. In \cref{ch:hw_definition_datasets} we give the definition of heatwaves used for this study, we present the two datasets used and the set of predictors. In \cref{sec:bayes_committor_composite_gaussianframe} we show with two theoretical examples that composite maps and committor functions are two different probabilistic objects. We then introduce the Gaussian approximation framework and we derive the formulae for computing composite maps and committor functions.
\Cref{sec:composite} and \cref{sec:committor} are dedicated to a methodological study of the Gaussian framework using the climate model PlaSim. Finally, in \cref{sec:ERA5results} we apply our methodology to ERA5.
In \cref{sec:conclusions} we summarize our findings and give perspectives for future works.

\section{Heatwave Definition, Datasets, and Predictors}
\label{ch:hw_definition_datasets}

In this section we provide the definition of heatwaves that will be used in the following (\cref{sec:heatwave_def}), we present the datasets  (\cref{sec:data}), and we identify the weather variables of interest (\cref{sec:X}).

\subsection{Heatwave Definition} \label{sec:heatwave_def}
    In the literature heatwaves have been defined in a plethora of different ways for different analysis purposes \cite{HeatwaveDefinitions}. Short and long-lasting heatwaves affect differently our society and environment, but long-lasting ones are the most detrimental \cite{barriopedroHotSummer20102011}.
    Despite this, most of the literature on heatwaves focuses on daily events \cite{IPCC-AR5-extremes}, as was pointed out in the last assessment report of the Intergovernmental Panel on Climate Change \cite{IPCC-AR6-11}.

    Having a definition which measures independently the persistence and the amplitude of heatwaves is thus of primary interest.
    The simplest way to achieve this is by monitoring the running average of the air temperature field, and this has been applied to the study of heatwaves of different duration (7 days, two weeks, one month) \cite{barriopedroHotSummer20102011,coumouDecadeWeatherExtremes2012,scharRoleIncreasingTemperature2004}.
    In this work, following the recent studies of \citeA{galfiLargeDeviationTheorybased2019,galfiFingerprintingHeatwavesCold2021, Ragone24, ragoneRareEventAlgorithm2021, jacques-dumasDatadrivenMethodsEstimate2023, miloshevichProbabilisticForecastsExtreme2023}, we use a definition which is based on a time and a spatial average of the \SI{2}{\meter} temperature anomaly.
    We believe that this viewpoint is complementary with the more common definitions \cite{HeatwaveDefinitions} and relevant for our analysis.
    Such an average-based definition has the advantage of carrying a natural measure of the heatwave amplitude, which can be easily adapted to heatwaves of different duration and intensity or over different regions of the globe. On the contrary, many classical heatwave definitions involve hard thresholds to be reached within specified time frames and are thus less flexible \cite{HeatwaveDefinitions}.

    Let $\tilde{T}_\text{2m}$ denote the daily-averaged \SI{2}{\meter} air temperature field, which depends on the location $\vec{r}$ and time $t$. Given that the statistics of $\tilde{T}_\text{2m}$ are affected by the seasonal cycle, we use the temperature anomaly $T_\text{2m} := \tilde{T}_\text{2m} - \mathbb{E}_y(\tilde{T}_\text{2m})$, where $\mathbb{E}_y(\bullet)$ denotes the climatology of the weather variable $\bullet$, i.e. the calendar-day-wise average over the whole dataset. We thus define the heatwave amplitude $A$ as the space and time average of the temperature anomaly:
    \begin{equation} \label{eq:heatwave}
        A(t) := \frac{1}{T}\int_{t}^{t+T} \left( \frac{1}{\mathcal{A}}\int_{\mathcal{A}}T_\text{2m}(\vec{r},u)d\vec{r}  \right)du,
    \end{equation}
    where $T$ is the duration in days of the heatwave and $\mathcal{A}$ is the spatial region of interest. Both parameters, $T$ and $\mathcal{A}$ can be changed according to the event one wishes to study. In this work, $T$ ranges from one day (short event) to one month (long event), but nothing prevents it from going even to longer, seasonal events. The region $\mathcal{A}$ typically extends over distances comparable to the synoptic scale, which, in the mid-latitudes, is about \SI{1000}{\kilo\meter}. This is the order of magnitude of the spatial correlations in tropospheric dynamics, corresponding to the size of cyclones and anticyclones, and of the jet stream meanders. In this study we choose $\mathcal{A}$ to be roughly the region of France, which is shown for instance in the last column of \cref{fig:composite}.
    Moreover, as summer heatwaves have higher impacts, we consider only the months of June, July and August.

    Following the studies \citeA{jacques-dumasDatadrivenMethodsEstimate2023, miloshevichProbabilisticForecastsExtreme2023}, we define an extreme heatwave as an event for which the amplitude $A$ exceeds a threshold $a$ corresponding to rare fluctuations.
    This threshold can be changed depending on the heatwaves of interest. In this work we will mainly focus on $a$ defined as the $95^\text{th}$ quantile of the distribution of $A$, i.e. we consider as heatwaves the 5\% most extreme events in our dataset.
    For a two-week heatwave, in the PlaSim model (see \cref{sec:data:plasim}), the threshold amounts to $a=\SI{2.76}{\kelvin}$.
    We will also comment briefly on heatwaves that are more or less rare than the 5\% most extreme ones.

\subsection{Datasets} \label{sec:data}
    In this work we use two datasets. The first is the output of the intermediate complexity climate model called PlaSim, and the second is the ERA5 reanalysis data.
    We use PlaSim to generate an extremely long dataset, over which to train, optimize and test our Gaussian approximation framework (introduced in \cref{sec:bayes_committor_composite_gaussianframe}) with little statistical error. On the other hand, the simplicity of this climate model means that our results may suffer from potentially large biases with respect to the real climate.
    Hence, after this validation step we also apply our new methods to ERA5 data, which can be expected to suffer from smaller biases and be a more faithful representation of the actual climate.

    \subsubsection{PlaSim} \label{sec:data:plasim}
        The Planet Simulator (PlaSim) \cite{Plasim,puma} is an intermediate complexity climate model that has a dynamical core that solves the moist primitive equations \cite{vallisAtmosphericOceanicFluid2017} in the atmosphere.
        The model has a T42 horizontal resolution in Fourier space, that in direct space corresponds to a $64\times 128$ grid of 2.8 degrees both in latitude and longitude, with 10 vertical layers and covering the whole globe.
        The model uses a relatively simplified parameterization of the sub-grid processes such as radiation, clouds, convection and hydrology over land.
        For the latter, in particular, PlaSim uses a single-layer bucket model \cite{manabeCLIMATEOCEANCIRCULATION1969}, with soil moisture increased by snow melt and precipitation and depleted by evaporation.
        Sea ice cover and ocean surface temperature are cyclically prescribed for each day of the year, acting as boundary conditions. By prescribing as well the greenhouse gases concentration and incoming solar radiation, the model is able to run in a steady state that reproduces a climate close to the one of the $1990$s.

        The fact that PlaSim lacks a dynamic ocean means that, in our study of heatwaves, we cannot investigate the effects of ocean related phenomena such as El Niño \cite{hafezRelationshipHeatWaves2017,zhouPossibleImpactsMegaEl2016}, or the North Atlantic Oscillation \cite{hafezRelationshipHeatWaves2017,liCollaborativeImpactNAO2020}.
        On the other hand, the representation of the atmosphere of PlaSim is sufficient to properly resolve the large scale dynamics of cyclones, anticyclones and the jet stream, including important teleconnection patterns relevant for heatwaves \cite{miloshevichRobustIntramodelTeleconnection2023,puma}.
        Moreover, the simplified parameterizations used in PlaSim allow it to run 100 times faster than the models used for CMIP studies, which makes it very suitable to obtain extremely long datasets.
        Here, we use a dataset consisting of $8000$ years.
        It is the same data that was used for previous work on probabilistic forecast of heatwaves using machine learning \cite{miloshevichProbabilisticForecastsExtreme2023}. Mode details on the model setup can be found in \citeA{miloshevichProbabilisticForecastsExtreme2023}.

        As we will show, our proposed method for studying heatwaves does not need such a long dataset to achieve good performances. However, we also want to perform comparisons with alternative deep learning methods, and those do require as much data as possible \cite{miloshevichProbabilisticForecastsExtreme2023}.

        PlaSim resolves the daily cycle and has an output frequency of 3 hours, but we are interested only in daily averages.
        In particular, we will focus on the anomalies (with respect to the daily, grid point-wise climatology) of \SI{2}{\meter} temperature ($T_\text{2m}$), \SI{500}{\hecto\pascal} geopotential height ($Z_\text{500}$) and soil moisture ($S$).

    \subsubsection{ERA5}\label{sec:data:era5}
        In this manuscript we also present an application of our methodology to the ERA5 dataset \cite{hersbachERA5GlobalReanalysis2020}. We use daily data from the publicly available dataset of the ECMWF service for summer seasons from 1940 to 2022.
        ERA5 has a resolution of 0.25 degrees in latitude and longitude. We use this fine resolution to compute the average \SI{2}{\meter} temperature anomaly over France and hence the heatwave amplitude $A$ \cref{eq:heatwave}.

        On the other hand, since the dataset is quite small, we reduce the number of predictors (see next section) by using only the \SI{500}{\hecto\pascal} geopotential height anomaly field and re-gridding it onto the coarser PlaSim grid.

        An important remark is that in our study of heatwaves we assume a stationary climate. We thus need to remove the global warming signal from ERA5 data. This is achieved by means of a parabolic detrending of the averaged temperature over France and of zonal averages of the geopotential height.
        More technical details on the detrending procedure are given in 
        Supporting Information S1.

\subsection{Predictors} \label{sec:X}
    To study heatwaves, we focus on a subset of climate variables that we call \emph{predictors} and denote it with $X$. In particular, for a heatwave that starts at time $t$, we will be interested in the predictors $\tau \geq 0$ days before the event, i.e. $X(t - \tau)$.

    For PlaSim, $X$ will be the stack of the anomalies of \SI{2}{\meter} temperature ($T_\text{2m}$), \SI{500}{\hecto\pascal} geopotential height ($Z_\text{500}$) and soil moisture ($S$).
    The choice of $T_\text{2m}$ is straightforward given its implication in heatwaves, and the potential of simple persistence and advection of temperature to be useful for prediction. The geopotential height anomaly at the middle of the troposphere ($Z_\text{500}$) is a good representation of the dynamical state of the atmosphere because of its relation with cyclones and anticyclones in the lower troposphere. At that height, the geostrophic approximation applies and thus $Z_\text{500}$ gives also a good insight into the wind flow. Finally, it has been shown that low soil moisture acts as an important preconditioning factor for the occurrence of extreme summer temperatures in the mid-latitudes, by limiting the evaporative cooling of the surface \cite{HeatwaveDefinitions,miloshevichProbabilisticForecastsExtreme2023,bensonCharacterizingRelationshipTemperature2021,dandreaHotCoolSummers2006,fischerSoilMoistureAtmosphere2007,hirschiObservationalEvidenceSoilmoisture2011,lorenzPersistenceHeatWaves2010,rowntreeSimulationAtmosphericResponse1983,schubertNorthernEurasianHeat2014,shuklaInfluenceLandSurfaceEvapotranspiration1982,stefanonHeatwaveClassificationEurope2012,vargaszeppetelloProjectedIncreasesMonthly2020,zeppetelloPhysicsHeatWaves2022,zhouLandAtmosphereFeedbacks2019,vautardSummertimeEuropeanHeat2007}.

    For the \SI{2}{\meter} air temperature and \SI{500}{\hecto\pascal} geopotential height fields we will focus on the whole Northern Hemisphere (latitude above 30 degrees North), while soil moisture, instead, is a local variable, and we care only about the values on our region of interest (France). Considering the resolution of the PlaSim model, this will amount to a total of $d=5644$ scalar predictors.

    On the other hand, for ERA5 we use only the \SI{500}{\hecto\pascal} geopotential height anomaly field, which yields a total of $d=2816$ pixels.

    For both datasets, as it is commonly done in the machine learning community, we normalize each field value at each grid point independently dividing by its standard deviation.
    This way, $X$ will be a collection of $d$ (correlated) dimensionless variables with zero mean and unitary standard deviation, which also allows us to easily compare fields with different physical units.

\section{Optimal Projection, Committor Functions, Composite Maps, and the Case of Gaussian Statistics}
\label{sec:bayes_committor_composite_gaussianframe}

    As climate scientists, concerned in understanding extreme events, we might ask two classes of questions. The first class is related to prediction or a priori statistics: given the current state of the system (the predictors $X$), what is the probability to observe an extreme event starting within $\tau$ days? The second class of question is related to a posteriori understanding: given that the extreme event actually occurred, what were the statistical properties of the system states leading to this event? For instance, composite maps defined as the averaged state given that the event occurred, widely used by climate scientists, are examples of a posteriori statistics. Both a priori and a posteriori statistics are useful and important for the sake of understanding, but only a priori statistics is useful for prediction.

    Indeed, the first goal of this section is to stress the difference between a priori and a posteriori statistics. For instance, it is key to understand that in general composite maps do not provide useful information for prediction.
    At the same time, we define some useful statistical quantities for prediction, namely the committor function (see a definition below).
    The second goal is to explain the difficulty to compute committor functions, motivating why they are not commonly used. The third and final goal is to devise predictive and simply interpretable statistical models, for instance the regression of the predictors (the state $X$) on the extreme event observable.

    \subsection{A Posteriori Statistics are Usually not Useful for Prediction}
    
        In this subsection we will stress the differences and the links between a posteriori and a priori statistics.

        Let's consider two events, $F$ and $G$, where $G$ happens after $F$. We will denote with $\mathbb{P}(F | G)$ the a posteriori probability of $F$ conditioned on the occurrence of the future event $G$. Vice versa, $\mathbb{P}(G | F)$ will be the a priori probability of $G$ conditioned on the past event $F$.
        In our case the past event will be the predictors being in a particular state: $X(t-\tau)=x$, while the future event will be the realization of a heatwave: $Y(t)=1$, where $Y$ is the binary random variable
        \begin{equation}\label{eq:Y}
            Y(t) :=
            \begin{cases}
                1 & \text{if } A(t) \geq a \\
                0 & \text{otherwise}
            \end{cases},
        \end{equation}
        where $a$ is the threshold which defines a heatwave and will be chosen based on the quantiles of the distribution of $A$.
        \subsubsection{Bayes' Formula}
            
            When comparing different conditional probabilities, we can make use of Bayes' formula:
            \begin{equation} \label{eq:bayes}
                \mathbb{P}(X(t - \tau)=x|Y(t)=1) \mathbb{P}(Y(t)=1) = \mathbb{P}(X(t - \tau)=x, Y(t)=1) = \mathbb{P}(Y(t)=1|X(t - \tau)=x) \mathbb{P}(X(t - \tau)=x),
            \end{equation}
            where
            \begin{itemize}
                \item $\mathbb{P}(X(t - \tau)=x, Y(t)=1)$ is the joint probability of being in state $x$ and experiencing a heatwave
                \item $\mathbb{P}(X(t - \tau)=x) =: P_S(x)$ is the \emph{stationary measure} of the predictors, namely the probability  of being in state $x$
                \item $\mathbb{P}(Y(t)=1|X(t - \tau)=x) =: q(x)$ is the \emph{committor function}: the a priori probability of observing a heatwave, conditioned on being in state $x$
                \item $\mathbb{P}(Y(t)=1) = \int q(x) P_S(x) dx =: p$ is the unconditional (or climatological) probability of having a heatwave, inversely proportional to its \emph{return time}, that tells us how extreme the event is.
                \item $\mathbb{P}(X(t - \tau)=x|Y(t)=1)$ is the a posteriori probability that the state of the predictors were $x$ given that the heatwave occurred.
            \end{itemize}

            Summarising, Bayes' formula clearly shows the difference and the relation between a priori and a posteriori statistics. In the next subsections we will illustrate a proper tool for the prediction task, namely the committor function, and we will illustrate for what composite maps can be used for, namely a posteriori statistics.

        \subsubsection{Definition of Committor Functions}
        \label{subsec:committor}
            If one is interested in a prediction task, the proper tool is the committor function $q(x)$, originally introduced in the field of stochastic processes
            (see Supporting Information S5)
            for studying transitions between attractors \cite{bolhuisTransitionPathSampling2000a,lucenteCommittorFunctionsClimate2022}. In our case we do not have two attractors, but rather a typical state of the climate with no heatwaves ($Y = 0$) and an atypical one ($Y = 1$). In this context the concept of transition gets a bit blurred, and the committor is simply the a priori conditional probability mentioned before.
            If we expand the notation, we can write it as
            \begin{equation} \label{eq:committor}
                q(x)= \mathbb{P}(A(t) \geq a \mid X(t - \tau)=x) ,
            \end{equation}
            where $a$ is the threshold used to define a heatwave.
            Ultimately all probabilistic predictions amount at providing an estimate of the committor function.
            However, as we will discuss later, for rare events in high-dimensional systems, committor functions are extremely difficult to compute.
            Here, we will make use of their mathematical structure to relate them to a posteriori statistics, before showing that, under a simple assumption, an analytical formula can be obtained, which ultimately provides a way to estimate the committor function, i.e. make a probabilistic prediction.

            \subsubsection{Definition of Composite Maps}
        \label{subsec:composite}
            On the other hand, a commonly used tool in the climate community to study a wide range of events, including the extreme ones, is the \emph{composite map} \cite{grotjahnCompositePredictorMaps2008,sillmannPresentFutureAtmospheric2009,tengProbabilityUSHeat2013,ratnamAnatomyIndianHeatwaves2016,miloshevichRobustIntramodelTeleconnection2023,noyelleInvestigatingTypicalityDynamics2024}. It is defined as the average state of the climate $\tau$ days before the heatwave happened:
            \begin{equation}{\label{eq:composite}}
                C := \mathbb{E}(X(t-\tau) | A(t) \geq a ) ,
            \end{equation}
            where $\mathbb{E}$ denotes an expectation over event realizations and $a$ is the threshold used to define a heatwave.
            In practice one would estimate such expectation with an empirical average over all the heatwave events in the dataset,
            which makes the composite one of the easiest objects to compute and hence motivates its popularity.

            It is important to point out that the empirical average will be a good estimate of the true composite provided that the number of heatwave events is enough. This means that, depending on the size of our dataset, a direct estimation of the composite map is useful only for not too rare (extreme) events, because of sampling errors.

            Going back to the simpler notation used earlier, we can interpret the composite as the mean of the a posteriori probability distribution
            \begin{equation}\label{eq:composite_th}
                C = \mathbb{E}(X | Y=1) := \int x \mathbb{P}(X=x|Y=1) dx ,
            \end{equation}
            and thus, through Bayes' theorem, we can relate it to the stationary measure $P_S$ and the committor function $q$.
            \begin{equation} \label{eq:comp_bayes}
                C = \int x \frac{\mathbb{P}(X=x)\mathbb{P}(Y=1|X=x)}{\mathbb{P}(Y=1)} dx = \frac{\int x P_S(x)q(x) dx}{\int P_S(x)q(x) dx} ,
            \end{equation}
            \Cref{eq:comp_bayes} clearly shows that the composite is the mean of a distribution proportional to $P_S(x)q(x)$ and thus not equivalent to $q(x)$.
            In particular, for rare events, we expect $q(x)$ to be peaked for very atypical values of $x$, namely in the tail of the stationary measure $P_S(x)$. Thus the composite map may differ significantly from the typical states $x$ associated with a high committor.

        \subsubsection{Two Simple Examples which Illustrate that Composites Might be Useless for Prediction}
        \label{subsec:committor_v_composite_examples}
            Now that we have defined the important quantities of interest, we will use some examples to highlight the difference between composites and committor, and in particular how the first may not give us any useful insights on the second.

            As a first example, let us assume that our predictor is one dimensional ($X \in \mathbb{R}$), with stationary measure given by a standard normal distribution $P_S(x) \propto \exp\left( -\frac{x^2}{2} \right)$. Similarly, let the committor be another Gaussian function centered at $x^* > 0$ and with standard deviation $\sigma$: $q(x) = \mathbb{P}(Y=1 | X=x) \propto \exp \left( -\frac{(x - x^*)^2}{2\sigma^2} \right)$. This means that the probability of a heatwave is maximum when we are in state $X = x^*$.
            We will now compute the composite, and show that it is different from $x^*$.

            From \cref{eq:comp_bayes} we know that the composite is the mean of a distribution proportional to $P_S(x)q(x)$, and with some trivial algebraic manipulations, we find that
            \begin{equation*}
                P_S(x)q(x) \propto \exp \left( -\frac{x^2}{2} - \frac{(x - x^*)^2}{2\sigma^2} \right) \propto \exp \left( -\frac{1}{2} \left( 1 + \frac{1}{\sigma^2} \right) \left( x - \frac{x^*}{\sigma^2 + 1} \right)^2 \right).
            \end{equation*}
            Hence, the composite is
            \begin{equation*}
                C = \frac{x^*}{\sigma^2 + 1} ,
            \end{equation*}
            which is strictly smaller than the condition where the heatwave probability is highest. An important consequence is that the probability of having a heatwave when we are in the composite state may be vanishingly small depending on the values of $x^*$ and $\sigma$, showing the low predictive power of the composite map:
            \begin{equation*}
                \frac{q(C)}{q(x^*)} = \exp \left( -\frac{1}{2} \left( \frac{x^*}{\sigma + \sigma^{-1}} \right)^2 \right).
            \end{equation*}

            As a second example, let us consider $X = (X_1, X_2) \in \mathbb{R}^2$ with $P_S(x)$ being a distribution that correlates the two components $X_1$ and $X_2$, for instance a bi-variate Gaussian with mean $(0,0)$ and covariance matrix $\begin{pmatrix}
                \sigma_1^2 & \phi \\
                \phi & \sigma_2^2
            \end{pmatrix}$. We will then consider a committor $q(x) = q(x_1)$ that depends only on the first component.
            Without going into the details
            (available in Supporting Information S3),
            it will be clear that the composite map will have a non-zero $x_2$ component, thanks to the correlation $\phi$ between $x_1$ and $x_2$. However, we know that the committor depends only on $x_1$, and so the composite will be misleading if we are interested in prediction, as it will draw our attention to variables that do not contain \emph{any} information about the probability of having a heatwave.

            \vspace{0.5cm}
            In conclusion, the composite map is an average that takes into account both the probability of having a heatwave starting from state $x$ \emph{and} the probability of \emph{being} in state $x$ (\cref{eq:comp_bayes}). This is good to study the statistics of our extreme event, but if we want to know if there is going to be a heatwave tomorrow, we do not care how rare it was to have had today's weather.

    \subsection{Committor Functions and Optimal Projection}
        Now that we have a clear mathematical understanding of committor functions as the proper tool for prediction, we can move to the problem of computing them in practice. In this subsection we will point out why this is such a complex task as well as provide a way to evaluate how good any approximation of the true committor is. Finally, we will propose the framework of optimal projection of the committor, which will mitigate the problem of high dimensionality as well as make the committor much more interpretable.

        \subsubsection{Complexity of Committor Functions}
        \label{sec:committor_complexity}
            The committor is a function that maps every point of the phase space $x$ to a number $q(x)$ between 0 and 1 that quantifies the likelihood of having a heatwave.
            A naive way of estimating the committor would be to initialize many trajectories at the point $x$ and count how many actually lead to a heatwave.
            This method is called direct numerical simulation, and, if rather inefficient, it is still doable for simple stochastic processes in low dimensional spaces.

            In our case, however, $x \in \mathbb{R}^{d}$, with $d=5644$  for PlaSim and $d=2816$ for ERA5 and the dynamics is described by a rather complex climate model. One could argue that we do not need to explore the whole $\mathbb{R}^{d}$ space, but only the much lower dimensional manifold of \emph{physical} states, which, under ergodic conditions, would be properly sampled by an extremely long trajectory.
            This argument is absolutely correct, but the task of a thorough and precise sampling of the committor still remains out of reach, even with the help of supercomputers.

            Given the importance of committor functions, there is incentive to find efficient ways to get a reasonable approximation of the committor, potentially also limiting the search to only the physical states that are most likely to yield a heatwave.
            This makes the task feasible, but far from simple, and attempts have been made using machine learning  \cite{miloshevichProbabilisticForecastsExtreme2023}, rare event algorithms \cite{Ragone24} or both \cite{lucenteCouplingRareEvent2022}.

            In this work, we strive to find an approach which is far simpler than all the aforementioned, yet still leads to a good enough approximation of the committor.
            
        \subsubsection{Evaluation of Approximations of the Committor Function}
            To quantify how good an approximation $\hat{q}$ of the true committor $q$ is, we need a sort of distance between the two. Since committors are probabilities, the natural object to use is the averaged Kullback-Leibler divergence
            \begin{align}\label{eq:KL}
                \mathcal{K}(q,\hat{q}) &= \int P_S(x) KL\left(\mathcal{B}(q),\mathcal{B}(\hat{q})\right) \\
                &= \int P_S(x) \left( q(x) \log \left( \frac{q(x)}{\hat{q}(x)} \right) + (1 - q(x)) \log \left( \frac{1 - q(x)}{1 - \hat{q}(x)} \right) \right) dx ,
            \end{align}
            where $\mathcal{B}(\bullet)$ is a binomial distribution with parameter $\bullet$.
            $\mathcal{K}$ thus quantifies the average amount of information lost when using $\hat{q}$ instead of $q$.
            Expanding the logarithm and removing the terms that depend only on the true committor, we are left with the cross entropy loss.
            \begin{equation}\label{eq:CE}
                CE(q, \hat{q}) = -\int P_S(x) \left( q(x) \log \hat{q}(x) + (1 - q(x)) \log (1 - \hat{q}(x)) \right) dx.
            \end{equation}
            Now, since we have access to neither the true committor $q$ nor the stationary measure $P_S(x)$, we can replace the first with the heatwave labels $Y$ and the integral over the second with the average over our dataset $\mathcal{D}$. We obtain then the empirical cross entropy loss
            \begin{equation}\label{eq:loss}
                \mathcal{L} = -\left\langle Y(t)\log \hat{q}(X(t)) + (1 - Y(t))\log \left( 1 -\hat{q}(X(t))\right)\right\rangle _{(X(t), Y(t)) \in \mathcal{D}} ,
            \end{equation}
            which is proven to be the only \emph{proper} score for a probabilistic forecast, where \emph{proper} means it satisfies the three basic desiderata defined in \citeA{ProperScores}.

            $\mathcal{L} = 0$ is the perfect prediction, but $\mathcal{L}$ can be arbitrarily large. To have a reference we can consider the climatological committor, that comes from assuming the only information we have is that we are studying the $p$-eth most extreme heatwave, for example setting the threshold $a$ to be the $95^\mathrm{th}$ quantile of the distribution of $A$ means $p = 0.05$. With only this information, the climatological committor is the constant $p$, and the associated empirical cross entropy is
            \begin{equation} \label{eq:clim-loss}
                \mathcal{L}_{\textrm{clim}} = -\left\langle Y(t)\log p + (1 - Y(t))\log ( 1 -p)\right\rangle _{(X(t), Y(t)) \in \mathcal{D}} = -p \log p - (1-p) \log (1 - p).
            \end{equation}
            Finally, we can define the \emph{normalized log score} $\mathcal{S}$ as in \citeA{miloshevichProbabilisticForecastsExtreme2023}, that will quantify the skill of our prediction:
            \begin{equation}
                \mathcal{S} := 1 - \frac{\mathcal{L}}{\mathcal{L}_{\textrm{clim}}}.
            \end{equation}
            A value $\mathcal{S} = 1$ will mean a perfect prediction, namely $\hat{q}(t) = Y(t) \,\, \forall t$, and $\mathcal{S} < 0$ will mean that our forecast is worse than the climatology.

        \subsubsection{Optimal Committor Projection}
        \label{sec:committor-proj}
            Now that we have the tools for evaluating committor approximations, we can tackle the problem of the high dimensionality of $q : \mathbb{R}^d \to [0,1]$. The key idea is to write a surrogate committor $q_\phi = \tilde{q} \circ \phi$, which first applies a projection $\phi : \mathbb{R}^d \to \mathbb{R}^m$ to a space with dimension $m \ll d$, and then represents the committor in this reduced space with function $\tilde{q} : \mathbb{R}^m \to [0,1]$. We want to perform this decomposition in an \emph{optimal} way, which means minimizing the cross entropy defined above, i.e., losing as little information as possible about the original committor.

            It is relatively easy to see that, for a given projection function $\phi$, the best committor representation $\tilde{q}$ that minimizes the cross entropy between $q$ and $\tilde{q} \circ \phi$ is the average of the original committor $q$ on the iso-levels of $\phi$
            \begin{equation}
                \tilde{q}^*(f) = \mathbb{E}_{x \in \phi^{-1}(f)} q(x) ,
            \end{equation}
            where the average is taken according to the stationary measure $P_S(x)$.

            Moreover, the information loss comes from mapping very different values of the original committor onto the same iso-level.
            Ideally, then, the optimal projection would be the one that has the same iso-levels of $q$, namely $q$ itself (up to any monotonic rescaling). Of course this is not desirable, as we simply shifted the problem from computing $q$ to computing $\phi$. To have something useful, we need to constrain the search space of $\phi$, for example to linear maps.

            Even with these simplifications, the general problem remains hard to treat in practice. In the next subsection, we will show the case of Gaussian statistics, which gives an analytic way to compute the optimal linear projection, as well as the reduced committor.

    \subsection{The Case of a Joint Gaussian Distribution}
    \label{sec:gaussian_theory}

        In this section we present the theory for what we call the Gaussian approximation.
        We describe the theoretical idea and derive analytically the expressions for the composite map and the committor function.

        \subsubsection{The Gaussian Approximation}

        The Gaussian approximation consists in assuming that the predictor $X$ at time $t-\tau$ and the heatwave amplitude $A$ at time $t$ follow a jointly Gaussian distribution
        \begin{equation}\label{eq:GA}
            \left(X(t-\tau),A(t)\right) \sim \mathcal{N}\left(0,\Sigma(T,\tau)\right),
        \end{equation}
        where $X$ is thought of as a $d$-dimensional vector, and represents all grid-point values of either a single field or stacked fields.
        The joint distribution has mean zero because both $X$ and $A$ are anomalies, and it is then solely characterized by the $d + 1$ dimensional covariance matrix $\Sigma(T,\tau)$, that depends on the heatwave duration and the lead time.
        To simplify the notation, we assume that we work at fixed $T$ and $\tau$, and thus drop the dependencies on them. We can then write $\Sigma$ as a block matrix of the form
        $\begin{bmatrix}
            \Sigma_{XX} & \Sigma_{XA}\\
            \Sigma_{AX} & \Sigma_{AA}
        \end{bmatrix},$
        where $\Sigma_{XX} = \mathbb{E}(XX^\top)$ is the $d\times d$ covariance matrix of $X$, $\Sigma_{XA} = \Sigma_{AX}^\top = \mathbb{E}(XA)$ is the $d\times 1$ correlation map between $X$ and $A$ and $\Sigma_{AA} = \mathbb{E}(A^2)$ is the scalar variance of $A$.

        The property expressed by \cref{eq:GA} arises naturally in a number of contexts.
        In particular, it holds as soon as $X(t)$ is a Gaussian process.
        An important family of models of this type are the \emph{vector autoregressive models} or \emph{Linear Inverse Models} (LIMs)~\cite{hasselmannPIPsPOPsReduction1988}.
        Such models assume a linear relation between $X(t-\tau)$ and $X(t)$, perturbed by noise, which allows to easily compute a Green function and thus perform predictions.
        This approach has been successfully applied to a number of climate problems~\cite{penlandPredictionNino31993,penlandOptimalGrowthTropical1995,kwasniokLinearInverseModeling2022}, even specifically for extreme events~\cite{tsengMappingLargeScaleClimate2021}.
        In the same line of thought, another common approximation is to assume that $(X_t,A_t)$ follows a Gaussian process, thus implying further linearity constraints.
        Here we do not make any assumption about the dynamics; in particular we do not require that the joint distribution of $X$ at arbitrary times is Gaussian, but only the joint distribution of $X$ at a given time and the observable $A$ at a fixed later time.
        This is in principle more general; for instance, components of $X$ that are not involved in calculating $A$, such as the geopotential height field, might evolve non-linearly with time, breaking the Gaussian process hypothesis, while the Gaussian approximation might still be valid.
        Our goal here is to show that making only the assumption of \cref{eq:GA} leads to simple formulas to compute composite maps and committor function, and to test these formulas against climate data.

        \subsubsection{Composite Maps Within the Gaussian Approximation}
        \label{subsec:gaussian_composite}
            Under the Gaussian assumption, the composite map can be computed analytically as
            \begin{equation}
                \label{eq:composite_gauss}
                 C_{\mathcal{G}} = \mathbb{E}[X|A \geq a] = \int x \frac{ \int_a^{+\infty} \mathbb{P}(x, A) dA}{\int_a^{+\infty} \mathbb{P}(A) dA} dx = \eta \left( \frac{a}{\sqrt{2\Sigma_{AA}}} \right) \frac{\Sigma_{XA}}{\sqrt{\Sigma_{AA}}} ,
            \end{equation}
            with
            \begin{equation}
                \eta(z) = \sqrt{\frac{2}{\pi}} \frac{e^{-z^2}}{\mathrm{erfc}(z)},
            \end{equation}
            where $\mathrm{erfc}(\bullet)$ is the complementary error function and the subscript $\mathcal{G}$ reminds that the composite is evaluated under the Gaussian assumption.
            The detailed computation is shown in
            Supporting Information S4.

            From \cref{eq:composite_gauss}, we can clearly see that the composite is directly proportional to the correlation map, with the proportionality constant depending only on the threshold $a$. This has the important implication that the average state of the climate $\tau$ days before a heatwave looks like the $\tau$-lagged correlation between the fields and the heatwave amplitude, \emph{regardless} of how extreme the heatwave is. In other words, the composite of a more extreme event has exactly the same pattern as a less extreme one, but amplified according to the function $\eta$.
            Already in \cref{tab:composite_era_pl_cesm} we observed a qualitative similarity between the composite maps of the 5\% and 3\% most extreme heatwaves. Later in \cref{sec:composite}, and particularly in \cref{fig:scaling}, we thoroughly test the validity of \cref{eq:composite_gauss}, checking both that composite maps are mostly proportional to the correlation map and that they closely follow the scaling predicted by the Gaussian approximation.
            This way, we gain access to composites of very extreme events, where the direct estimation as the average over the (very small) heatwave set would suffer from huge sampling errors. On the other hand, the correlation map $\Sigma_{XA}$ is estimated on the whole dataset and thus does not have this issue.

            The function $\eta$ is plotted in
            fig. S3 in Supporting Information S9,
            and has the interesting property that $\eta(z) \sim \sqrt{2} z$ as $z \to \infty$, which means that for very extreme heatwaves the composite map tends to the simple linear regression of $X$ against $A$.
            \begin{equation} \label{eq:extreme_composite_gaus}
                C_\mathcal{G} \xrightarrow[a \gg \sqrt{\Sigma_{AA}}]{} a \frac{\Sigma_{XA}}{\Sigma_{AA}} = a \xi, \quad \xi = \arg \min_{\xi} \mathbb{E} \left( (X - A\xi)^2 \right).
            \end{equation}

        \subsubsection{Committor Functions Within the Gaussian Approximation}
        \label{subsec:gaussian_committor}
            By definition, the committor is the integral of the a priori distribution of $A$ conditioned on knowing $X$:
            \begin{equation}\label{eq:q_int}
                q(x) = \mathbb{P}(A \geq a | X = x) = \int_{a}^{+\infty} \mathbb{P}(A = a^\prime | X = x) da^\prime.
            \end{equation}
            Under the assumption of a joint Gaussian distribution for $(X, A)$, the conditional distribution of $A$ given $X$ is also Gaussian. In particular, it has mean $\mu(x)$ that scales linearly with $x$ and constant variance $\sigma^2$:
            \begin{equation}\label{eq:mu-sigma}
                \mu(x) = \Sigma_{XX}^{-1}\Sigma_{XA} \cdot x, \qquad \sigma^2 = \Sigma_{AA} - \Sigma_{AX}\Sigma_{XX}^{-1}\Sigma_{XA} .
            \end{equation}
            For the details of this computation see
            Supporting Information S4.
            In fact, $\mu(x) = \tilde{M}^\top x$ is precisely the linear regression of $A$ against $X$:
            \begin{equation}\label{eq:Mtilde}
                \tilde{M} := \Sigma_{XX}^{-1} \Sigma_{XA} = \arg\min_M \left( M^\top \Sigma_{XX} M - 2M^\top \Sigma_{XA} \right) = \arg\min_M \mathbb{E}\left((A - M^\top X)^2\right) .
            \end{equation}
            Then, to obtain the full committor, we just have to compute the Gaussian integral in \cref{eq:q_int}, which gives
            \begin{equation} \label{eq:q-gaus}
                q_{\mathcal{G}}(x) = \frac{1}{2} \mathrm{erfc} \left( \frac{a - \tilde{M}^\top x}{\sqrt{2}\sigma} \right).
            \end{equation}

            This result can be viewed in light of the framework of optimal committor projection presented in \cref{sec:committor-proj}. In this case, the optimal projection of the high dimensional committor is onto the normalized projection pattern
            \begin{equation} \label{eq:M}
                M = \frac{\Sigma_{XX}^{-1}\Sigma_{XA}}{|\Sigma_{XX}^{-1}\Sigma_{XA}|} ,
            \end{equation}
            which condenses all the important information of the high dimensional vector $x$ into the scalar variable $f = M^\top x$. Then the committor in the projected space is simply
            \begin{equation} \label{eq:qtilde-gaus}
                \tilde{q}(f) = \frac{1}{2} \mathrm{erfc} \left( \alpha + \beta f \right) ,
            \end{equation}
            with
            \begin{gather} \label{eq:ab}
                \alpha = \frac{a}{\sqrt{2}\sigma}, \quad
                \beta = -\frac{|\tilde{M}|}{\sqrt{2}\sigma}.
            \end{gather}

            The two operations of linear projection and reduced committor can also be viewed as the architecture of a simple one layer perceptron with the custom activation function $\tilde{q}$. In comparison to other neural network architectures (such as convolutional ones) that may be trained on the same task \cite{miloshevichProbabilisticForecastsExtreme2023}, this approach is far simpler, and depends on a much smaller number of parameters.

            In addition, we would like to stress that the method is \emph{interpretable} by design: with complex neural networks one may need sophisticated explainable AI techniques to understand \emph{why} they are outputting a particular probability \cite{McGovern2019,tomsPhysicallyInterpretableNeural2020,delaunayInterpretableDeepLearning2022}, while in our case the answer is straightforward, namely, it is computing the optimal index $f$.
            Furthermore, since the projection pattern $M$ has the same dimension as the predictor $X$, we can plot it as a map, representing the relative importance of each pixel in our predictor, and providing potential insight in the physical dynamics leading to extreme heatwaves.

            Another interesting point to pay attention to is the difference between the two linear regressions for the composite (\cref{eq:extreme_composite_gaus}) and for the committor (\cref{eq:Mtilde}). In the first case, we are doing $d$ independent linear regressions of each pixel in $X$ against the heatwave amplitude $A$, while for the committor we have a single optimization, regressing $A$ against $X$. This shows once again the fundamental difference between a posteriori and a priori statistics.

        In the following sections, we apply the Gaussian approximation to actual data, see to what extent the assumption of gaussianity holds and what useful information we are able to extract.


\section{Validation of the Gaussian Approximation for the Computation of Composite Maps for Extreme Heatwaves}
\label{sec:composite}


Composite maps are very interesting to understand weather situations that actually led to extreme events (a-posteriori statistics). They are defined as the average of weather variables conditioned on the future occurrence of the extreme event.

In \cref{subsec:comp_gauss_comp_data} we show and compare qualitatively composite maps evaluated empirically and using the Gaussian approximation. In \cref{subsec:error_composite} we quantify the error made under the Gaussian approximation, and we distinguish systematic and sampling errors.
In \cref{subsec:comp_gauss_ind_a}, using the Gaussian approximation, we give an explanation of the puzzling independence of the empirical composite maps patterns from the threshold $a$ used to define an extreme heatwave.
Finally, in \cref{subsec:generalization_power} we discuss in more detail the effect on the quality of the Gaussian approximation of both the dataset length and the threshold defining extreme events, and conclude that the Gaussian approximation is the best way to estimate composite maps in a regime of lack of data.

In \cref{sec:physics}, we will use these results to make a physical analysis of extreme events, by varying the heatwave duration $T$ and the lead time $\tau$.

In this section we use the PlaSim dataset with 8000 years of data and predictors $X=\left( T_{2m},Z_{500},S\right)$ (see \cref{sec:data:plasim}).
We show an application of our methodology to the ERA5 dataset in \cref{sec:ERA5results}.

    \subsection{Comparing Empirical Composite Maps with Composite Maps Computed Within the Gaussian Approximation}
    \label{subsec:comp_gauss_comp_data}

    We now compare the composite maps computed either directly from the data or using the Gaussian approximation. We show that the two are qualitatively very similar, with a relative error of the order of 20\%.
    We consider 14-day heatwaves ($T=14$), looking at the composites for the first day of the heatwave (lead time $\tau=0$). We first focus on the 5\% most extreme events ($a = \SI{2.76}{\kelvin}$).

\begin{figure}
        \centering
        \includegraphics[width=0.8\textwidth]{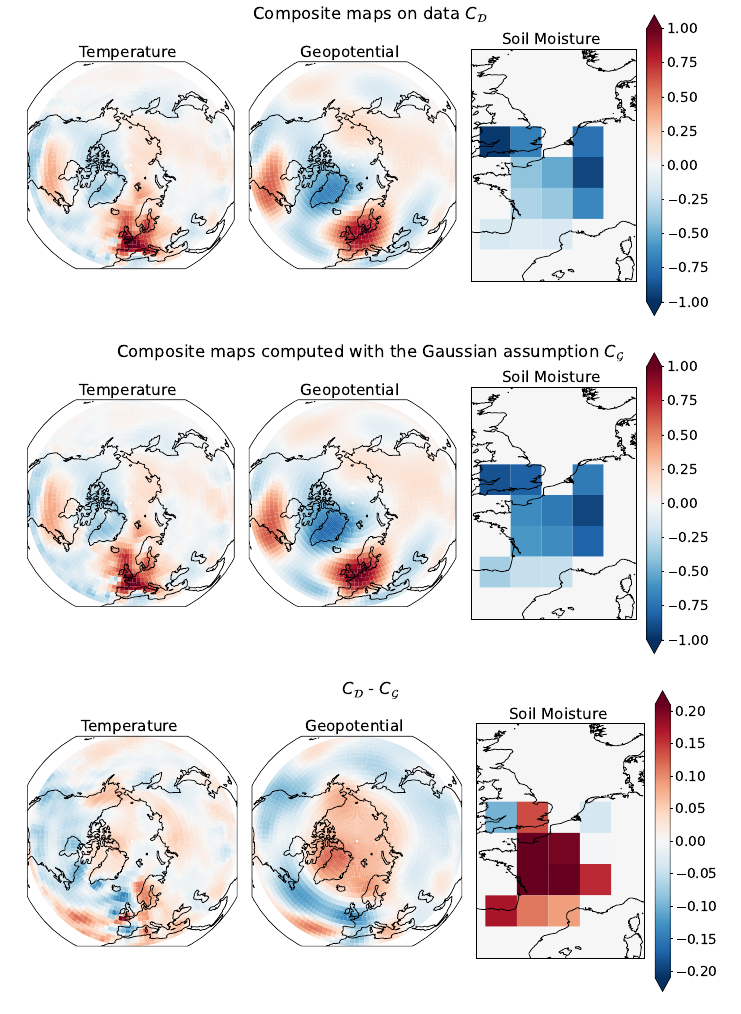}
        \caption{Composite maps of normalized \SI{2}{\meter} temperature, \SI{500}{\hecto\pascal} geopotential height, and soil moisture anomalies, conditioned on events with the 5\% most extremes 14-day temperature over France.   Composite maps are computed either directly from PlaSim data (first line), or under the Gaussian approximation (second line). The third line shows the difference between the two. The salient features of both temperature and geopotential are well captured by the Gaussian approximation, with errors of the order of 20\% at most.}
        \label{fig:composite}
    \end{figure}

    Composite maps $C$ are averages of the predictors $X$ conditioned on the occurrence of a heatwave: $C = \mathbb{E}[X(t-\tau)|A(t)\geq a]$ (see \cref{subsec:composite}).
        We first estimate this conditional expectation as an empirical average $C_\mathcal{D} = \frac{1}{N}\sum_{\mu=1}^N x_\mu $, where $\{x_\mu\}_{\mu=1}^N = \{X(t-\tau) | A(t) \geq a\}$.
    \Cref{fig:composite} shows the empirical composite maps for the three predictors $X$ (top row). We observe a positive anomaly of both \SI{2}{\meter} temperature and \SI{500}{\hecto\pascal} geopotential height over France and Western Europe, which is expected since we are conditioning over events that are happening over the French region. In the PlaSim grid, France is identified as the 12 pixels shown for the soil moisture field.
    Soil moisture anomaly displays negative values, as the soil tends to be drier than usual when heatwaves happen. In the rest of the Northern Hemisphere, we see teleconnection patterns in the temperature and geopotential height field, in particular a cyclone over Greenland and an anticyclone over the Central and Eastern United States.

    All these important features are also visible in the composite map $C_{\mathcal{G}}$ computed with the Gaussian approximation (using \cref{eq:composite_gauss}), represented in \cref{fig:composite} (middle row), to the point that the only visible discrepancy with the empirical map is slightly darker shades of soil moisture.
    Indeed, if we take the difference between the two estimates of the composite (\cref{fig:composite}, bottom row), most of the weight is concentrated on the soil moisture field. However, non-trivial patterns are also visible in the temperature and geopotential fields. The latter, in particular, shows a wave zero pattern, with positive values around the polar region and negative ones in the mid-latitudes.
    The amplitude of the difference, read on the color bar, is on the order of 20\% of the amplitude of the composite.
    To have a more quantitative measure, we compute the ratio $\mathcal{R}$ between the L2 norms of the difference between the two composites and the empirical one:
    \begin{equation}\label{eq:norm_ratio}
      \mathcal{R} = \frac{|C_\mathcal{D} - C_\mathcal{G}|}{|C_\mathcal{D}|}.
    \end{equation}
    In evaluating the norms, we took into account that we consider grid-cells of different areas.
    For the parameters considered in this section, the norm ratio is $\mathcal{R} = 0.21$, in agreement with our visual estimate (in \cref{sec:physics} we will investigate how this metric varies with the heatwave duration $T$ and the lead time $\tau$).

    In the next section, we analyze in more detail the sources of the difference between the two estimates.
    We will then give an explanation of the striking independence of the pattern from the extreme event threshold $a$ in \cref{subsec:comp_gauss_ind_a}.

    \subsection{Quantification of the Quality of the Gaussian Approximation for Composite Maps of Extreme Heatwaves}
    \label{subsec:error_composite}
        In the previous section, we showed that the empirical composite map $C_{\mathcal{D}}$ and the Gaussian composite map $C_{\mathcal{G}}$ differ at most by 20\% (\cref{fig:composite}, bottom row).
        A natural interpretation of this difference is that it is an error due to the fact that the Gaussian assumption is not exactly satisfied, and therefore the Gaussian composite map is only an approximation of the true composite map.
        Indeed, we can investigate the validity of this assumption by visualizing the joint and marginal distributions of the heatwave amplitude $A$ and the predictors at the grid-point level, for regions of low or high error (see
        Supporting Information S9).
        For instance, we show in
        fig. S3 in Supporting Information S9
        that the assumption is poorly satisfied for soil moisture at a grid point over France, where the error is large, while it is a much better assumption for geopotential over Greenland, where the error is small.

        However, another source of discrepancy between the two composites is the sampling error affecting $C_{\mathcal{D}}$ due to the limited number of heatwaves in the dataset over which we perform the empirical average.
        Indeed, if we focus on a single pixel $i$, and call $\{x_\mu\}_{\mu=1}^N = \{X^i(t-\tau) | A(t) \geq a\}$ the subset of heatwave events, the central limit theorem tells us that
        \begin{equation}
          \label{eq:central_limit_theorem}
          \sqrt{N_\mathrm{eff}} \frac{C^i - C_\mathcal{D}^i}{\sigma(C_\mathcal{D}^i)} \xrightarrow[N \to \infty]{} \mathcal{N}(0,1),
        \end{equation}
        where $C^i$ is the true composite, $C_\mathcal{D}^i = \frac{1}{N}\sum_{\mu=1}^N x_\mu$ is the empirical one, $\sigma(C_\mathcal{D}^i) = \sqrt{\frac{1}{N}\sum_{\mu=1}^N(x_\mu - C_\mathcal{D}^i)^2}$ is the standard deviation of the heatwave set and $N_\mathrm{eff}$ is the number of \emph{effectively} independent heatwaves. If all the $x_\mu$ were actually independent, we would have $N_\mathrm{eff} = N$, but from our definition of heatwave (\cref{eq:heatwave}), it is very likely that a series of consecutive days will be all heatwave events, and thus far from independent. In this paper we decide to fix $N_\mathrm{eff}$ to the number of years with at least one heatwave (equals to 2627 years for  $5 \%$ most extreme heatwaves of duration $T=14$ days and lead time $\tau = 0$). The motivation beside this choice can be found in
        Supporting Information S8.

        \Cref{eq:central_limit_theorem} tells us, then, that the distance between the empirical composite and the true one at grid point $i$ will be of the order of $\frac{\sigma(C_\mathcal{D}^i)}{\sqrt{N_\mathrm{eff}}}$, and thus if the Gaussian composite $C_\mathcal{G}^i$ falls much farther than $\frac{\sigma(C_\mathcal{D}^i)}{\sqrt{N_\mathrm{eff}}}$ from the empirical one, we can safely say that it is also far from the true composite.
        In other words, to quantify the statistical significance of the error we make, we can define a grid-point-wise test statistic as
        \begin{equation}
          \label{eq:significance}
          s_i = \frac{\sqrt{N_\mathrm{eff}}|C_\mathcal{G}^i- C_\mathcal{D}^i| }{\sigma(C_\mathcal{D}^i)}.
        \end{equation}
        To obtain a global metric for the whole composite map, we can consider the fraction of area $\mathcal{F}$ where this test statistic is above $2$, which conventionally corresponds to a 95\% exceedance probability. This allows us to say that a fraction $\mathcal{F}$ of the region of interest has, with 95\% confidence, a systematic error, not explainable by the finite size effect of the empirical composite.
        For the parameters studied here, we obtain the value $\mathcal{F}=0.37$. In \cref{sec:physics} we will investigate how this metric varies with the heatwave duration $T$ and the lead time $\tau$.

        This allows us to conclude that the Gaussian composite suffers from a statistically significant error over roughly a third of the domain.
        In spite of this, it gives a reasonable approximation of the empirical composite, within an error of order 20\%.
        However, having 8000 years of data to work with is not common in the climate community, especially when working with observational data or complex model simulations, and we can expect that when data is scarce, the error due to the Gaussian approximation becomes smaller than the sampling error in the empirical composite.
        In \cref{subsec:generalization_power} we will address this point on the dataset length and identify a regime where the Gaussian composite gives a better estimation of the true one than the empirical composite.

        We acknowledge that our analysis of the statistical significance does not take into account the correlation between pixel values. Thus, while $\mathcal{F}$ is defined rigorously, it cannot be interpreted as a global significance test for the whole composite map. Estimating this would require a more precise analysis, for instance by comparison with simulated generative statistical models for spatially smooth composite maps. However, this goes beyond the scope of this manuscript and thus will be left for further future studies.

    \subsection{Composite Maps do not Depend Much on the Extreme Event Threshold}
    \label{subsec:comp_gauss_ind_a}
        This section aims firstly at giving an explanation for the striking independence of composite maps pattern from the threshold $a$.
        Secondly, we show how the norm of the empirical composite maps scales with the threshold $a$ and that this scaling is very close to the one predicted from the Gaussian composite.

        In \cref{subsec:gaussian_composite} we explained that the composite map pattern does not depend on the extreme event threshold $a$. The independence of the pattern of the empirical composite maps from the threshold $a$ is explained by the Gaussian composite, \cref{eq:composite_gauss}.
        In this equation we see that the threshold intervenes only in the scaling of the pattern and not on the structure of the pattern itself, which is precisely what we observe in the estimated composite maps.
        Indeed, in 
        fig. S4 in Supporting Information S10
        we show the difference between the empirical composite and the Gaussian one (evaluated using \cref{eq:composite_gauss}) for the three fields, namely \SI{2}{\meter} air temperature anomaly, \SI{500}{\hecto\pascal} geopotential height anomaly and soil moisture anomaly evaluated for $a$ corresponding to the $1\%$ most extreme heatwaves for the PlaSim dataset.
        As predicted by the theory, the observed \SI{500}{\hecto\pascal} geopotential height pattern is the same as the one from \cref{fig:composite}.

\begin{figure}[h]
            \centering
            \includegraphics[width=0.9\textwidth]{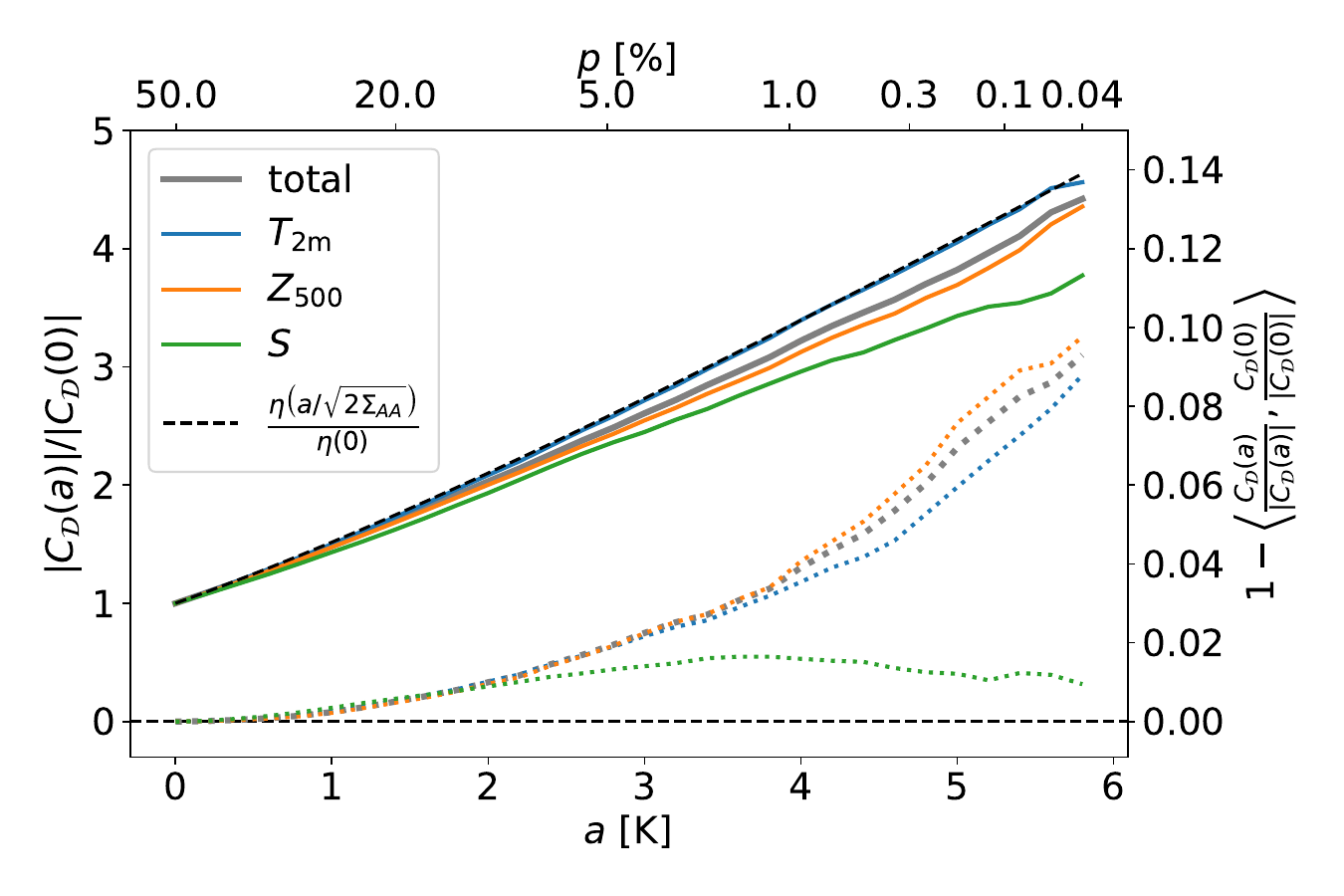}
            \caption{Normalized norm (solid lines, left y-axis) and misalignment (dotted lines, right y-axis) of the empirical composite map as function of $a$, the threshold to define a heatwave. The thick gray lines refer to the full composite map, while the thin colored ones to each of three weather fields individually. The normalization and the misalignment are computed with respect to the empirical composite evaluated at $a=0$. The dashed black line represents the theoretical scaling of the composite maps using the Gaussian approximation, see \cref{eq:composite_gauss}. The scaling of the empirical composite (gray curve) is not far from the Gaussian one (black dashed line) and it is quite precise for the 2 meter air temperature field (blue curve). Similarly, the misalignment is at most 10\%, and that only for very extreme heatwaves.
            The bottom x-axis $a$ is the threshold value used to define a heatwave event from the distribution of the temperature anomaly over France, $A$, for 14-day heatwaves. On the top x-axis, $p$ is the exceedance probability corresponding to a given $a$.}
            \label{fig:scaling}
        \end{figure}

        To make this discussion more quantitative, we computed the \emph{misalignment} between the empirical composite evaluated at variable threshold $a$, $C_\mathcal{D}(a)$, and a reference composite $C^0_\mathcal{D} = C_\mathcal{D}(a=0)$. The misalignment is then defined as $1 - \frac{C_\mathcal{D}(a)}{|C_\mathcal{D}(a)|} \cdot \frac{C_\mathcal{D}^0}{|C_\mathcal{D}^0|}$, which has value 0 if $C_\mathcal{D}(a)$ is proportional to $C^0_\mathcal{D}$, and 1 if $C_\mathcal{D}(a)$ and $C_\mathcal{D}^0$ are orthogonal.
        We plotted it as the dotted curves of \cref{fig:scaling}, where the thick gray line considers all three weather fields at once, while the colored lines show the misalignment for each field separately. As we can see, the hypothesis that the composite maps do not depend on the extreme event threshold $a$ is well satisfied, with a maximum misalignment of 10\% reached for the 0.04\% most extreme heatwaves. Interestingly, if we look only at the soil moisture field, the misalignment is much smaller, never exceeding 2\%. This can be explained by the fact that soil moisture has only 12 pixels, while the other two fields have $22\times 128 = 2816$, leading to more ways of being misaligned.

        Now that we verified that the composite maps of more extreme events are scaled versions of less extreme ones, we can question whether this scaling is the one predicted by the Gaussian approximation.
        To do so, we can plot the norm of the empirical composite map $C_\mathcal{D}(a)$ normalized by the norm of $C_\mathcal{D}(a=0)$, which, if the Gaussian hypothesis holds, should follow $\eta \left( a/\sqrt{2\Sigma_{AA}} \right)/\eta(0)$.
        Looking at the solid lines in \cref{fig:scaling}, we see that the normalized norm closely follows the prediction from the Gaussian approximation. In this case, soil moisture shows the worst performance, due to its highly non-Gaussian distribution (see
        fig. S3 in Supporting Information S9).

        Finally, we can give an overall metric of goodness of the Gaussian approximation by using the norm ratio defined in \cref{eq:norm_ratio}.
        In
        fig. S5 in Supporting Information S10,
        we provide its values as a function of the heatwave threshold $a$, which again shows a monotonic increase of the error for increasing values of $a$, with soil moisture showing the worst performance. For the 5\% most extreme heatwaves, the error measured by the norm ratio is around 20\%, which is in agreement with the qualitative observation of \cref{fig:composite}.

    \subsection{Effect of Dataset Length on Estimation of Composite Maps}
    \label{subsec:generalization_power}
        This section aims at motivating the usage of the Gaussian composite when the estimation of the true composite is highly affected by  sampling issues, i.e. when we are in a regime of scarcity of data.
        For datasets of length 200 years, the same order of magnitude as ERA5, the Gaussian composite performs much better than the empirical one, for events more extreme than 5\%.

        Firstly, we use the empirical composite $C_\mathcal{D}$ computed on the whole 8000 years dataset as an estimate of the true composite. Then we take a subset $\mathcal{P}$ of our data and compute over it the empirical composite $C_\mathcal{P}$ and the Gaussian one $C_\mathcal{G}^\mathcal{P}$. 

\begin{figure}
            \centering
            \includegraphics[width=0.8\textwidth]{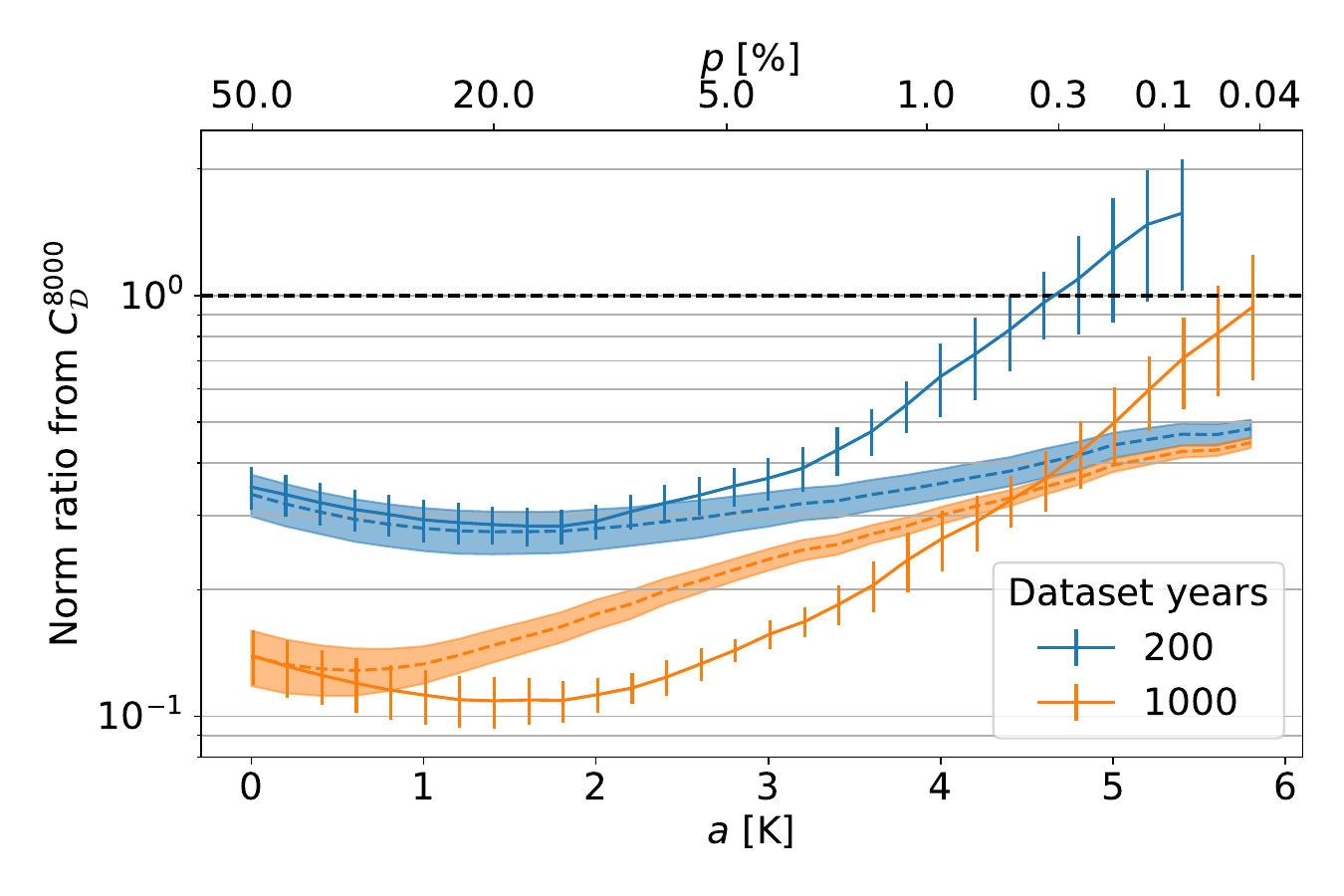}
            \caption{Norm of the relative error of the conditional average (composite map) using  the Gaussian approximation (dashed lines with shading) and of the composite maps using only a part of the full PlaSim dataset (solid lines with error bars).
            The error is relative to the empirical conditional average (composite map) evaluated over the full PlaSim dataset of 8000 years. In blue the 200-year-long datasets and in orange the 1000-year-long ones.
            Shading or error bars indicate the one standard deviation spread obtained from 8 independent batches of either 200 or 1000 years. On the bottom x-axis, $a$ is the threshold value used to define a heatwave event from the distribution of the \SI{2}{\meter} temperature anomaly over France, $A$.
            On the top x-axis, $p$ is the exceedance probability corresponding to a given $a$. The lower its value, the higher the value of the threshold $a$, the more extreme the heatwaves considered.
            The relative error for dataset of 1000 years is always lower than the one obtained for 200 years simply because of higher amount of available data. However, the effect of the amount of data differs for the two methods. For the empirical composite, the gap between the relative error obtained with 200 years of data (solid blue curve) and 1000 years (solid orange curve) remains essentially constant as $a$ increases. For  the Gaussian approximation (dashed blue and orange curves), instead it decreases, almost vanishing for very extreme events.
            All the curves show an increase in the relative error as $a$ increases due to the lack of data.
            When we are in this regime, the relative error obtained with the Gaussian composite is lower than the one obtained with the empirical composite. This happens for a $p$ value of around $0.2\%$ for 1000-year-long datasets, and of $5\%$ for 200-year-long datasets. For less extreme events, the Gaussian composite performs worse or similarly to the empirical one.}
            \label{fig:generalization_power}
        \end{figure}

        In \cref{fig:generalization_power} we see the values of the empirical norm ratio $\mathcal{R}_\mathcal{P} = \frac{|C_\mathcal{P} - C_\mathcal{D}|}{|C_\mathcal{D}|}$ (solid lines) and the Gaussian one $\mathcal{R}_\mathcal{G} = \frac{|C_\mathcal{G}^\mathcal{P} - C_\mathcal{D}|}{|C_\mathcal{D}|}$ (dashed lines), for datasets $\mathcal{P}$ of different lengths. To get confidence intervals, we repeat the experiment 8 times for each dataset length, with 8 independent batches of data.

        The Gaussian composites over 1000 years and over 200 years of data show an almost monotonic increase as function of the heatwave threshold $a$. The latter shows a plateau for values of $p$ ranging from $50 \%$ to $1\%$ (with a non-significant minimum at around $a=\SI{1.5}{\kelvin}$), meaning that the error made for typical events is comparable to the one made on fairly extreme ones. This is not valid for the composite over 1000 years as there is a constant and more rapid worsening of the Gaussian norm ratio.
        It is interesting to notice that in the very tail of the distribution of $A$, i.e. for small values of $p$, we achieve very similar values of the norm ratio in both datasets. The spread of the norm ratio among the batches is more pronounced for less extreme events than for the most extreme ones. In the case of the Gaussian composite, the main source of error is systematic, as we use the full dataset $\mathcal{P}$ to evaluate the Gaussian composite and not a small subset which depends on the threshold (\cref{eq:composite_gauss}).

        The empirical composite norm ratio for 200 years of data stays almost constant until $p = 5\%$ (with a shallow minimum around $a=\SI{1.5}{\kelvin}$ similar to the Gaussian composite), after which it starts increasing both in the mean and in the spread of data. For the empirical composite norm ratio over 1000 years we see a less evident constant behavior and a more pronounced minimum of the norm ratio, both in the mean and in the standard deviation. Similar to the 200 years line, there is a worsening of the norm ratio as $a$ increases. In fact, there is an almost constant gap between the two solid lines. For very extreme events, the norm ratio of the empirical composites is much higher than their Gaussian counterparts, showing that in this regime the Gaussian approximation provides a much better estimate of the true composite map than the empirical composite.

        Focusing on both composites estimated for 200 years datasets, until $p=5\%$ both the empirical and the Gaussian have the same values of the norm ratio. For more extreme events, the norm ratio of the empirical one increases drastically, mostly due to the more and more limited data available in the tail, reaching 100\% of error at the $0.3\%$ most extreme heatwaves. This is not the case for the Gaussian approximation, whose values of the norm ratios still increase but much more slowly.

        Indeed, when we are in a regime of scarcity of data, which naturally arises when one wants to study very extreme heatwaves, calculating composite maps using empirical data poses a sampling issue. Our methodology overcomes this issue by relying on an estimate of the composite map which uses the whole dataset.
        To confirm this, we see that on longer datasets, such as the 1000-year-long one, where we already have a sufficient amount of data to have a good estimate of the empirical composite, the Gaussian approximation is not a better estimate than computing the composite directly. This holds only for events up to $a=\SI{4.5}{\kelvin}$, after which, due to the sampling issue, the Gaussian estimation performs, once again, better than the empirical composite.

\vspace{0.5cm}
In this section we investigated in detail the effect of changing the heatwave threshold $a$, showing that empirical composites of very extreme heatwaves are rescaled versions of milder events. Moreover, the Gaussian approximation is able to capture the correct scaling and can thus be used as a valuable surrogate for the empirical composite map when the latter suffers from sampling issues, either due to short datasets or when dealing with very extreme events.
So far we focused on the composite maps of 14-day heatwaves at lead time $\tau=0$; later we will expand our study to other lead times, and also to other heatwave durations.


\section{Validation of the Gaussian Approximation for Computing Committor Functions on Climate Datasets}
\label{sec:committor}

In \cref{sec:bayes_committor_composite_gaussianframe} we defined committor functions and optimal projection patterns, both generally and within the Gaussian approximation. In this section, we apply the Gaussian approximation of the committor on climate data, the PlaSim dataset described in \ref{sec:data:plasim}, and compare its skill with the prediction from a neural network.
We then proceed to study the optimal projection pattern, which is given by \cref{eq:M}. However, we will see in this section that the mathematical expression \cref{eq:M}, is not directly applicable to high dimensional climate data, where the datasets are usually too short.
Indeed, in \cref{sec:q:regularization} we show that regularization is necessary to have physically meaningful projection patterns.
In \cref{sec:q:less-data,sec:q:more-extreme} we will show the effect of lack of data on the performance. In the first case lack of data will come from reduced dataset lengths, and in the second from more extreme events.

We illustrate this for the task of predicting heatwaves, but we assume it will generalize well to other prediction problems in climate.

\subsection{Skill of the Gaussian Approximation Compared to Prediction with Neural Networks}
\label{sec:q:nonregularized-ga}
    We first apply the Gaussian approximation of the committor, defined in \cref{eq:q-gaus}, to the forecast of the 5\% most extreme two week heatwaves ($T=14$), predicted at lead time $\tau = 0$, using the full PlaSim dataset.
    To have a robust estimate of the performance of our method, we repeat the experiment 10 times in a k-fold cross validation process (see
    Supporting Information S2).
    Doing so we get an average validation normalized log score of $0.455 \pm 0.010$.
    We can say that the score is much better than the climatology ($S = 0$), but it is very tricky to quantify the maximum achievable score, as $S = 1$ is absolutely unrealistic due to the chaotic nature of the climate system.

    However, we can compare to other methods, for instance the prediction using a deep convolutional neural network \cite{miloshevichProbabilisticForecastsExtreme2023}. This network takes as input the stack of predictors and produces an estimate of the committor. It is trained on a probabilistic binary classification of the labels $Y$, i.e. it directly minimizes the loss $\mathcal{L}$ defined in \cref{eq:loss}. More details about the network's architecture can be found in \citeA{miloshevichProbabilisticForecastsExtreme2023}.
    Such a network yields a validation score of $S_{CNN} = 0.465 \pm 0.007$.

    This is a remarkable result, as the Gaussian approximation is much simpler than a deep neural network, but is able to achieve a result that is only 2\% (or less than a standard deviation) worse.

\subsection{Regularization of the Projection Pattern}
\label{sec:q:regularization}
    The simplicity of the Gaussian committor comes with the added benefit of being an interpretable forecast, as we can look at the projection pattern $M$ to obtain some insight into the dynamics leading to a heatwave.

\begin{figure}[t]
        \centering
        \includegraphics[width=0.8\textwidth]{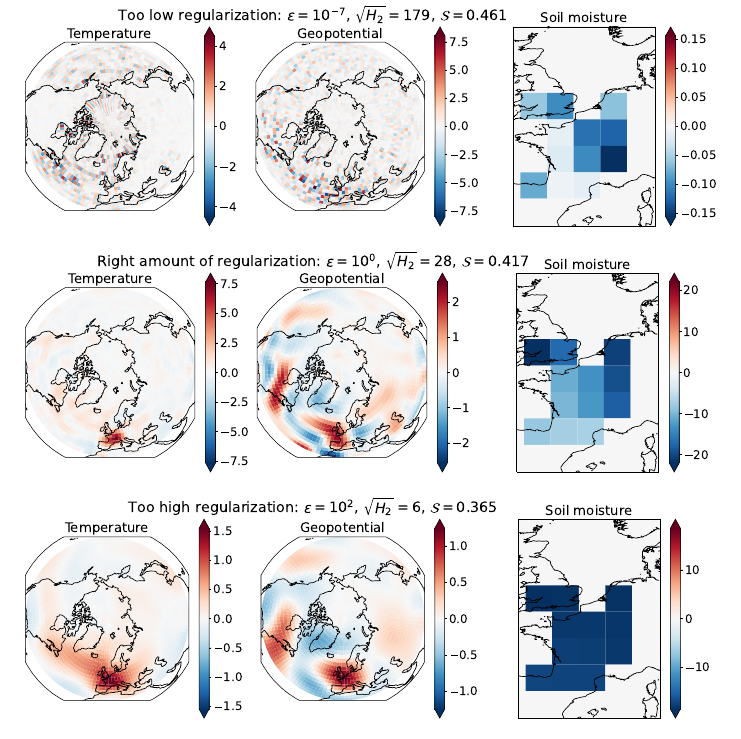}
        \caption{Choice of regularization for optimal physical content of the projection map $M$, using PlaSim data. Each line features the projection map $M$ at different values of the regularization coefficient $\epsilon$. Each map $M$ is represented as its three field components: \SI{2}{\meter} air temperature, geopotential height at \SI{500}{\hecto\pascal} and soil moisture anomalies (trained on 7200 years of data, for one of the 10 folds). On top of the figures we report the values of $\epsilon$, of the norm of the gradient $\sqrt{H_2}$ and of the normalized log score $\mathcal{S}$. The intermediate value, $\epsilon = 1$, is the best compromise with a very high predictive skill and an excellent readability of the physical fields.}
        \label{fig:Ms}
    \end{figure}

    Unfortunately, a direct plot of $M$ looks like the first row of \cref{fig:Ms}, from which we cannot extract any meaningful information as no well-defined patterns emerge.
    This is due to the fact that the covariance matrix $\Sigma_{XX}$ is very high dimensional ($d^2 \sim 10^7)$ and is estimated with a relatively low number of data points ($8000\times0.9\times (90 - T + 1) \sim 10^6$).
    Hence, it will be nearly singular, causing problems when we compute the inverse in \cref{eq:M}.

    A simple solution is the standard Tikhonov regularization, that corresponds to adding an $L_2$ penalty to the minimization problem:
    \begin{equation}\label{eq:l2-reg}
        M_\epsilon \propto (\Sigma_{XX} + \epsilon \mathbb{I})^{-1} \Sigma_{XA} = \arg \min_M \left( (A - M^\top X)^2 + \epsilon |M|^2 \right) ,
    \end{equation}
    where $\mathbb{I}$ is the identity matrix.

    However, in our case we can better enforce interpretability of the pattern $M$ by requiring it to be spatially smooth. Namely, we will penalize the squared norm of the spatial gradient, $H_2$, that we can compute as the weighted sum of the square differences between values of adjacent pixels in the map $M$. We can then write $H_2(M) = M^\top W M$
    (see Supporting Information S6
    for the exact formula of matrix $W$), and hence the regularized pattern will be
    \begin{equation}\label{eq:h2-reg}
        M_\epsilon \propto (\Sigma_{XX} + \epsilon W)^{-1} \Sigma_{XA} = \arg \min_M \left( (A - M^\top X)^2 + \epsilon H_2(M) \right).
    \end{equation}
    Note that if we tweak the projection pattern $M$, we should also update the formulas for the coefficients $\alpha$ and $\beta$ in \cref{eq:ab}. This is relatively straightforward and is discussed in
    Supporting Information S7.

    Varying $\epsilon$ yields the different maps shown in \cref{fig:Ms}, where indeed we see that the regularization makes the patterns progressively smoother. Unsurprisingly, we note that a higher regularization comes at the price of a lower skill score $S$ (see also \cref{tab:gacnn-yr_eps}).
    It is then up to the user to decide what is a good compromise between performance and interpretability of the pattern. In our case, we argue that the best pattern is the one in the center row of \cref{fig:Ms} ($\epsilon = 1$), as it is smooth enough that we can see some clear structures in the \SI{500}{\hecto \pascal} geopotential height field, while a higher regularization does not improve its physical understanding.
    At this value of the regularization coefficient, the average validation score is $0.418 \pm 0.006$: three standard deviations or 8\% worse than the non-regularized case, and five standard deviations or 10\% worse than the neural network.

    It is important to point out that after proper regularization the skill of the prediction is still much better than climatology, while providing  physical insight on the dynamics leading to heatwaves. This latter point is further discussed in \cref{sec:physics}.

\subsection{Performance on Smaller Datasets}
\label{sec:q:less-data}
    So far we have applied the Gaussian approximation to an extremely long 8000 year dataset. Such datasets are uncommon in the climate community, especially when dealing with observations or high resolution simulations. To study the effect of the amount of data on the performance of our method, we apply it to gradually smaller and smaller subsets of our climate model output.

    In the left panel of \cref{tab:gacnn-yr_eps}, we can see the behavior of the normalized log score $S$ of the Gaussian committor, as a function of the regularization coefficient and the size of the training set. The first important thing to notice is that the score is not very sensitive to the amount of training data, showing that our method is well suited also for small datasets. By looking at the dependence with respect to $\epsilon$, we see that when we have a lot of data, a stronger regularization means a poorer prediction skill. On the other hand for small datasets the best performance is achieved at a finite value of $\epsilon$. This can be explained by the fact that as we have less and less data to estimate a constant size covariance matrix, it will become more and more singular, thus requiring a stronger regularization (see
    fig. S8 in Supporting Information).
    Also, a smoother pattern is more likely to generalize well when training and validation data are very small.

    In any case, we remind the reader that choosing the proper regularization coefficient is not just a matter of score, but also of physical interpretability of the projection pattern, as explained in the previous section. From a qualitative look at projection maps at different values of $T$, $\tau$ and $\epsilon$, $\epsilon = 1$ seemed to be a universally good compromise for the PlaSim dataset. Hence, if not specified differently, in the remainder of this work we will always consider $\epsilon = 1$.

    On the right panel of \cref{tab:gacnn-yr_eps} we see the comparison with the skill of the neural network in the form $1 - \mathcal{S}/\mathcal{S}_{CNN}$, which shows that as the dataset gets smaller, the CNN loses its advantage, being outperformed when crossing the 1000 years threshold. An important caveat here is that the many hyperparameters of the CNN where optimized for the biggest dataset \cite{miloshevichProbabilisticForecastsExtreme2023}, and then kept constant for the experiments when training on less data. This potentially makes the comparison between the neural network and our method not completely fair. In fact, some experiments (not shown in this work), suggest that by optimizing hyperparameters such as the learning rate and batch size used for training the neural network allow it to prevail even when training only on 450 years of data. The Gaussian approximation, however, is still better when working with 200 years or less, even considering the optimization. So, the qualitative behavior displayed in \cref{tab:gacnn-yr_eps} still holds, and can be ultimately attributed to the higher complexity of the CNN (roughly a million parameters) with respect to the Gaussian approximation (roughly a few thousands of parameters).

    \begin{table}[h]
        \centering
        \begin{tabular}{c|c|c|c|c|c|c|}
            \multicolumn{2}{c}{} & \multicolumn{5}{c}{Normalized log score} \\
            \cline{3-7}
            \multicolumn{2}{c|}{} & \multicolumn{5}{c|}{$\epsilon$} \\
            \cline{3-7}
            \multicolumn{2}{c|}{} & $10^{-2}$ & $10^{-1}$ & $10^{0}$ & $10^{1}$ & $10^{2}$ \\
            \cline{2-7}
            \multirow{6}{*}{\rotatebox[origin=c]{90}{years of training}}
             & 7200 & \cellcolor[rgb]{0.96470588,0.98235294,0.40000000}0.43 & \cellcolor[rgb]{0.88235294,0.94117647,0.40000000}0.43 & \cellcolor[rgb]{0.75294118,0.87647059,0.40000000}0.42 & \cellcolor[rgb]{0.51764706,0.75882353,0.40000000}0.40 & \cellcolor[rgb]{0.16862745,0.58431373,0.40000000}0.37 \\
            \cline{2-7}
             & 3600 & \cellcolor[rgb]{0.89803922,0.94901961,0.40000000}0.43 & \cellcolor[rgb]{0.81176471,0.90588235,0.40000000}0.42 & \cellcolor[rgb]{0.68235294,0.84117647,0.40000000}0.41 & \cellcolor[rgb]{0.45098039,0.72549020,0.40000000}0.39 & \cellcolor[rgb]{0.11372549,0.55686275,0.40000000}0.37 \\
            \cline{2-7}
             & 1800 & \cellcolor[rgb]{0.83921569,0.91960784,0.40000000}0.42 & \cellcolor[rgb]{0.76862745,0.88431373,0.40000000}0.42 & \cellcolor[rgb]{0.64313725,0.82156863,0.40000000}0.41 & \cellcolor[rgb]{0.40784314,0.70392157,0.40000000}0.39 & \cellcolor[rgb]{0.05490196,0.52745098,0.40000000}0.36 \\
            \cline{2-7}
             & 900 & \cellcolor[rgb]{1.00000000,1.00000000,0.40000000}0.44 & \cellcolor[rgb]{0.96470588,0.98235294,0.40000000}0.43 & \cellcolor[rgb]{0.86274510,0.93137255,0.40000000}0.43 & \cellcolor[rgb]{0.63921569,0.81960784,0.40000000}0.41 & \cellcolor[rgb]{0.28235294,0.64117647,0.40000000}0.38 \\
            \cline{2-7}
             & 450 & \cellcolor[rgb]{0.85490196,0.92745098,0.40000000}0.43 & \cellcolor[rgb]{0.83921569,0.91960784,0.40000000}0.42 & \cellcolor[rgb]{0.74901961,0.87450980,0.40000000}0.42 & \cellcolor[rgb]{0.52549020,0.76274510,0.40000000}0.40 & \cellcolor[rgb]{0.18823529,0.59411765,0.40000000}0.37 \\
            \cline{2-7}
             & 180 & \cellcolor[rgb]{0.00000000,0.50000000,0.40000000}0.36 & \cellcolor[rgb]{0.29019608,0.64509804,0.40000000}0.38 & \cellcolor[rgb]{0.39607843,0.69803922,0.40000000}0.39 & \cellcolor[rgb]{0.35294118,0.67647059,0.40000000}0.39 & \cellcolor[rgb]{0.15294118,0.57647059,0.40000000}0.37 \\
            \cline{2-7}
        \end{tabular}
        \begin{tabular}{c|c|c|c|c|c|}
            \multicolumn{1}{c}{} & \multicolumn{5}{c}{$1 - \mathcal{S}/\mathcal{S}_{CNN}$} \\
            \cline{2-6}
             & \multicolumn{5}{c|}{$\epsilon$} \\
            \cline{2-6}
             & $10^{-2}$ & $10^{-1}$ & $10^{0}$ & $10^{1}$ & $10^{2}$ \\
            \cline{1-6}
              & \cellcolor[rgb]{0.91872357,0.83383314,0.62475971}0.07 & \cellcolor[rgb]{0.89042676,0.78708189,0.53863899}0.08 & \cellcolor[rgb]{0.83267974,0.67581699,0.38562092}0.10 & \cellcolor[rgb]{0.70196078,0.46159170,0.14417532}0.14 & \cellcolor[rgb]{0.42414456,0.24405998,0.02806613}0.20 \\
            \cline{1-6}
              & \cellcolor[rgb]{0.96378316,0.92179931,0.81084198}0.03 & \cellcolor[rgb]{0.95409458,0.89227220,0.73241061}0.05 & \cellcolor[rgb]{0.90811226,0.81630142,0.59246444}0.07 & \cellcolor[rgb]{0.80315263,0.61584006,0.31180315}0.11 & \cellcolor[rgb]{0.56862745,0.33610150,0.05267205}0.17 \\
            \cline{1-6}
              & \cellcolor[rgb]{0.96163014,0.94978854,0.91849289}0.01 & \cellcolor[rgb]{0.96270665,0.93579393,0.86466744}0.02 & \cellcolor[rgb]{0.96470588,0.90980392,0.76470588}0.04 & \cellcolor[rgb]{0.87204921,0.75578624,0.48404460}0.09 & \cellcolor[rgb]{0.65490196,0.41730104,0.11188005}0.15 \\
            \cline{1-6}
              & \cellcolor[rgb]{0.82283737,0.92779700,0.91280277}-0.03 & \cellcolor[rgb]{0.85113418,0.93456363,0.92264514}-0.03 & \cellcolor[rgb]{0.92895040,0.95317186,0.94971165}-0.01 & \cellcolor[rgb]{0.96378316,0.92179931,0.81084198}0.03 & \cellcolor[rgb]{0.84252211,0.69580930,0.41022684}0.10 \\
            \cline{1-6}
              & \cellcolor[rgb]{0.50742022,0.80615148,0.75963091}-0.09 & \cellcolor[rgb]{0.52925798,0.81507113,0.77070358}-0.08 & \cellcolor[rgb]{0.63844675,0.85966936,0.82606690}-0.07 & \cellcolor[rgb]{0.87235679,0.93963860,0.93002691}-0.02 & \cellcolor[rgb]{0.96116878,0.90396002,0.75394079}0.05 \\
            \cline{1-6}
              & \cellcolor[rgb]{0.96163014,0.94978854,0.91849289}0.01 & \cellcolor[rgb]{0.72579777,0.89534794,0.87035755}-0.05 & \cellcolor[rgb]{0.58385236,0.83737024,0.79838524}-0.07 & \cellcolor[rgb]{0.63844675,0.85966936,0.82606690}-0.07 & \cellcolor[rgb]{0.86528258,0.93794694,0.92756632}-0.02 \\
            \cline{1-6}
        \end{tabular}
        \caption{Left table: normalized log score of the Gaussian approximation (the higher the better), versus training dataset length and the regularization coefficient $\epsilon$. With small datasets, intermediate $\epsilon$ values are optimal, while vanishing ones are for large datasets. The apparent peak in performance for 900 years of training is not significant. Right table: comparison with the skill of the neural network ($\epsilon$ affects only the Gaussian approximation). Brown colors mean the CNN performs better, while blue hues mean the Gaussian approximation is better. When the neural network has much data to learn, it can leverage its expressivity potential to outperform the Gaussian approximation. With small datasets, the added complexity of neural networks is detrimental to its score. Both panels are based on PlaSim data.}
        \label{tab:gacnn-yr_eps}
    \end{table}

    Summarizing, our method is well suited to work in a regime of lack of data due to short datasets, where complex neural networks struggle.

\subsection{More Extreme Heatwaves}
\label{sec:q:more-extreme}
    A question complementary to the one of smaller datasets is the one of more extreme heatwaves, as they both result in very few samples of the event of interest.

    First, the Gaussian approximation provides a committor that depends on the heatwave threshold $a$ only through the parameter $\alpha$. This means that, similarly to the composite maps, the projection pattern $M$ will be the same for all heatwaves independently on how extreme they are. It is thus extremely easy and cheap to get a new committor estimate for a different value of $a$.

\begin{figure}[h]
        \centering
        \includegraphics[width=0.8\textwidth]{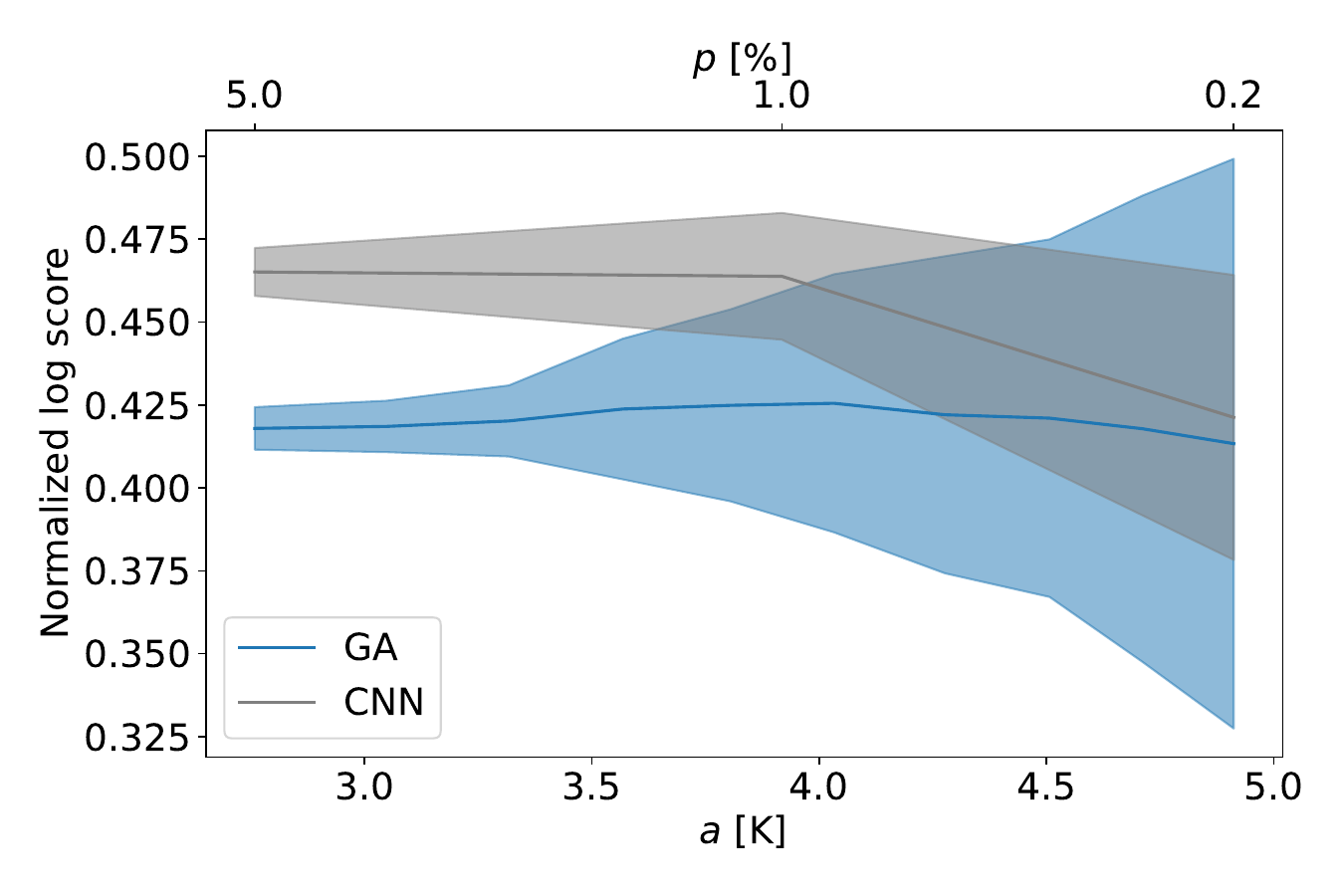}
        \caption{Normalized log score of the Gaussian approximation (blue) and CNN (gray) when varying the heatwave threshold $a$ (bottom horizontal axis) or, equivalently, its climatological probability $p$ (top horizontal axis). The solid line is the mean over the 10-fold cross validation process, while the shaded area represents one standard deviation. The experiment is performed with 7,200 years of training with PlaSim data. The regularization coefficient for the Gaussian approximation is kept at the optimal $\epsilon=1$ value. The CNN is always best but because of the lack of data for rare events, its relative skill decreases with $a$. The skill of the Gaussian approximation is not much sensitive to the rareness of the event.}
        \label{fig:Svpercent}
    \end{figure}

    On the other hand, since the neural network we consider is trained on a classification task, as we change $a$, the labels $Y(t)$ change as well, and hence the whole network needs to be retrained every time. Although transfer learning can reduce the computational cost and avoid retraining from scratch, it still a more complex task than computing a new Gaussian committor. Furthermore, as we focus on more and more extreme heatwaves, the imbalance between the $Y=0$ and $Y=1$ classes becomes more and more relevant, eventually hindering the performance of the network (see the gray error-band in \cref{fig:Svpercent}).
    On the contrary the smaller size of the heatwave class affects the performance of the Gaussian approximation only in its variance, while the mean normalized log score $\mathcal{S}$ has a very weak dependence on the amplitude of the heatwave (blue line in \cref{fig:Svpercent}). This, in turn, suggests that our Gaussian approximation is sufficient to capture well the relationship between the predictors and the heatwave amplitude $A$ even in the most extreme tails of the distribution.

\vspace{0.5cm}

In this section we showed that the Gaussian approximation can be a simple, but powerful, tool for the prediction of extreme heatwaves. Compared to other methods, such as deep neural networks, it does not need as much data to be properly trained. This makes it particularly suited for short datasets, which is typically the case in the climate community. This direction is further expanded in \cref{sec:ERA5results}, where we apply our method to ERA5.
Furthermore, and crucially, it is usually very hard to interpret the prediction performed by a deep neural network, while the Gaussian approximation, through the optimal projection pattern, is interpretable \emph{by design}. The study of the projection pattern opens the possibility for insight on the physical processes behind the event under study, and we expand on this in \cref{sec:physics}.


\section{Committor Function and Optimal Projection for Extreme Heatwaves}
\label{sec:physics}

In \cref{sec:composite,sec:committor} we computed composite maps and committor functions for extreme heatwaves. However, in these sections the focus was mainly methodological, with attention to performance and the technical details that influence it. In this section we complement the previous analysis by focusing instead on the physical insight that our method provides on extreme heatwaves. To do so we will compare composite maps and optimal projection patterns at different values of the heatwave duration $T$ and the lead time $\tau$.

\subsection{Comparison Between Composite Maps and Projection Patterns}
\begin{figure}[tbhp]
        \centering
        \includegraphics[width=1.\textwidth]{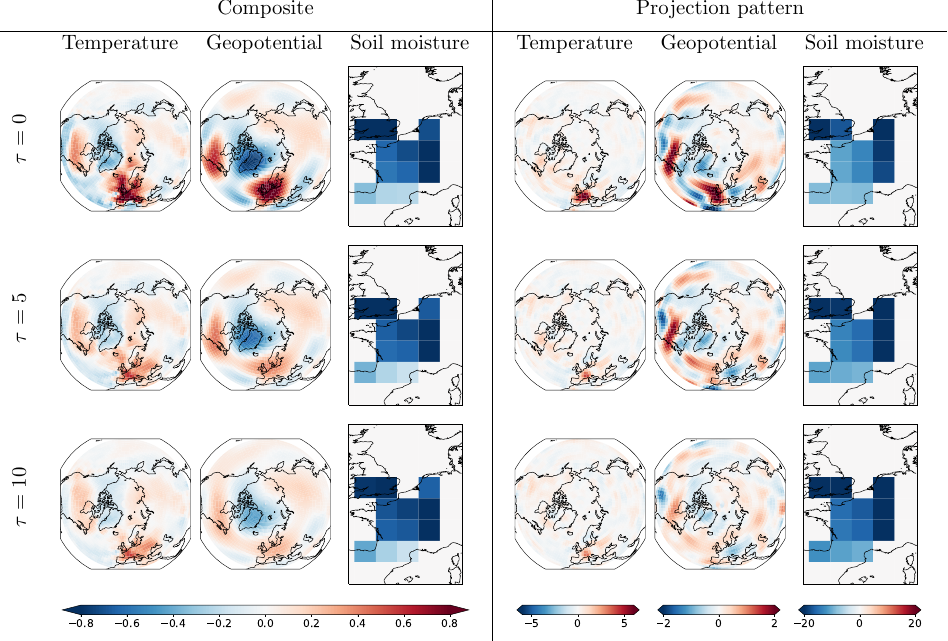}
        \caption{Left columns: Gaussian composite maps, temperature, geopotential height at 500hPa and soil moisture, for three different values of the lag time $\tau$. Right column: the optimal projection pattern for prediction, within the Gaussian approximation ($\epsilon=1$). As expected the two sets of maps are different, characterizing either a-posteriori statistics or best prediction patterns. The composite features hemispheric scale patterns dominated by zonal wave-number zero and zonal wave number three modes. For long lead times, the zonal wave-number zero pattern clearly dominates. The soil moisture composite pattern does not change much with the lag time. The information needed for making an optimal projection, as seen through the projection pattern, is at a finer scale, less global, with a strong meridional structure. Temperature contributes weakly and only through its local values to the projection pattern.}
        \label{tab:CvM}
    \end{figure}

    In \cref{sec:bayes_committor_composite_gaussianframe} we showed that a-priori and a-posteriori statistics are fundamentally different. Here we proceed to further include some physical reasoning that arises when comparing the two types of statistics.
    In \cref{tab:CvM} we have the side by side comparison, at different values of the lead time $\tau$, of the Gaussian composite map $C_\mathcal{G}$ with the projection pattern $M$ needed for the computation of the committor.
    As explained in \cref{subsec:committor_v_composite_examples}, the composite map captures the \emph{correlations} between the heatwave amplitude $A$ and the predictors $X$, while the committor, and thus the projection pattern $M$, focuses on what is really important for the \emph{prediction}.

    A clear example of this is the difference between the \SI{2}{\meter} temperature anomaly field in the composite map and in the projection pattern. From \cref{tab:CvM}, we can see that the composite shows many teleconnection features, for example over North America, while in the projection map virtually all the weight is over France.
    This shows that the information encoded in the temperature correlations is not actionable for prediction, regardless of the lead time, at least within the framework of the Gaussian approximation.
    Similarly, the \SI{500}{\hecto\pascal} geopotential height field anomaly shows a very strong anticyclone over Greenland in the composite maps, which is not present in the projection patterns.
    In the case of geopotential height, however, there remains part of the large-scale correlations which actually yield some predictive skill for the Gaussian committor, such as the positive anomaly over the storm tracks region.

    Another remarkable difference between $C_\mathcal{G}$ and $M$ is the relative magnitude of the fields.
    By looking at the colorbars at the bottom of the figure, we see that, in the composite, all the fields have roughly the same order of magnitude, and this makes sense as we work with normalized data and the composite is representative of the typical heatwave event. On the other hand, from the projection patterns we observe that the values of soil moisture are 4 to 10 times higher than the ones of temperature and geopotential, showing that the soil moisture anomaly field is far more important for prediction than one might assume by just looking at the composite.

    If we now focus on what happens when we change the lead time $\tau$, we see that in the composites there is essentially just a fading of the structure of the \SI{2}{\meter} temperature and \SI{500}{\hecto\pascal} geopotential height anomalies apparent at $\tau=0$, with some minor qualitative changes, such as the connection of the two high pressure systems over the Atlantic at $\tau=5$. On the other hand, the soil moisture anomaly component remains almost unchanged.
    This increased prominence of soil moisture as the lead time increases is even more pronounced for the projection pattern $M$, showing that soil moisture is the key factor for long term heatwave forecast.

    Finally, from the evolution of the projection map for the \SI{500}{\hecto\pascal} geopotential height field, we see a clear shift of focus from the North-eastern Atlantic at $\tau=0$ to the United States at $\tau=5$. At $\tau=10$ the most prominent feature in the \SI{500}{\hecto\pascal} geopotential height projection pattern is a small cyclone over the continental US, something which can barely be seen at all in the composite.
    These changes in the projection pattern give us insight into the dynamics of atmospheric circulation that leads to heatwaves over France, in particular the dynamics of the jet stream.

\subsection{Effects of Changing $T$ and $\tau$}
    In this section we analyze more quantitatively how the performance of the Gaussian approximation is affected by the heatwave duration $T$ and the lead time $\tau$, and what physical conclusions we can derive from it. We will first perform this sensitivity analysis on the composite maps (a-posteriori statistics) in \cref{subsubsec:comp_phys} and then for committor functions (a-priori statistics) in \cref{subsubsec:committor_phys}.

    \subsubsection{Composites}
    \label{subsubsec:comp_phys}

        \begin{table}
            \centering
            \begin{tabular}{c|c|c|c|c|c|c|c|c|c|c|c|c|}
                \multicolumn{2}{c}{} & \multicolumn{11}{c}{Fraction of area with error above $2\sigma$} \\
                \cline{3-13}
                \multicolumn{2}{c|}{} & \multicolumn{11}{c|}{$\tau$ [days]} \\
                \cline{3-13}
                \multicolumn{2}{c|}{} & 0 & 3 & 6 & 9 & 12 & 15 & 18 & 21 & 24 & 27 & 30 \\
                \cline{2-13}
                \multirow{5}{*}{\rotatebox[origin=c]{90}{$T$ [days]}}
                 & 1 &\cellcolor[rgb]{1.00000000,1.00000000,0.61397020}0.52 & \cellcolor[rgb]{1.00000000,1.00000000,0.49044067}0.50 & \cellcolor[rgb]{1.00000000,1.00000000,0.25882279}0.46 & \cellcolor[rgb]{1.00000000,0.75048993,0.00000000}0.34 & \cellcolor[rgb]{1.00000000,0.52401971,0.00000000}0.27 & \cellcolor[rgb]{1.00000000,0.21519669,0.00000000}0.19 & \cellcolor[rgb]{1.00000000,0.07107927,0.00000000}0.15 & \cellcolor[rgb]{0.83430299,0.00000000,0.00000000}0.08 & \cellcolor[rgb]{0.66958549,0.00000000,0.00000000}0.04 & \cellcolor[rgb]{0.59752158,0.00000000,0.00000000}0.02 & \cellcolor[rgb]{0.58722673,0.00000000,0.00000000}0.02 \\
                \cline{2-13}
                 & 3 &\cellcolor[rgb]{1.00000000,1.00000000,0.59852901}0.52 & \cellcolor[rgb]{1.00000000,1.00000000,0.16617564}0.44 & \cellcolor[rgb]{1.00000000,0.99754836,0.00000000}0.41 & \cellcolor[rgb]{1.00000000,0.55490202,0.00000000}0.28 & \cellcolor[rgb]{1.00000000,0.30784360,0.00000000}0.22 & \cellcolor[rgb]{1.00000000,0.04019697,0.00000000}0.14 & \cellcolor[rgb]{0.91666174,0.00000000,0.00000000}0.11 & \cellcolor[rgb]{0.71076486,0.00000000,0.00000000}0.05 & \cellcolor[rgb]{0.60781642,0.00000000,0.00000000}0.02 & \cellcolor[rgb]{0.56663705,0.00000000,0.00000000}0.01 & \cellcolor[rgb]{0.58722673,0.00000000,0.00000000}0.02 \\
                \cline{2-13}
                 & 7 &\cellcolor[rgb]{1.00000000,1.00000000,0.24338160}0.45 & \cellcolor[rgb]{1.00000000,1.00000000,0.01176372}0.41 & \cellcolor[rgb]{1.00000000,0.74019583,0.00000000}0.34 & \cellcolor[rgb]{1.00000000,0.46225511,0.00000000}0.26 & \cellcolor[rgb]{1.00000000,0.18431438,0.00000000}0.18 & \cellcolor[rgb]{1.00000000,0.06078517,0.00000000}0.15 & \cellcolor[rgb]{0.88577721,0.00000000,0.00000000}0.10 & \cellcolor[rgb]{0.72105971,0.00000000,0.00000000}0.05 & \cellcolor[rgb]{0.62840611,0.00000000,0.00000000}0.03 & \cellcolor[rgb]{0.61811127,0.00000000,0.00000000}0.02 & \cellcolor[rgb]{0.56663705,0.00000000,0.00000000}0.01 \\
                \cline{2-13}
                 & 14 &\cellcolor[rgb]{1.00000000,0.86372504,0.00000000}0.37 & \cellcolor[rgb]{1.00000000,0.62696072,0.00000000}0.30 & \cellcolor[rgb]{1.00000000,0.34902000,0.00000000}0.23 & \cellcolor[rgb]{1.00000000,0.13284388,0.00000000}0.17 & \cellcolor[rgb]{0.92695659,0.00000000,0.00000000}0.11 & \cellcolor[rgb]{0.81371330,0.00000000,0.00000000}0.08 & \cellcolor[rgb]{0.76223908,0.00000000,0.00000000}0.06 & \cellcolor[rgb]{0.71076486,0.00000000,0.00000000}0.05 & \cellcolor[rgb]{0.63870096,0.00000000,0.00000000}0.03 & \cellcolor[rgb]{0.56663705,0.00000000,0.00000000}0.01 & \cellcolor[rgb]{0.56663705,0.00000000,0.00000000}0.01 \\
                \cline{2-13}
                 & 30 &\cellcolor[rgb]{1.00000000,0.06078517,0.00000000}0.15 & \cellcolor[rgb]{0.90636690,0.00000000,0.00000000}0.10 & \cellcolor[rgb]{0.80341846,0.00000000,0.00000000}0.08 & \cellcolor[rgb]{0.78282877,0.00000000,0.00000000}0.07 & \cellcolor[rgb]{0.73135455,0.00000000,0.00000000}0.05 & \cellcolor[rgb]{0.65929064,0.00000000,0.00000000}0.04 & \cellcolor[rgb]{0.62840611,0.00000000,0.00000000}0.03 & \cellcolor[rgb]{0.56663705,0.00000000,0.00000000}0.01 & \cellcolor[rgb]{0.56663705,0.00000000,0.00000000}0.01 & \cellcolor[rgb]{0.57693189,0.00000000,0.00000000}0.01 & \cellcolor[rgb]{0.59752158,0.00000000,0.00000000}0.02 \\
                \cline{2-13}
            \end{tabular}
            \caption{Fraction $\mathcal{F}$ of area with a test statistic $s_i$ above $2$ (see \cref{eq:significance}), i.e. with an error that is systematic rather than caused by sampling with 95\% confidence, as a function of the heatwave duration $T$ and the lead time $\tau$. We consider the 5\% most extreme heatwaves, which means the threshold $a$ changes with $T$. Low values (dark red) mean smaller areas beyond two standard deviations from the empirical composite, which in turns means that differences between the empirical and Gaussian composites are more due to sampling issues of the empirical composite rather than systematic errors in the Gaussian estimation. $\mathcal{F}$ is monotonically decreasing as $T$ and $\tau$ increase.}
            \label{tab:significance}
        \end{table}

        \begin{table}
            \centering
            \begin{tabular}{c|c|c|c|c|c|c|c|c|c|c|c|c|}
                \multicolumn{2}{c}{} & \multicolumn{11}{c}{Norm ratio} \\
                \cline{3-13}
                \multicolumn{2}{c|}{} & \multicolumn{11}{c|}{$\tau$ [days]} \\
                \cline{3-13}
                \multicolumn{2}{c|}{} & 0 & 3 & 6 & 9 & 12 & 15 & 18 & 21 & 24 & 27 & 30 \\
                \cline{2-13}
                \multirow{5}{*}{\rotatebox[origin=c]{90}{$T$ [days]}}
                 & 1 &\cellcolor[rgb]{1.00000000,0.68872533,0.00000000}0.25 & \cellcolor[rgb]{1.00000000,1.00000000,0.16617564}0.28 & \cellcolor[rgb]{1.00000000,1.00000000,0.61397020}0.29 & \cellcolor[rgb]{1.00000000,0.87401915,0.00000000}0.26 & \cellcolor[rgb]{1.00000000,0.92548965,0.00000000}0.27 & \cellcolor[rgb]{1.00000000,1.00000000,0.02720491}0.27 & \cellcolor[rgb]{1.00000000,1.00000000,0.33602875}0.28 & \cellcolor[rgb]{1.00000000,0.84313684,0.00000000}0.26 & \cellcolor[rgb]{1.00000000,0.14313798,0.00000000}0.22 & \cellcolor[rgb]{0.77253393,0.00000000,0.00000000}0.21 & \cellcolor[rgb]{0.80341846,0.00000000,0.00000000}0.21 \\
                \cline{2-13}
                 & 3 &\cellcolor[rgb]{1.00000000,0.52401971,0.00000000}0.24 & \cellcolor[rgb]{1.00000000,0.61666662,0.00000000}0.25 & \cellcolor[rgb]{1.00000000,0.75048993,0.00000000}0.26 & \cellcolor[rgb]{1.00000000,0.43137281,0.00000000}0.24 & \cellcolor[rgb]{1.00000000,0.53431382,0.00000000}0.25 & \cellcolor[rgb]{1.00000000,0.65784303,0.00000000}0.25 & \cellcolor[rgb]{1.00000000,0.80196044,0.00000000}0.26 & \cellcolor[rgb]{1.00000000,0.30784360,0.00000000}0.23 & \cellcolor[rgb]{0.83430299,0.00000000,0.00000000}0.21 & \cellcolor[rgb]{0.56663705,0.00000000,0.00000000}0.19 & \cellcolor[rgb]{0.56663705,0.00000000,0.00000000}0.19 \\
                \cline{2-13}
                 & 7 &\cellcolor[rgb]{0.95784112,0.00000000,0.00000000}0.22 & \cellcolor[rgb]{1.00000000,0.26666719,0.00000000}0.23 & \cellcolor[rgb]{1.00000000,0.50343151,0.00000000}0.24 & \cellcolor[rgb]{1.00000000,0.70931353,0.00000000}0.25 & \cellcolor[rgb]{1.00000000,0.98725426,0.00000000}0.27 & \cellcolor[rgb]{1.00000000,1.00000000,0.58308782}0.29 & \cellcolor[rgb]{1.00000000,1.00000000,0.18161683}0.28 & \cellcolor[rgb]{1.00000000,0.45196101,0.00000000}0.24 & \cellcolor[rgb]{1.00000000,0.16372618,0.00000000}0.23 & \cellcolor[rgb]{1.00000000,0.04019697,0.00000000}0.22 & \cellcolor[rgb]{0.56663705,0.00000000,0.00000000}0.19 \\
                \cline{2-13}
                 & 14 &\cellcolor[rgb]{0.70047002,0.00000000,0.00000000}0.20 & \cellcolor[rgb]{1.00000000,0.21519669,0.00000000}0.23 & \cellcolor[rgb]{1.00000000,0.84313684,0.00000000}0.26 & \cellcolor[rgb]{1.00000000,1.00000000,0.18161683}0.28 & \cellcolor[rgb]{1.00000000,1.00000000,0.19705802}0.28 & \cellcolor[rgb]{1.00000000,0.97696016,0.00000000}0.27 & \cellcolor[rgb]{1.00000000,0.95637195,0.00000000}0.27 & \cellcolor[rgb]{1.00000000,0.89460735,0.00000000}0.26 & \cellcolor[rgb]{1.00000000,0.48284331,0.00000000}0.24 & \cellcolor[rgb]{1.00000000,0.09166748,0.00000000}0.22 & \cellcolor[rgb]{1.00000000,0.06078517,0.00000000}0.22 \\
                \cline{2-13}
                 & 30 &\cellcolor[rgb]{0.88577721,0.00000000,0.00000000}0.21 & \cellcolor[rgb]{1.00000000,0.36960820,0.00000000}0.24 & \cellcolor[rgb]{1.00000000,0.84313684,0.00000000}0.26 & \cellcolor[rgb]{1.00000000,1.00000000,0.18161683}0.28 & \cellcolor[rgb]{1.00000000,1.00000000,0.22794040}0.28 & \cellcolor[rgb]{1.00000000,1.00000000,0.15073444}0.27 & \cellcolor[rgb]{1.00000000,1.00000000,0.07352849}0.27 & \cellcolor[rgb]{1.00000000,0.72990173,0.00000000}0.26 & \cellcolor[rgb]{1.00000000,0.54460792,0.00000000}0.25 & \cellcolor[rgb]{1.00000000,0.62696072,0.00000000}0.25 & \cellcolor[rgb]{1.00000000,0.58578432,0.00000000}0.25 \\
                \cline{2-13}
            \end{tabular}
            \caption{Norm ratio $\mathcal{R}$ (see \cref{eq:norm_ratio}), i.e. relative error of the Gaussian composite map with respect to the empirical one, as a function of the heatwave duration $T$ and the lead time $\tau$, computed on 8000 years of PlaSim data. As for \cref{tab:significance}, we consider the 5\% most extreme heatwaves. The higher the value (bright yellow) the worst the Gaussian composite approximates the empirical one. There is a non-monotonic trend which is due to the different atmospheric fields used in the conditional average. Events which are long-lasting and far in time behave more closely to Gaussian distributed events.}
            \label{tab:norm_ratio}
        \end{table}

        In \cref{tab:significance}, we see the fraction $\mathcal{F}$ of area of the composite map for which the test statistic $s_i$ defined in \cref{eq:significance} overcomes the value of $2$.
        The table shows a monotonic trend in $\mathcal{F}$, with fast and imminent heatwaves having more non-Gaussian features with respect to long and delayed ones.
        Indeed, for higher values of the heatwave duration $T$, we expect the statistics of $A$ to be more Gaussian, as we average over a larger number of days.
        Instead, when we increase the lead time we can think that the chaotic nature of the weather makes the states that led to a heatwave more different from one another. So, both the empirical and the Gaussian composite will tend to $0$ as $\tau$ increases. Moreover, the higher differences between the states over which we take the empirical average increase the standard deviation. Thus, the values of the test statistic $s_i$ of each pixel naturally decrease with $\tau$ (see \cref{eq:significance}).

        On the other hand if we look at the values for the norm ratio (\cref{eq:norm_ratio}) displayed in \cref{tab:norm_ratio}, we see a rather non-monotonic behavior.
        In fact, we can gain more understanding if we plot the norm ratio for the three climate variables independently
        (tables S1 to S3 and Supporting Information S11),
        which shows that the main contribution to the norm ratio comes from the \SI{500}{\hecto\pascal} geopotential height field.

        This overall non-monotonic trend can be explained as a competition between the non-linear chaotic dynamics of the weather, that makes the real composite stray more from its Gaussian approximation as $\tau$ increases, with the loss of memory that averages out the non-linear effects, bringing the empirical composite closer to the Gaussian one. This also would explain why geopotential dominates the norm ratio, as, of the three fields, it is the one with the most non-linear dynamics.

    \subsubsection{Committor}
    \label{subsubsec:committor_phys}
        Similarly to what has been done for the composite maps, we can look at how the skill of the prediction is affected by the heatwave duration $T$ and the lead time $\tau$.
        In the left panel of \cref{tab:SvTtau}, we can see that the prediction skill decreases monotonically with $\tau$ at any level of $T$. For shorter lead times the skill is best when dealing with shorter heatwaves, while for longer delays, the skill is higher for longer-lasting events.
        In the limit of $T=1$ and $\tau=0$, we are forecasting a one-day heatwave that starts today, so we might just look outside the window and see if it is hot. And indeed there is perfect correlation between the temperature anomaly over France and the heatwave amplitude $A$. However, one day heatwaves are very erratic events, which become very hard to predict for longer lead times.
        On the other hand, longer lasting events are non-trivial to predict for very short delays, but are more influenced by processes with long timescales such as the dynamics of soil moisture, and hence maintain some predictability at higher values of $\tau$ \cite{miloshevichProbabilisticForecastsExtreme2023}.

        On the right panel of \cref{tab:SvTtau}, we see the skill comparison with the neural network, which is able to capture non-linear and non-Gaussian structures in the data.
        We can see that our Gaussian committor struggles the most for shorter heatwaves and, more importantly, around $\tau = 5$. We can interpret this region of struggle as the one where the prediction is most \emph{dynamical}, rather than \emph{statistical}. Namely, where mere linear correlations are not enough and the complex and non-linear dynamics of the atmosphere plays a significant role.

        \begin{table}
            \begin{tabular}{c|c|c|c|c|c|c|c|}
                \multicolumn{2}{c}{} & \multicolumn{6}{c}{Normalized log score} \\
                \cline{3-8}
                \multicolumn{2}{c|}{} & \multicolumn{6}{c|}{$\tau$ [days]} \\
                \cline{3-8}
                \multicolumn{2}{c|}{} & 0 & 5 & 10 & 15 & 20 & 30 \\
                \cline{2-8}
                \multirow{4}{*}{\rotatebox[origin=c]{90}{$T$ [days]}}
                 & 1 & \cellcolor[rgb]{1.00000000,1.00000000,0.40000000}0.89 & \cellcolor[rgb]{0.35686275,0.67843137,0.40000000}0.27 & \cellcolor[rgb]{0.12156863,0.56078431,0.40000000}0.14 & \cellcolor[rgb]{0.05490196,0.52745098,0.40000000}0.11 & \cellcolor[rgb]{0.01568627,0.50784314,0.40000000}0.09 & \cellcolor[rgb]{0.00000000,0.50000000,0.40000000}0.08 \\
                \cline{2-8}
                 & 7 & \cellcolor[rgb]{0.86666667,0.93333333,0.40000000}0.53 & \cellcolor[rgb]{0.32941176,0.66470588,0.40000000}0.25 & \cellcolor[rgb]{0.18431373,0.59215686,0.40000000}0.18 & \cellcolor[rgb]{0.11764706,0.55882353,0.40000000}0.14 & \cellcolor[rgb]{0.08627451,0.54313725,0.40000000}0.13 & \cellcolor[rgb]{0.07058824,0.53529412,0.40000000}0.12 \\
                \cline{2-8}
                 & 14 & \cellcolor[rgb]{0.65098039,0.82549020,0.40000000}0.42 & \cellcolor[rgb]{0.34117647,0.67058824,0.40000000}0.26 & \cellcolor[rgb]{0.23137255,0.61568627,0.40000000}0.20 & \cellcolor[rgb]{0.18431373,0.59215686,0.40000000}0.18 & \cellcolor[rgb]{0.16078431,0.58039216,0.40000000}0.17 & \cellcolor[rgb]{0.14509804,0.57254902,0.40000000}0.16 \\
                \cline{2-8}
                 & 30 & \cellcolor[rgb]{0.49411765,0.74705882,0.40000000}0.34 & \cellcolor[rgb]{0.34117647,0.67058824,0.40000000}0.26 & \cellcolor[rgb]{0.27843137,0.63921569,0.40000000}0.23 & \cellcolor[rgb]{0.25098039,0.62549020,0.40000000}0.21 & \cellcolor[rgb]{0.23921569,0.61960784,0.40000000}0.21 & \cellcolor[rgb]{0.21960784,0.60980392,0.40000000}0.20 \\
                \cline{2-8}
            \end{tabular}
            \begin{tabular}{c|c|c|c|c|c|c|}
                \multicolumn{1}{c}{} & \multicolumn{6}{c}{$1 - \mathcal{S}/\mathcal{S}_{CNN}$} \\
                \cline{2-7}
                 & \multicolumn{6}{c|}{$\tau$ [days]} \\
                \cline{2-7}
                 & 0 & 5 & 10 & 15 & 20 & 30 \\
                \cline{1-7}
                  & \cellcolor[rgb]{0.95724721,0.95993849,0.95955402}-0.00 & \cellcolor[rgb]{0.42414456,0.24405998,0.02806613}0.26 & \cellcolor[rgb]{0.57647059,0.34348328,0.05805459}0.22 & \cellcolor[rgb]{0.70980392,0.46897347,0.14955786}0.19 & \cellcolor[rgb]{0.74117647,0.49850058,0.17108804}0.18 & \cellcolor[rgb]{0.79331027,0.59584775,0.28719723}0.15 \\
                \cline{1-7}
                  & \cellcolor[rgb]{0.88688966,0.78123799,0.52787389}0.11 & \cellcolor[rgb]{0.61568627,0.38039216,0.08496732}0.21 & \cellcolor[rgb]{0.75886198,0.52587466,0.20107651}0.17 & \cellcolor[rgb]{0.82775855,0.66582084,0.37331795}0.14 & \cellcolor[rgb]{0.89750096,0.79876970,0.56016917}0.10 & \cellcolor[rgb]{0.92579777,0.84552095,0.64628989}0.08 \\
                \cline{1-7}
                  & \cellcolor[rgb]{0.89750096,0.79876970,0.56016917}0.10 & \cellcolor[rgb]{0.74901961,0.50588235,0.17647059}0.17 & \cellcolor[rgb]{0.84744329,0.70580546,0.42252980}0.13 & \cellcolor[rgb]{0.92226067,0.83967705,0.63552480}0.09 & \cellcolor[rgb]{0.95763168,0.89811611,0.74317570}0.06 & \cellcolor[rgb]{0.96362937,0.92379854,0.81853133}0.04 \\
                \cline{1-7}
                  & \cellcolor[rgb]{0.93994617,0.86889658,0.68935025}0.07 & \cellcolor[rgb]{0.91872357,0.83383314,0.62475971}0.09 & \cellcolor[rgb]{0.96455210,0.91180315,0.77239523}0.05 & \cellcolor[rgb]{0.96255286,0.93779316,0.87235679}0.03 & \cellcolor[rgb]{0.96101499,0.95778547,0.94925029}0.00 & \cellcolor[rgb]{0.95724721,0.95993849,0.95955402}-0.00 \\
                \cline{1-7}
            \end{tabular}
            \caption{Left table: normalized log score of the Gaussian approximation (the higher, the better), versus heatwave duration $T$ and lag time $\tau$. In all cases, we focus on the 5\% most extreme heatwaves. As the prediction task gets harder, the skill decreases monotonically with the lag time, faster for shorter heatwaves. Right table: comparison with the skill of the neural network. The CNN is always better, but more so for shorter heatwaves and around $\tau=5$. This is the regime where the dynamics is more non-linear, and thus the neural network complexity has a better opportunity to make a difference.}
            \label{tab:SvTtau}
        \end{table}


\section{Application to the ERA5 Reanalysis Dataset}
\label{sec:ERA5results}

In this article we presented a methodology for estimating composite maps and committor functions using a theoretical framework that we called the Gaussian approximation (see \cref{sec:bayes_committor_composite_gaussianframe}) and we tested it over a very long simulation dataset obtained from the climate model PlaSim. The results are really promising.

A key point that we showed in the previous sections is that our Gaussian framework is particularly suited for short datasets. In the case of the composite map (see \cref{subsec:generalization_power}), the empirical average is performed over too few samples to be very accurate. For the committor (see \cref{sec:q:less-data}), the alternative approach of deep neural networks struggles with the lack of data.
It is then natural to try to apply our method ERA5~\cite{hersbachERA5GlobalReanalysis2020}, and in this section we show that indeed for this dataset the Gaussian approximation is the best option.

\subsection{Composites}
\label{subsec:era_comp}
    In this subsection we compute composite maps on the ERA5 dataset, using both the empirical average and the Gaussian approximation.
    Because the dataset is much shorter than the PlaSim dataset, we do not know the ground truth as precisely as in \Cref{sec:composite}.
    We can nevertheless compare the two estimates and see if they qualitatively agree.
    
\begin{figure}
        \centering
        \includegraphics[width=1.0\textwidth]{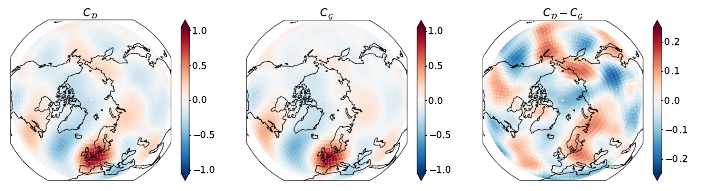}
        \caption{Composite maps of normalized \SI{500}{\hecto\pascal} geopotential height for 5\% most extremes 14-day temperature anomaly over France ($T=14$). Composite maps are computed either directly from ERA5 data (left map), or under the Gaussian approximation (central map). The right map shows the difference between the first two. The salient features of geopotential are well captured by the Gaussian approximation, with errors of the order of 25\% at most.}
        \label{fig:ERA-comp}
    \end{figure}

    \Cref{fig:ERA-comp} shows the empirical composite and the composite evaluated within the Gaussian framework for the geopotential height anomaly at \SI{500}{\hecto\pascal}. They are both evaluated for $T=14$, $\tau =0$ and for heatwaves corresponding to the $5\% $ most extreme value in the distribution of $A$.
    The two composites look qualitatively very similar.
    In both cases we see a clear wave train which starts from the western part of the United States and Canada with an anticyclonic anomaly, continues over the North Atlantic Ocean and finally terminates over Western Europe with another anticyclonic anomaly, stronger than the rest of the wave pattern. This is consistent with the fact that we condition on the temperature anomaly over France. The overall wave structure is well represented by the Gaussian composite, even if it puts a higher weight over the Western Europe anticyclone (\cref{fig:ERA-comp}, right panel).
    The difference between the two composites is larger over Asia and over the Pacific Ocean.
    Unlike the case of PlaSim data, the difference does not have an annular mode structure but contains a visible wavenumber 6 component.
    The largest differences between the two composites are on the order of 20\%.

    As for PlaSim data, we analyzed how the two indices $\mathcal{R}$ (norm ratio, defined in \cref{eq:norm_ratio}) and $\mathcal{F}$ (fraction of area where differences between Gaussian and empirical composites are significant, defined in \cref{eq:significance}) vary with the parameters $T$ and $\tau$ for ERA5 data.
    \Cref{tab:ERA_norm_ratio} shows the norm ratio as a function of $T$ and $\tau$ for the $5\%$ most extreme heatwaves. We see that, similarly to PlaSim, there is a non-monotonic trend with lowest values for $T$ between 1 and 14 and $\tau$ between 0 and 6. Outside this range, the norm ratio has rather high values which are the sign of a great mismatch between the two composites.
    However, $\mathcal{F}$ (table not shown) assumes values which are almost never above 1\% and very often below 0.1\%, meaning that we cannot rule out that any discrepancy between the Gaussian and empirical composite is simply a sampling error.

    \begin{table}[h]
        \centering
        \begin{tabular}{c|c|c|c|c|c|c|c|c|c|c|c|c|}
            \multicolumn{2}{c}{} & \multicolumn{11}{c}{Norm ratio} \\
            \cline{3-13}
            \multicolumn{2}{c|}{} & \multicolumn{11}{c|}{$\tau$ [days]} \\
            \cline{3-13}
            \multicolumn{2}{c|}{} & 0 & 3 & 6 & 9 & 12 & 15 & 18 & 21 & 24 & 27 & 30 \\
            \cline{2-13}
            \multirow{5}{*}{\rotatebox[origin=c]{90}{$T$ [days]}}
             & 1 &\cellcolor[rgb]{0.56663705,0.00000000,0.00000000}0.29 & \cellcolor[rgb]{0.82400815,0.00000000,0.00000000}0.38 & \cellcolor[rgb]{1.00000000,0.11225568,0.00000000}0.47 & \cellcolor[rgb]{1.00000000,0.47254921,0.00000000}0.59 & \cellcolor[rgb]{1.00000000,1.00000000,0.19705802}0.81 & \cellcolor[rgb]{1.00000000,1.00000000,0.53676424}0.89 & \cellcolor[rgb]{1.00000000,1.00000000,0.49044067}0.88 & \cellcolor[rgb]{1.00000000,1.00000000,0.49044067}0.88 & \cellcolor[rgb]{1.00000000,1.00000000,0.04264610}0.78 & \cellcolor[rgb]{1.00000000,0.92548965,0.00000000}0.74 & \cellcolor[rgb]{1.00000000,0.90490145,0.00000000}0.73 \\
            \cline{2-13}
             & 3 &\cellcolor[rgb]{0.66958549,0.00000000,0.00000000}0.33 & \cellcolor[rgb]{0.89607206,0.00000000,0.00000000}0.40 & \cellcolor[rgb]{1.00000000,0.13284388,0.00000000}0.48 & \cellcolor[rgb]{1.00000000,0.78137224,0.00000000}0.69 & \cellcolor[rgb]{1.00000000,1.00000000,0.58308782}0.90 & \cellcolor[rgb]{1.00000000,1.00000000,0.47499947}0.87 & \cellcolor[rgb]{1.00000000,1.00000000,0.42867590}0.86 & \cellcolor[rgb]{1.00000000,1.00000000,0.19705802}0.81 & \cellcolor[rgb]{1.00000000,1.00000000,0.21249921}0.82 & \cellcolor[rgb]{1.00000000,0.86372504,0.00000000}0.72 & \cellcolor[rgb]{1.00000000,1.00000000,0.21249921}0.81 \\
            \cline{2-13}
             & 7 &\cellcolor[rgb]{0.75194424,0.00000000,0.00000000}0.35 & \cellcolor[rgb]{0.87548237,0.00000000,0.00000000}0.39 & \cellcolor[rgb]{1.00000000,0.29754949,0.00000000}0.53 & \cellcolor[rgb]{1.00000000,0.88431325,0.00000000}0.73 & \cellcolor[rgb]{1.00000000,1.00000000,0.08896968}0.79 & \cellcolor[rgb]{1.00000000,1.00000000,0.16617564}0.80 & \cellcolor[rgb]{1.00000000,1.00000000,0.30514636}0.83 & \cellcolor[rgb]{1.00000000,1.00000000,0.33602875}0.84 & \cellcolor[rgb]{1.00000000,1.00000000,0.32058756}0.84 & \cellcolor[rgb]{1.00000000,0.97696016,0.00000000}0.76 & \cellcolor[rgb]{1.00000000,1.00000000,0.19705802}0.81 \\
            \cline{2-13}
             & 14 &\cellcolor[rgb]{0.74164940,0.00000000,0.00000000}0.35 & \cellcolor[rgb]{1.00000000,0.07107927,0.00000000}0.46 & \cellcolor[rgb]{1.00000000,0.53431382,0.00000000}0.61 & \cellcolor[rgb]{1.00000000,0.90490145,0.00000000}0.73 & \cellcolor[rgb]{1.00000000,1.00000000,0.22794040}0.82 & \cellcolor[rgb]{1.00000000,1.00000000,0.42867590}0.86 & \cellcolor[rgb]{1.00000000,1.00000000,0.35146994}0.85 & \cellcolor[rgb]{1.00000000,1.00000000,0.47499947}0.87 & \cellcolor[rgb]{1.00000000,1.00000000,0.28970517}0.83 & \cellcolor[rgb]{1.00000000,1.00000000,0.10441087}0.79 & \cellcolor[rgb]{1.00000000,1.00000000,0.18161683}0.81 \\
            \cline{2-13}
             & 30 &\cellcolor[rgb]{1.00000000,0.55490202,0.00000000}0.62 & \cellcolor[rgb]{1.00000000,1.00000000,0.21249921}0.81 & \cellcolor[rgb]{1.00000000,1.00000000,0.55220543}0.89 & \cellcolor[rgb]{1.00000000,1.00000000,0.61397020}0.90 & \cellcolor[rgb]{1.00000000,1.00000000,0.33602875}0.84 & \cellcolor[rgb]{1.00000000,1.00000000,0.16617564}0.81 & \cellcolor[rgb]{1.00000000,1.00000000,0.22794040}0.82 & \cellcolor[rgb]{1.00000000,1.00000000,0.24338160}0.82 & \cellcolor[rgb]{1.00000000,1.00000000,0.21249921}0.82 & \cellcolor[rgb]{1.00000000,1.00000000,0.19705802}0.81 & \cellcolor[rgb]{1.00000000,1.00000000,0.30514636}0.83 \\
            \cline{2-13}
        \end{tabular}
        \caption{Norm ratio $\mathcal{R}$ (see \cref{eq:norm_ratio}), i.e. relative error of the Gaussian composite map with respect to the empirical one, as a function of the heatwave duration $T$ and the lead time $\tau$, computed on the ERA5 dataset. We consider the 5\% most extreme heatwaves. The higher the value (bright yellow) the worst the Gaussian composite approximates the empirical one. Lower values (dark red) denote a better agreement between the two.}
        \label{tab:ERA_norm_ratio}
    \end{table}

\subsection{Committor}
    In this subsection we deal with the computation of committor functions on the reanalysis dataset. After the necessary technical adaptations to work on this dataset, we compare the prediction skill of the Gaussian approximation with the one of neural networks, which shows the first is clearly better.

    Before discussing any result, we need to define a protocol for the choice of the proper regularization coefficient $\epsilon$, the only hyperparameter of our method. When working with 8000 years of PlaSim data, we had to choose empirically $\epsilon=1$ to have interpretability in the projection patterns, and this interpretability came at the cost of a lower skill score.
    On the other hand, on ERA5, and more generally when working with small datasets (\cref{tab:gacnn-yr_eps}), the value $\epsilon_\mathrm{best}$ of the regularization coefficient that yields the highest skill score also provides an interpretable projection map.

    The ERA5 dataset contains 83 years of data. To have a meaningful cross validation we take the 80 years from 1943 to 2022 and split them in 5 balanced folds
    (see Supporting Information S2).
    This way we train on 64 years and validate on 16.

    With this choice, for the 5\% most extreme two-week heatwaves ($T=14$) at $\tau=0$, we obtain a skill score of $\mathcal{S} = 0.16 \pm 0.07$. This number is considerably lower than the skill we have measured for PlaSim (\cref{sec:q:nonregularized-ga}). To understand why, we can investigate the impact on the skill score for the PlaSim dataset of the reduced number of predictors (using only the \SI{500}{\hecto\pascal} geopotential height field as for ERA5) and of the amount of data (training on a subset of the same size as the ERA5).
    \begin{table}[h]
        \centering
        \begin{tabular}{c|cc}
            & \multicolumn{2}{c}{Predictor fields} \\
            years of data &  $T_\mathrm{2m}, Z, S$ & $Z$ \\
            \hline
            8000 & $0.418 \pm 0.006$ & $0.23 \pm 0.01$ \\
            80 & $0.33 \pm 0.07$ & $0.18 \pm 0.04$
        \end{tabular}
        \caption{Skill score on PlaSim, when using different amount of data and different predictor fields.}
        \label{tab:Svyear-field_PlaSim}
    \end{table}

    As can be seen from \cref{tab:Svyear-field_PlaSim}, both the reduced number of predictor fields and the smaller dataset severely impact the skill score. However, even combining the two effects, the performance remains slightly better than for the reanalysis data. This suggests that the more realistic data of ERA5 have more complexity and variability than PlaSim, and thus it is harder to make a skillful prediction.

\begin{figure}
        \centering
        \includegraphics[width=0.8\textwidth]{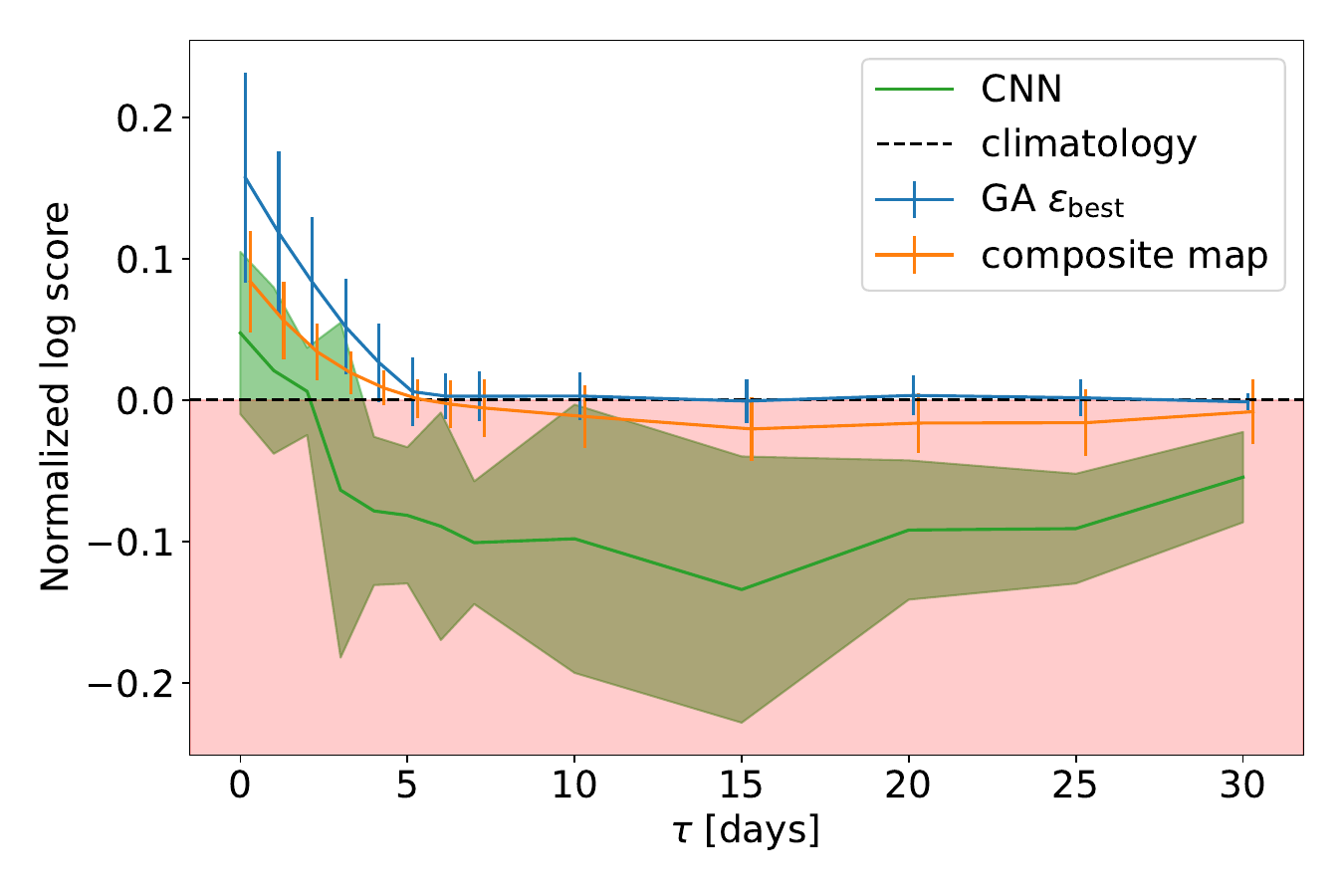}
        \caption{Skill score of different prediction techniques for reanalysis data (using geopotential height at \SI{500}{\hecto\pascal} anomaly as the only predictor, $T=14$) changing the lead time. In green the convolutional neural network, in blue the Gaussian approximation, both at their best values for hyperparameters. In orange the Gaussian approximation when using the composite map as projection pattern. Error bars or shaded area indicates the standard deviation among the 5 folds. The red shaded zone below 0 indicates where the prediction is worse than the climatology. The Gaussian approximation is always the best, and gives results better than the climatology only for $\tau \leq 5$.}
        \label{fig:ERA_S_T14}
    \end{figure}

    Nevertheless, we argue that the result achieved for ERA5, albeit humble, is the best we can do. To support this claim, in \cref{fig:ERA_S_T14} we compare it to the skill of other methods at different values of the lead time $\tau$. In green is the performance of a convolutional neural network with a similar architecture to the one used for PlaSim. It always performs worse than the Gaussian approximation (in blue), and already at $\tau=3$ days it is consistently below the climatology. On the other hand, the Gaussian approximation manages to extend the predictability margin a few more days.
    For $\tau \geq 6$ days the latter becomes useless too, and, interestingly, $\epsilon_\mathrm{best} \to \infty$, yielding a uniform projection pattern (see
    tab. S4 of the Supporting Information).

    In the regime where the prediction is still skillful, the projection patterns look remarkably similar to the composite maps (\cref{tab:PEMC}), so it is natural to try to project onto the composite itself. This is the orange line in \cref{fig:ERA_S_T14}, which, despite having a smaller error bar than the optimal projection pattern $M$, on average yields a worse performance.
    This once again highlights the fact that composite maps are not the proper tool for prediction.

\begin{figure}
        \centering
        \includegraphics[width=0.8\textwidth]{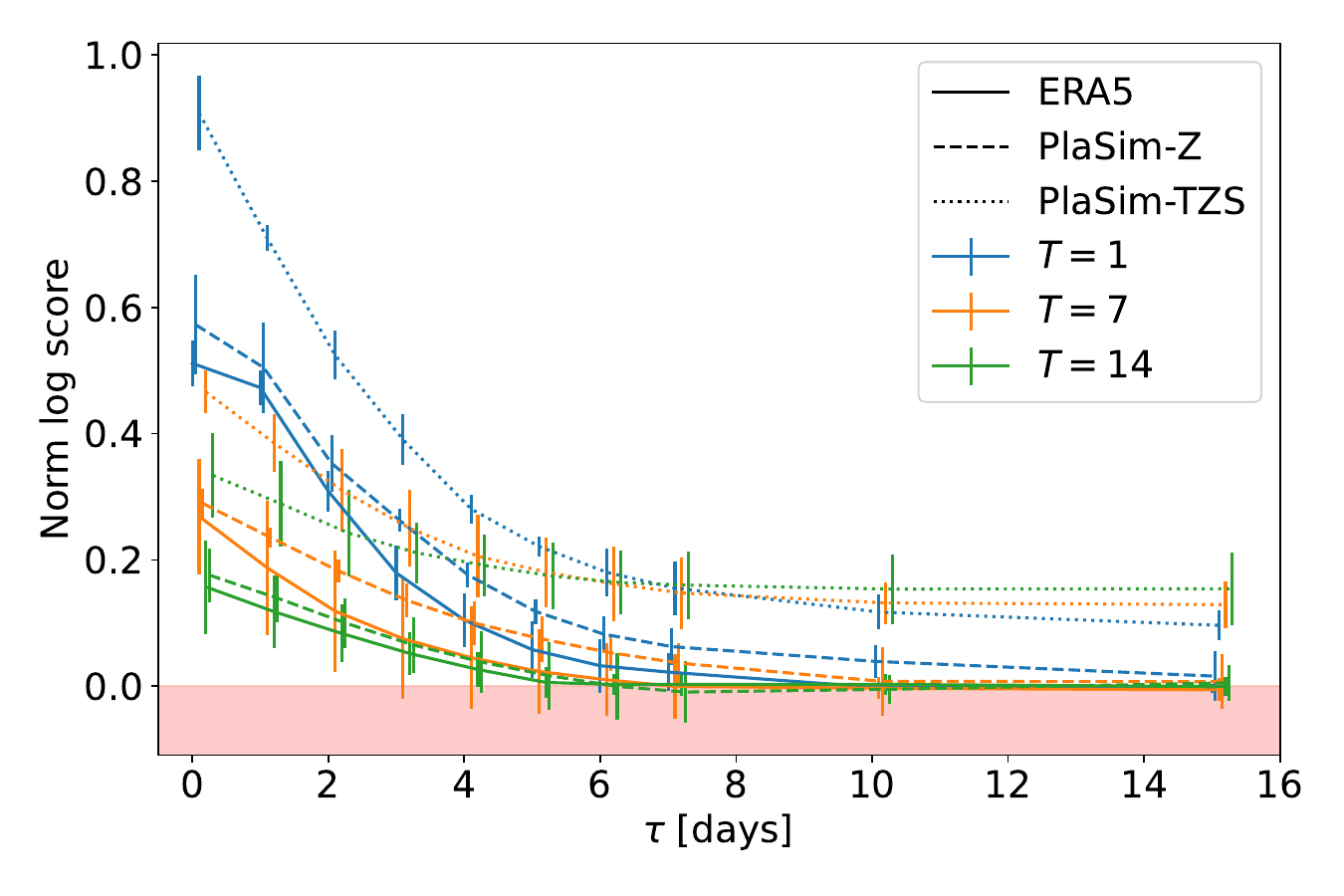}
        \caption{Skill score of the Gaussian committor as a function of $\tau$ for different values of the heatwave duration $T$, and three different datasets: ERA5 with only geopotential height at \SI{500}{\hecto\pascal} (solid line), 80 years of PlaSim data with \SI{2}{\meter} temperature, geopotential height at \SI{500}{\hecto\pascal} and soil moisture (dotted lines) and only with geopotential height at \SI{500}{\hecto\pascal} (dashed line). PlaSim has a consistently higher predictability than ERA5, and the addition of the slow evolving soil moisture greatly extends the predictability horizon. In the absence of this slow variable, predictability decreases with the heatwave duration.}
        \label{fig:SvTtau}
    \end{figure}

    Now that we showed that the Gaussian approximation is the best option for very small datasets, we can investigate what happens when we vary the heatwave duration. From \cref{fig:SvTtau} we see that, at any fixed value of $\tau$, the prediction skill decreases with increasing heatwave duration $T$ (solid lines), with shorter heatwaves having a longer predictability horizon.
    The comparison with 80 years of PlaSim data with only the \SI{500}{\hecto \pascal} geopotential height as predictor (dashed lines), shows that predicting heatwaves is harder on the more realistic data. This can be an effect of the lower spatial resolution of the PlaSim model, which yields a more sluggish and less chaotic atmospheric dynamics, and, hence, better predictability. This hypothesis is further reinforced by the fact that, on average, training on PlaSim requires a lower regularization coefficient than the one on ERA5 (\cref{tab:PEMC}).

    Finally, the dotted lines in \cref{fig:SvTtau} represent the skill when still training on 80 years of PlaSim data, but with all three predictors. For short lead times and heatwave duration, the increase in skill comes mainly from the direct information of the \SI{2}{\meter} temperature field, but the more interesting effect is for longer delays. Here, almost all the predictive power resides in the soil moisture field, and is able to extend the predictability horizon significantly. This effect is enhanced for longer lasting heatwaves.
    As was already pointed out in \citeA{miloshevichProbabilisticForecastsExtreme2023}, soil moisture acts as a slow modulator of the chance of a heatwave, that is still able to give some useful information when the fast predictors, such as the \SI{500}{\hecto\pascal} geopotential height, are beyond their de-correlation timescale.
    
    Summarizing, the higher complexity of the ERA5 dataset, its reduced length, and the absence of soil moisture as a slow predictor, all diminish the forecast skill relative to PlaSim. However, the prediction performed by the Gaussian approximation proved to be the best available option, with results that are still remarkable.

\subsection{Physical discussion}
    Now that we discussed composite maps and committor functions from the point of view of performance, we proceed to focus more on the physics-oriented analysis of composite maps and optimal projection patterns.

\begin{figure}[tbhp]
        \centering
        \includegraphics[width=1.0\textwidth]{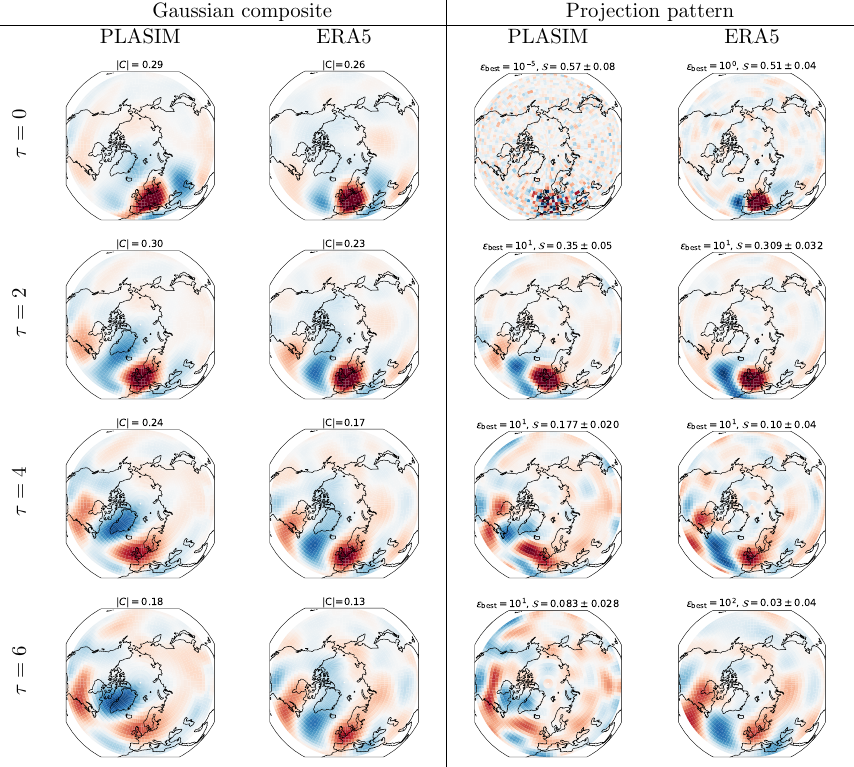}
        \caption{Comparison between composite maps and projection patterns of ERA5 and PlaSim (80 years, geopotential height at \SI{500}{\hecto\pascal} only), at $T=1$ and different values of $\tau$. All maps are shown as normalized to unitary L2 norm. The L2 norm of the actual composite maps is reported on top of them, while for the projection pattern we display the regularization coefficient and the skill score. For ERA5 composites and projection patterns \emph{look} qualitatively similar. However, this is a result of the small size of the dataset, which forces us to use high values of the regularization coefficient.}
        \label{tab:PEMC}
    \end{figure}

    In \cref{tab:PEMC}, we show the comparison between composite maps and projection patterns computed on ERA5 and on 80 years of PlaSim data using only the \SI{500}{\hecto\pascal} geopotential height as a predictor.
    Interestingly, we observed that both composite maps and projection patterns do not change much with respect to the heatwave duration $T$ (not shown). This is only partially explained by consecutive days with high temperature contributing both to short and long heatwaves. In the future it may be worth investigating this further, but in this work we simply exploit it to discuss the patterns only for a single value of $T$ and still provide a relatively comprehensive picture.
    In particular, we show the results for 1-day heatwaves, as they display a clearer evolution of both composites and projection patterns with the lead time $\tau$.

    As already mentioned before, one of the main differences between PlaSim and ERA5 is a generally higher signal-to-noise ratio in PlaSim, that manifests itself in higher norms of the composite maps (left columns) and less smooth projection patterns (right column).
    At $\tau = 0$, most of the weight of both composite maps and projection patterns is concentrated around France. More precisely, there is an anticyclone over France and Central Europe to ensure clear skies and a cyclone north of Portugal to advect warm African air northward. This cyclone is more localized in the ERA5 projection map than in composite maps. As the lead time increases, this dipole structure stretches westward into the Atlantic Ocean. In the reanalysis dataset, there is a clear emergence of a wave-train pattern, and composite maps and projection patterns look rather similar. On the other hand, PlaSim's projection patterns stray considerably from the composite maps, and the physics that they hint at is harder to explain.
    For both datasets, and in both composite maps and projection patterns, a rather strong anticyclonic anomaly is always present over France, getting fainter as $\tau$ increases, but remaining always a prominent feature.
    This suggests that even for very short $T=1$ heatwaves, the most \emph{common} (composites) and the most \emph{informative} (committor) causes of the extreme events are connected with quasi-stationary weather states.

    Concerning the reanalysis dataset, the similarity between composite maps and projection patterns may tempt us to use the composite as a prediction tool. However, although both composites and projection maps display the dynamics of a stationary Rossby wave, a careful examination shows a different weight distribution in the projection patterns, for example at $\tau=6$ the focus is more over North America than in the composite map.
    The lower prediction skill achieved with the composite map compared to the optimal projection pattern, already discussed above (\cref{fig:ERA_S_T14}), suggests that such differences matter for prediction even if they appear small at first sight.
    Furthermore, the similarity between the two maps can most likely be attributed to the need for a relatively high regularization coefficient, required to have a prediction that generalizes well when trained on such a short dataset. More technical details are available in
    Supporting Information S12.

This section has three main conclusions. Firstly, given the size of the ERA5 dataset (and of any other real-data dataset), it is hard to go beyond the Gaussian approximation for both analysis of the averaged weather conditions that led to heatwave events (composite maps) and for a probabilistic forecast of heatwaves, as shown from \cref{fig:ERA-comp} and \cref{fig:ERA_S_T14}.
Secondly, the difference between the empirical and the Gaussian composite maps, shown in \cref{fig:ERA-comp} (right panel) has a different wave number with respect to the one observed for PlaSim, for heatwaves of the same duration and intensity, (see \cref{fig:composite} bottom row, central map, which exhibits a wave zero pattern). However, we cannot exclude that this mismatch is due to sampling error.
Thirdly, the reduced size of the dataset forces us to strongly regularize the optimal projection patterns, which makes them visually similar to the composite map. However, even if they do not provide any additional qualitative information, they do provide more precise quantitative information, leveraged for prediction skill.
Finally, the comparison with the data from PlaSim suggests the importance of predictor fields other than the \SI{500}{\hecto \pascal} geopotential height, which can significantly improve the prediction skill. For the ERA5, this opens the possibility to use also ocean variables, like sea surface temperature.

\section{Conclusions and Perspectives}
\label{sec:conclusions}

In this work, we discussed two ways to characterize the physical processes associated with extreme events.
On the one hand, one might consider the statistics of climate and weather conditions preceding an extreme event conditioned on the fact that the event does occur in the future (a-posteriori statistics, for instance composite maps).
On the other hand, one might want to know what is the probability that en extreme event occurs in the future, conditioned on some knowledge (predictors) of the state of the climate system (a-priori statistics, for instance committor functions).
The two quantities might not coincide, and in principle only the latter is relevant for actual prediction.
Here, we have shown the theoretical connection between the two quantities, and discussed the important problem of their estimation for rare events.
We have proposed a simple framework to compute these two quantities, based on the assumption that the predictors and the observable characterizing the event are jointly Gaussian, which is effective even with short datasets of length of the order of several decades to several centuries.
In the context of extreme heatwaves over France, we have evaluated our method on a very long time series of climate model output data before applying it to a reanalysis dataset.
In both cases we have compared the information contained in \emph{a posteriori} and \emph{a priori} statistics.

Regarding the a-posteriori statistics, our Gaussian framework provides an explanation of the observed fact that composite maps for very extreme heatwaves look qualitatively similar to less extreme ones.
We made this statement quantitative, showing that composite maps are the same up to a rescaling by a non-linear function of the threshold that defines heatwaves.
This opens the possibility to estimate composite maps of extremely rare events, even ones that have never been observed in the dataset.
For PlaSim data, the computation of composite maps using the Gaussian approximation gives results which are valid up to an error (in L2 norm) of the order of 20 to 30\%. We also stress that the deviations from the Gaussian prediction are statistically significant, showing that the statistics is actually not Gaussian and that information beyond the Gaussian approximation can be computed with dataset length of the order of a thousand years or more.
On the other hand, on the much shorter reanalysis dataset, errors are larger, but entirely compatible with the imperfect sampling of the empirical composite, and one cannot compute statistically significant deviations from the composite map obtained within the Gaussian approximation.

However, the information contained in composite maps does not in full generality provide the best predictor of heatwaves.
Making a prediction amounts at estimating the committor function for the observed state of the system.
We have shown that the Gaussian approximation provides an analytical formula for the committor function, which involves a linear projection of the fields used for prediction.
This method gives very good prediction skill, and is particularly competitive with more complex alternatives, such as neural networks, when working with small datasets, which are very common in climate research.
In fact, for the 80-year long ERA5 dataset, the Gaussian approximation proved to be the method with the highest predictive skill.
As demonstrated in
\citeA{miloshevichProbabilisticForecastsExtreme2023}, too short datasets prevent optimal use of neural networks in many applications in climate sciences.
This issue is particularly salient for rare events.
In this respect, the Gaussian framework developed in this paper could play an important role to make a first relevant prediction.
Going beyond the results of the Gaussian approximation may require to have datasets with more rare events. One way is to sample exceptionally rare extreme events using the recently developed rare event simulation techniques, that are able to multiply by several orders of magnitude the number of observed heatwaves with PlaSim model~\cite{Ragone24} and with CESM~\cite{ragoneRareEventAlgorithm2021}.
A perspective is to couple these rare event simulations with the Gaussian framework presented in this paper or other machine learning forecasts.
We have already coupled machine learning approaches with rare event algorithms, for simple academic models~\cite{lucenteCouplingRareEvent2022}.
Doing so for state-of-the-art models is a very interesting perspective to solve the key fundamental issue of lack of data in the science of climate extremes.

Beyond the prediction skill question, the Gaussian approximation introduces an optimal projection, associated to an optimal index for prediction, which, once properly regularized, makes it easy to interpret the prediction, giving insight in the dynamics behind our subject of study.
This optimal prediction map is one of the key results of this paper.
It makes the Gaussian approximation appealing even for applications on datasets long enough that neural networks have more skill, as they are often hard to understand.
From the point of view of understanding the underlying physics, in the case of extreme heatwaves over France, we found that both composite maps and optimal projection maps display a quasi-stationary pattern, that does not depend much on the lead time.
In particular, the development of a Rossby wave-train over the Atlantic Ocean plays an important role for the short term prediction.
This appears very clearly in ERA5, while PlaSim has a strong competing contribution from a wave number 0 pattern.
For longer lead times, instead, the analysis on PlaSim data and the comparison with ERA5 confirmed the key importance of slow drivers, such as soil moisture.
The natural next step is then to include these slow drivers in the study on ERA5, maybe even using ocean-based variables like sea surface temperature.

As further perspectives, we argue that a deeper analysis at the physical level of optimal projection patterns is needed, turning the qualitative insights presented in this work to more quantitative statements.
Moreover, we used here the example of extreme heatwaves over France as a test bed for the proposed method.
Given the generality of the formulation of the Gaussian approximation, it would be very interesting to investigate its wider applicability, first for heatwaves at different geographical locations, but also for different types of extreme events altogether.
It can be expected that it has the potential to provide good performance in particular for events which are aggregated over large spatial and/or temporal scales, but the exact conditions where it is useful should be confirmed on a case-by-case basis.
Another very interesting direction is, in the cases where the Gaussian approximation is outperformed by neural networks, to interpret where this extra skill comes from.
Finally, we suggest that our method can be used as a better baseline than the mere climatology when testing more sophisticated tools for probabilistic prediction.

\section*{References From the Supporting Information}
    \cite{hannachiEmpiricalOrthogonalFunctions2007,hersbachERA5GlobalReanalysis2020,santerStatisticalSignificanceTrends2000,hastieElementsStatisticalLearning2001}

%



%
%

\section*{Open Research Section}
In this study, we use the ERA5 reanalysis dataset \cite{hersbachERA5GlobalReanalysis2020}, in particular \citeA{ERA5-hourly-pressure-levels} and \citeA{ERA5-hourly-single-levels}. We also use data from a long simulation (8000 years) of the PlaSim climate model \cite{Plasim}. The full model output is too heavy to publish, but details of the model setup can be found in \citeA{miloshevichProbabilisticForecastsExtreme2023}. The software (version 2.0.1) associated with this study is available on GitHub under the MIT license at \url{https://github.com/AlessandroLovo/gaussian-approximation-zenodo} \cite{ga-zenodo}, together with all the processed data necessary to reproduce the figures of this paper.
Figures were made with Matplotlib version 3.6.2 \cite{matplotlib}, available under the Matplotlib license at \url{https://matplotlib.org} and with Cartopy version 0.21.1 \cite{cartopy}, available under the BSD 3-Clause license at \url{https://scitools.org.uk/cartopy/}.

\section*{Acknowledgements}
V.M. and A.L. have contributed equally to the work performed in this study and should both be considered as the main author. V.M. and A.L. have received funding from the European Union’s Horizon 2020 research and innovation program under the Marie Skłodowska-Curie grant agreement 956396 (EDIPI) and 956170 (Critical Earth), respectively. F.B. acknowledges funding by the ClimTip project under the European Union’s Horizon Europe research and innovation programme under grant agreement no. 101137601. The authors thank the computer resources provided by the Centre Blaise Pascal at ENS de Lyon. We are grateful to Emmanuel Quemener for his help with the platform. This work was granted access to the HPC/AI resources of CINES under the allocations 2018-A0050110575, 2019-A0070110575, 2020-A0090110575 and 2021-A0110110575 made by GENCI. We thank B. Cozian, C. Le Priol, and A. Lancelin for scientific discussions and suggestions.
The results of this paper contain modified Copernicus Climate Change Service information 2024. Neither the European Commission nor ECMWF is responsible for any use that may be made of the Copernicus information or data it contains.

We thank the referees for their constructive comments and for their careful reading of the manuscript.

All the authors declare no conflict of interest.




%
%

\bibliography{actual}

%
%
%
%
%



\section*{Supplementary material}

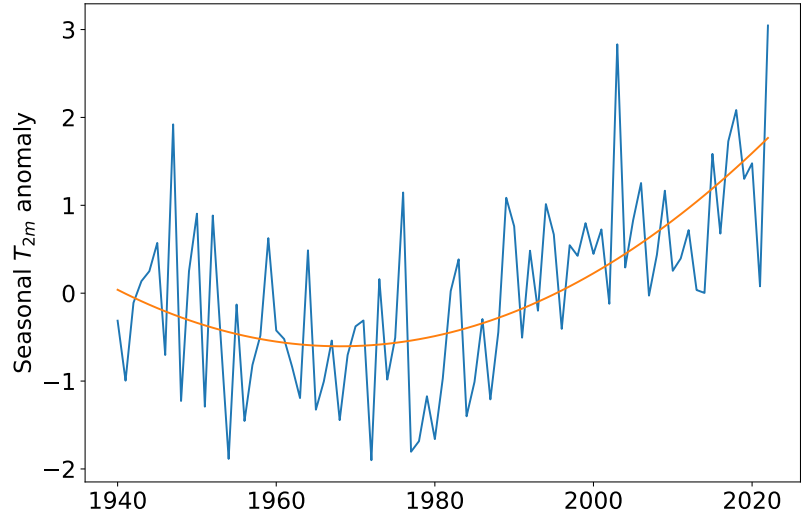

**Figure S1.** Seasonal $T_{2m}$ anomaly averaged over France for the ERA5 dataset. The orange curve is the trend fitted via a second order polynomial.

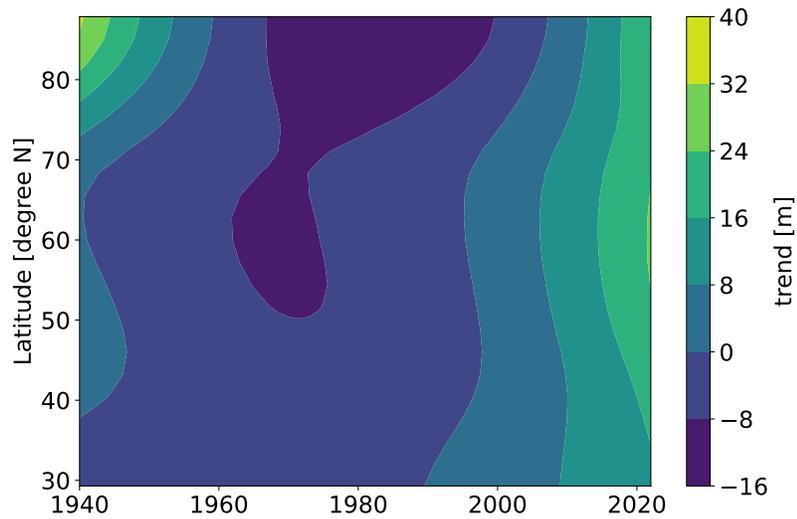

**Figure S2.** Contour plot of the 500 hPa geopotential height trend for ERA5 dataset as function of years and latitude. At latitudes the trend is non-monotonic, while it is monotonically increasing in time at lower latitudes.

## S2 Balanced $K$-fold Cross Validation

The standard $K$-fold cross validation process consists in splitting the dataset $\mathcal{D}$ into $K$ disjoint subsets of equal length $\{\mathcal{F}_k\}_{k=1}^{K}$, that can be called *folds*. Then, for each $k = 1, \ldots, K$ we define training and validation sets as

$$\mathcal{T}_k = \bigcup_{i \neq k} \mathcal{F}_i, \quad \mathcal{V}_k = \mathcal{F}_k. \tag{S2}$$

To make a *balanced* $K$-fold cross validation, we ask that the $\mathcal{F}_k$ all contain the same amount of heatwaves. This is essentially equivalent to the classical technique of stratified $K$-fold cross validation (Hastie et al., 2001), but in our case, to avoid contamination between the different folds, we force data belonging to the same summer to end up in only one of the folds.

## S3 Detailed Calculations of the Composite Map in a 2-Dimensional Gaussian System with Committor Depending Only on One Variable

In the second example of section 3.1.4, we use intuition to say that if we have two correlated variables but the committor depends only on one, the composite will still be non-zero for the variable upon which the committor does *not* depend. Here we give a formal proof.

According to the assumption of zero mean and Gaussianity for the two variables, we can write the stationary measure as

$$P_S(x_1, x_2) \propto \exp\left(-\frac{1}{2}(x_1, x_2)\begin{pmatrix}\sigma_1^2 & \phi \\ \phi & \sigma_2^2\end{pmatrix}^{-1}\begin{pmatrix}x_1 \\ x_2\end{pmatrix}\right) = \exp\left(-\frac{1}{2}(ax_1^2 + bx_2^2 - 2cx_1x_2)\right), \tag{S2}$$

where

$$(a, b, c) = \frac{1}{\sigma_1^2\sigma_2^2 - \phi^2}(\sigma_2^2, \sigma_1^2, \phi). \tag{S2}$$

Then, according to eq. (7), the composite value for $X_2$ is

$$C_2 \propto \int x_2 P_S(x_1, x_2) q(x_1) dx_1 dx_2, \tag{S2}$$

$$\propto \int dx_1 q(x_1) \int dx_2 x_2 e^{-\frac{1}{2}(ax_1^2 + bx_2^2 - 2cx_1x_2)}, \tag{S2}$$

$$= \int dx_1 q(x_1) e^{-\frac{1}{2}\left(a - \frac{c^2}{b}\right)x_1^2} \int dx_2 x_2 e^{-\frac{1}{2}b\left(x_2 - \frac{c}{b}x_1\right)^2}, \tag{S2}$$

$$\propto \int dx_1 q(x_1) e^{-\frac{1}{2}\left(a - \frac{c^2}{b}\right)x_1^2} \frac{c}{b} x_1. \tag{S2}$$

Now, first we notice that $C_2 \propto c \propto \phi$, so if there is no correlation between $x_1$ and $x_2$ we get the expected result that the composite is zero. Otherwise, for a generic committor $q$, $C_2 \neq 0$. A particular case for which $C_2 = 0$ is when $q$ is an even function. However, this means that the committor must give equal probability to $x_1$ and $-x_1$, and thus cannot focus on a single tail of the distribution of $x_1$.

## S4 Detailed Calculation of the Composite Maps and of the Committor Function in the Gaussian Approximation Framework

As already presented in the main text, the Gaussian approximation relies on the hypothesis that, at each pixel, the field and the heatwave amplitude follow a jointly Gaussian distribution, namely

$$(X, A) \sim \mathcal{N}(0, \Sigma), \tag{S2}$$

with $\Sigma$ being the covariance matrix, $\Sigma = \begin{bmatrix} \Sigma_{XX} & \Sigma_{XA} \\ \Sigma_{AX} & \Sigma_{AA} \end{bmatrix}$. The joint multivariate Gaussian distribution, for given values of $X$ and $A$, is generally written in the form:

$$\mathbb{P}(x,a) = \frac{1}{Z}\exp\left(-\frac{\left(x^T\Lambda_{XX}x + 2x^T\Lambda_{XA}a + \Lambda_{AA}a^2\right)}{2}\right), \tag{S2}$$

where $Z = \sqrt{(2\pi)^d \det(\Sigma)}$ is the normalization constant $d$ is the dimension of the stack of $X$ and $A$, $\Lambda = \Sigma^{-1}$, $\Lambda_{XA} = \Lambda_{AX}$. We took advantage of the fact that $a$ is a scalar quantity. Equation (15), can be obtained via the following calculation:

$$C = \mathbb{E}[X|A \geq a] = \frac{1}{\mathbb{P}(A \geq a)} \int_a^{+\infty} \left(\int x \mathbb{P}(x,a')\mathrm{d}x\right) \mathrm{d}a', \tag{S2}$$

$$= \frac{1}{\mathbb{P}(A \geq a)} \int_a^{+\infty} \mathbb{P}(a')\mathbb{E}[X|A = a']\mathrm{d}a', \tag{S2}$$

$$= \frac{\int_a^{+\infty} \mathbb{P}(a')a'\mathrm{d}a'}{\mathbb{P}(A \geq a)} \frac{\mathbb{E}[XA]}{\Sigma_{AA}}, \tag{S2}$$

$$= \eta\left(\frac{a}{\sqrt{2\Sigma_{AA}}}\right) \frac{\mathbb{E}[XA]}{\Sigma_{AA}}, \tag{S2}$$

with

$$\eta(z) = \sqrt{\frac{2}{\pi}} \frac{e^{-z^2}}{\mathrm{erfc}(z)}, \tag{S2}$$

where $\mathrm{erfc}(\bullet)$ is the complementary error function. Previously, we have used that:

$$\mathbb{E}[X|A = a] = \int x\mathbb{P}(x|a)\mathrm{d}x = \frac{\int x\mathbb{P}(x,a)\mathrm{d}x}{\mathbb{P}(a)} = \frac{a}{\Sigma_{AA}}\mathbb{E}[XA], \tag{S2}$$

and this completes the proof.

For the committor, the important point is finding the expression for the conditional probability $\mathbb{P}(A = a|X = x)$. This is done by taking a slice at $X = x$ from of $\mathbb{P}(X, A)$ and expressing it as a function of $a$:

$$\mathbb{P}(A = a|X = x) \propto \exp\left(-\frac{1}{2}\left(2(\Lambda_{XA}^\top x)a + \Lambda_{AA}a^2\right)\right), \tag{S2}$$

$$\propto \exp\left(-\frac{1}{2}\Lambda_{AA}\left(a + \Lambda_{AA}^{-1}\Lambda_{XA}^\top x\right)^2\right), \tag{S2}$$

which is the expression for a one dimensional Gaussian distribution with variance $\sigma^2 = \Lambda_{AA}^{-1}$ and mean $\mu(x) = -\Lambda_{AA}^{-1}\Lambda_{XA}^\top x$. Now from the expressions for inverting a block matrix like $\Sigma$, we know that

$$\Lambda_{AA} = \left(\Sigma_{AA} - \Sigma_{AX}\Sigma_{XX}^{-1}\Sigma_{XA}\right)^{-1}, \tag{S2}$$

and

$$\Lambda_{XA} = -\Sigma_{XX}^{-1}\Sigma_{XA}\Lambda_{AA}. \tag{S2}$$

Remembering again that $\Lambda_{AA}$ is a scalar, we immediately get eq. (19),

$$\mu(x) = \Sigma_{XX}^{-1}\Sigma_{XA} \cdot x, \qquad \sigma^2 = \Sigma_{AA} - \Sigma_{AX}\Sigma_{XX}^{-1}\Sigma_{XA}. \tag{S2}$$

After this, getting the full committor is a simple one dimensional Gaussian integral, which is already well explained in the main text.

## S5 Committor Function for a Stochastic Process

Let's consider a stochastic process $X(t)$ on a phase-space $\Omega$. The *first hitting time* $\tau'_{\mathcal{V}}$ of the set $\mathcal{V} \subset \Omega$, given that the trajectory started at $x$, is defined as:

$$\tau'_{\mathcal{V}}(x) := \inf\{t : X(t) \in \mathcal{V} \mid X(0) = x\}. \quad (S2)$$

The committor function $q$ is defined as the probability that the first hitting time of the set $\mathcal{C}$ is smaller than the first hitting time of set $\mathcal{B}$, given the initial conditions $x$, where $\mathcal{B}, \mathcal{C} \subset \Omega$, $\mathcal{B} \cap \mathcal{C} = \emptyset$:

$$q(x) := \mathbb{P}(\tau'_{\mathcal{B}}(x) > \tau'_{\mathcal{C}}(x)). \quad (S2)$$

Sets $\mathcal{B}$ and $\mathcal{C}$ can be two attractors of the system or for instance one could correspond to a typical state of the system around which it fluctuates and another one to an atypical state which is visited when rare fluctuations arise. In the context of this paper, we are interested in the second case, where we define the fluctuations of interest based on an observable, namely the heatwave amplitude, defined in eq. (1), reaching a given threshold $a$. It is then natural to rewrite the definition of the committor function as in eq. (4).

## S6 Spatial Gradient Regularization

To compute the spatial gradient of the projection pattern $M$, we need to consider that we are working in a spherical geometry, which has two effects. If $\Lambda$ and $\Phi$ are respectively latitude and longitude, the gradient in the local flat geometry $x$-$y$ (with $x$ pointing eastward and $y$ northward) is

$$\begin{cases} \frac{\partial}{\partial x} = \frac{\partial \Phi}{\partial x} \frac{\partial}{\partial \Phi} = \frac{1}{\cos \Lambda} \times \frac{\partial}{\partial \Phi} \\ \frac{\partial}{\partial y} = \frac{\partial \Lambda}{\partial y} \frac{\partial}{\partial \Lambda} = 1 \times \frac{\partial}{\partial \Lambda} \end{cases}. \quad (S2)$$

The other effect is that the area of a grid cell is

$$d\mathcal{A} = dxdy = \cos \Lambda \, d\Lambda \, d\Phi. \quad (S2)$$

If for simplicity we assume we are dealing with only one climate variable, the total squared spatial gradient of $M$ is

$$H_2(M) = \int \left( \left(\frac{\partial M}{\partial x}\right)^2 + \left(\frac{\partial M}{\partial y}\right)^2 \right) dxdy, \quad (S2)$$

$$= \int \cos \Lambda \left( \left(\frac{1}{\cos \Lambda} \frac{\partial}{\partial \Phi}\right)^2 + \left(\frac{\partial}{\partial \Lambda}\right)^2 \right) d\Lambda d\Phi, \quad (S2)$$

$$= \int \left( \frac{1}{\cos \Lambda} \left(\frac{\partial}{\partial \Phi}\right)^2 + \cos \Lambda \left(\frac{\partial}{\partial \Lambda}\right)^2 \right) d\Lambda d\Phi. \quad (S2)$$

In our case, however, $M$ spans three climate variables, and is sampled on a uniform grid in latitude and longitude. This means we can write the projection pattern as a tensor $M^{\lambda \phi f}$, with indices $\lambda = 1, \ldots, n_\lambda = 22$ for latitude, $\phi = 1, \ldots, n_\phi = 128$ for longitude and $f = 1, \ldots, n_f = 3$ for distinguishing the fields. The discrete version of the gradient is thus

$$H_2(M) = \sum_{f=1}^{n_f} \left[ \sum_{\lambda=1}^{n_\lambda - 1} (\cos \Lambda_\lambda) \sum_{\phi=1}^{n_\phi} (M^{(\lambda+1)\phi f} - M^{\lambda \phi f})^2 \right.$$
$$\left. + \sum_{\lambda=1}^{n_\lambda} (\cos \Lambda_\lambda) \sum_{\phi=1}^{n_\phi} \left( \frac{M^{\lambda((\phi \bmod n_\phi)+1)f} - M^{\lambda \phi f}}{\cos \Lambda_\lambda} \right)^2 \right], \quad (S2)$$

where the first row is the meridional gradient and the second row the zonal one, considering also the periodic term. To be precise, we should add the multiplicative term $\Delta \Lambda \Delta \Phi$,

but since it is a constant that we can include in the regularization coefficient $\epsilon$, we can ignore it

If we now collapse all the indices of $M$ into a single one $i = i(\lambda, \phi, f)$, it is quite obvious that we can write

$$H_2(M) = M^\top W M = \sum_{ij} W_{ij} M_i M_j. \tag{S2}$$

To get the expression for $W$, we can first notice that it is symmetric: $W_{ij} = U_{ij} + U_{ji}$, and, by matching terms, we get

$$\begin{aligned} U_{ij} = &\left( \frac{\cos \Lambda_\lambda + \cos \Lambda_{\lambda-1}}{2} + \frac{1}{\cos \Lambda_\lambda} \right) \delta_{i(\lambda,\phi,f)j(\lambda,\phi,f)} + \\ &- (\cos \Lambda_\lambda) \delta_{i(\lambda+1,\phi f)j(\lambda,\phi,f)} - \frac{1}{\cos \Lambda_\lambda} \delta_{i(\lambda,(\phi \mod n_\phi)+1,f)j(\lambda,\phi,f)}. \end{aligned} \tag{S2}$$

For simplicity of notation, we assumed a null contribution when one of the indices goes out of range or (in the case of soil moisture) points to a grid cell with no data.

### S7  Regularized Gaussian Committor

To have the proper coefficients $\alpha$ and $\beta$ when we deal with a regularized pattern, we can notice that the assumption that $X$ and $A$ follow a jointly Gaussian distribution implies that, for *any* $M$, $F = M^\top X$ and $A$ also follow a jointly Gaussian distribution. We can then use the same formulas of eqs. (19) and (24), but applied to the 2 by 2 covariance matrix between $F$ and $A$

$$\hat{\Sigma} = \begin{pmatrix} \sigma_F^2 & \mathbb{E}[FA] \\ \mathbb{E}[FA] & \sigma_A^2 \end{pmatrix} =: \hat{\Lambda}^{-1}, \tag{S2}$$

and simply

$$q_{\mathcal{G}}(x) = \frac{1}{2} \mathrm{erfc}\left( \hat{\alpha} + \hat{\beta} M^\top x \right), \tag{S2}$$

with

$$\hat{\sigma}^2 = \sigma_A^2 - \left( \frac{\mathbb{E}[FA]}{\sigma_F} \right)^2, \quad \hat{\alpha} = \frac{a}{\sqrt{2}\hat{\sigma}}, \quad \hat{\beta} = \frac{\mathbb{E}[FA]}{\sqrt{2}\hat{\sigma}\sigma_F^2}. \tag{S2}$$

### S8  Effective Number of Independent Heatwaves

As we said in section 4.2, estimating the effective number of independent heatwaves can be challenging. The standard way of computing an effective data size for a time-series is the one presented in (Santer et al., 2000), where one uses the lag-1 autocorrelation coefficient $r$ to rescale the total number of data points: $N_{\mathrm{eff}} = N(1-r)/(1+r)$. However, when we consider heatwave events, they are not evenly spaced in time, so the whole approach does not make sense.

We can, tough, easily provide some bounds by observing that surely $N_{\mathrm{y}} \leq N_{\mathrm{eff}} \leq N_{\mathrm{all}}$, where $N_{\mathrm{all}} = N$ is the total number of heatwaves and $N_{\mathrm{y}}$ is the number of years that have at least a heatwave. Assuming that heatwaves at least a year apart are independent is definitely reasonable, if rather conservative. In fact, if we indeed compute the lag-1 autocorrelation coefficient for the time-series of $A(t)$, which gives $r = 0.9896$, and then estimate the decorrelation time of $A$ as $\tau_{\mathrm{decorr}} = (1+r)/(1-r)$, we get $\tau_{\mathrm{decorr}} = 191$ days. Namely, it takes half a year to lose memory of the heatwave amplitude, and thus $N_y$ is not only a lower bound for $N_{\mathrm{eff}}$, but likely also very close to it.

If we apply this to our study of 14-day heatwaves we have $N_{\mathrm{y}} = 2627 \lesssim N_{\mathrm{eff}} \leq N_{\mathrm{all}} = 30800$. Considering that we work with 8000 years of data, $N_{\mathrm{y}}$ tells us that there is a heatwave at least once every three years, and a year with a heatwave, on average, has $N_{\mathrm{all}}/N_{\mathrm{y}} \approx 12$ days for which $A(t) \geq a$.

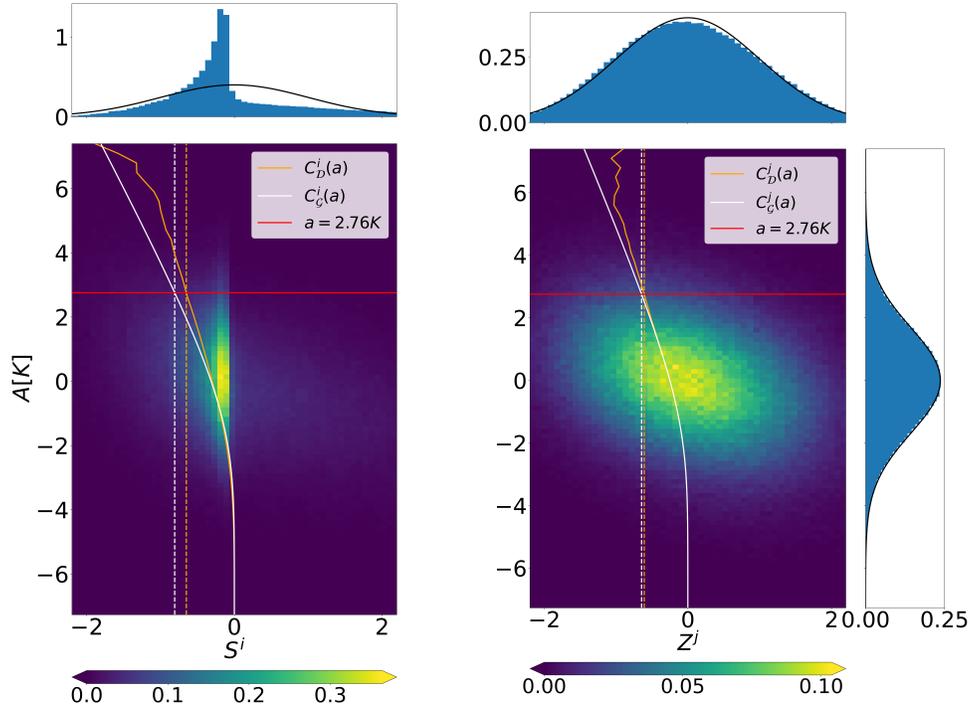

**Figure S3.** Comparison of the quality of the Gaussian and empirical composite map, for two grid-points and at different values of the heatwave threshold $a$. We show the results for a pixel over France of soil moisture $S^i$ (left) and one over Greenland of 500 hPa geopotential height $Z^j$ (right). For each panel, the main plot is the joint probability density function (PDF) of $S^i$ or $Z^j$ and the heatwave amplitude $A$, with marginal distributions displayed on top and to the side. In these plots the black line is a Gaussian fit. In the main plots, the orange line is the empirical composite as a function of $a$, while the white line is the one estimated through the Gaussian approximation. The red line is the threshold value of $a = 2.76K$ corresponding to the 5% most extreme heatwaves. The dotted vertical white and orange lines indicate the values of the empirical and Gaussian composites at this particular value of $a$. For $Z^j$ the error is much smaller.

## S9 Visualization of the Error Between Empirical and Gaussian Composites on Two Grid-Points

From fig. 2, we see that the biggest error we make when using the Gaussian composite is for the soil moisture variable. To investigate why this is the case, we show in fig. S3 (left) the joint and marginal distributions of the heatwave amplitude $A$ (on the y axis) and of one pixel of soil moisture $S^i$ (on the x axis). For comparison, we show the same for a pixel of the 500 hPa geopotential height $Z^j$ in fig. S3 (right). While the marginal distributions of $A$ and $Z^j$ are approximately Gaussian (as it is shown from the black curve on the marginal plots of the figure), the one of $S^i$ is clearly not: it is strongly skewed towards negative values of soil moisture anomaly, and exhibits fat tails. This is also reflected in the plot of the joint distribution of $A$ and $S^i$ on the one hand and $Z^j$ on the other hand (heat maps in fig. S3), as only the latter has the shape of an ellipsoid.

In both panels of fig. S3, the orange curve shows the behavior of the empirical composite $C_{\mathcal{D}}$ as we change the threshold $a$, while the white curve is the behavior of the Gaussian composite $C_{\mathcal{G}}$. By construction, they both tend to 0 as $a \to -\infty$, since soil moisture $S^i$ and 500 hPa geopotential height $Z^j$ both have zero mean as they are anomalies.

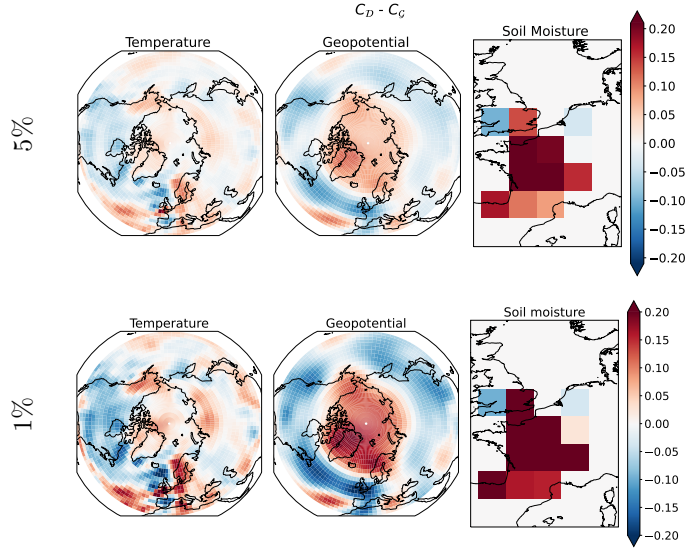

**Figure S4.** Maps of the difference between empirical composite maps and the ones estimated from the Gaussian approximation, for the 5% and 1% most extreme 14-day heatwaves over France. The patterns of this difference do not depend on the threshold $a$, varying only in intensity. The salient features of both temperature and geopotential height are well captured by the Gaussian approximation, with errors of the order of 30% at most.

When $a$ is very small, then, the Gaussian composite provides a good, yet useless, approximation of the empirical composite as both are very close to zero. As the threshold increases beyond this trivial region, for soil moisture the two curves start to diverge already at $a \approx 0$, and thus show a significant distance when they reach the $95^{\text{th}}$ quantile of $A$, $a = 2.76\,\text{K}$ (red line). On the other hand, the approximation holds quite well in the case of geopotential height, and the two curves separate significantly only when the sampling error kicks in (around $a = 5\,\text{K}$) and throws off the empirical estimate.

### S10 Error Between Empirical and Gaussian Composites at Different Heatwave Thresholds

In fig. S4 we show the difference between the empirical composite and the Gaussian one (evaluated using eq. (15)) for the three fields of 2 meter air temperature, $500\,\text{hPa}$ geopotential height and soil moisture evaluated for $a$ corresponding to the 5% and 1% most extreme 14-day heatwaves. The striking result is that the pattern observed changes only in magnitude between extreme and very extreme events.

To give a quantitative measure of the error, in fig. S5, we evaluate the error using the norm ratio defined in eq. (25), for different threshold level $a$, showing that the error is around the 20% for the 5% most extreme events. Figure S5 gives more details on the behavior of the norm of the error shown in fig. S4 for different values of $a$. On the y-axis we represent the norm ratio introduced in section 4.1 (eq. (25)), which measures how distant (in norm) the Gaussian composite is from the empirical composite (normalized by the norm of the empirical composite). We calculated this norm for the three fields independently, and for the whole stack of them (gray line). The norm ratios of the $500\,\text{hPa}$ geopotential height and the one of the $2\,\text{m}$ air temperature stay pretty close to the norm of the stack, showing values below 0.3 even for events which represents the 1% most extreme ones in the dataset (the x-axis on the top shows the respective percentile of rareness

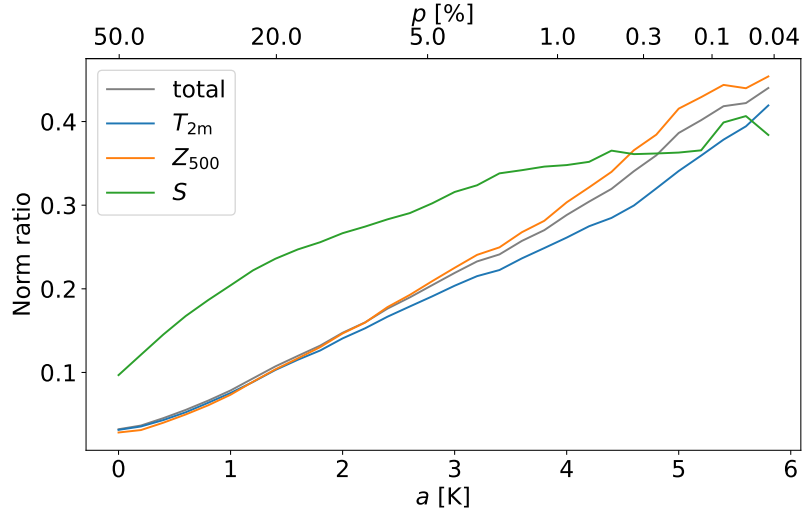

**Figure S5.** Norm ratio (see eq. (25) of the main text) of the difference between the empirical composite map and the Gaussian approximated one as a function of the threshold value used to define a heatwave event $a$. The total norm ratio is in gray, while the colors represent the norm ratio for each of the three fields (namely 2 m temperature anomaly, 500 hPa geopotential height anomaly and soil moisture anomaly). For events which are the 1% most extreme ones of the PlaSim dataset, the relative error is always below the 30%. The bottom x-axis $a$ is the threshold value used to define an heatwave event from the distribution of the 14-day heatwave amplitude $A$. On the top x-axis, $p$ is the respective percentile value corresponding to a given $a$.

of the $a$, which is on the bottom x-axis). Soil moisture has a different behavior, showing higher values of the norm ratio for much less extreme events. This is possibly due to a violation of the Gaussian approximation assumption, as we showed for a single pixel in fig. S3.

### S11 Field-Wise Norm Ratio of Composite Maps at Different Values of $T$ and $\tau$

In tables S1 to S3, we show the norm ratio as defined in eq. (25), but computed independently for the three climate variables of the PlaSim dataset. Values related to temperature peak at $T = 1$ and for small delay time. Soil moisture exhibits a clear monotonic trend with respect to $T$, while not being very sensitive to the lead time $\tau$. Finally, the geopotential height field shows the more complex structure, with the highest errors happening at intermediate values of both $T$ and $\tau$.

Soil moisture has only 12 pixels, compared to the 2816 of the other two fields, so its contribution to the total norm ratio (table 3) is negligible. Outside the small region of low values of $T$ and $\tau$, the norm ratio of the temperature field is almost constant, so the structure visible in table 3 is mostly due to the geopotential height field.

### S12 Asymptotic Behavior of the Regularized Projection Pattern

In this section we discuss the effect of the regularization coefficient $\epsilon$ on the optimal projection pattern $M_\epsilon$, and in particular why a highly regularized projection pattern may look similar to the composite map.

## Temperature

| | | τ [days] | | | | | | | | | |
|---|---|---|---|---|---|---|---|---|---|---|---|
| | | 0 | 3 | 6 | 9 | 12 | 15 | 18 | 21 | 24 | 27 | 30 |
| T [days] | 1 | 0.26 | 0.30 | 0.31 | 0.22 | 0.21 | 0.22 | 0.22 | 0.21 | 0.19 | 0.19 | 0.19 |
| | 3 | 0.25 | 0.29 | 0.26 | 0.19 | 0.19 | 0.20 | 0.20 | 0.19 | 0.18 | 0.17 | 0.18 |
| | 7 | 0.22 | 0.24 | 0.21 | 0.19 | 0.19 | 0.20 | 0.20 | 0.19 | 0.18 | 0.18 | 0.16 |
| | 14 | 0.19 | 0.20 | 0.20 | 0.19 | 0.20 | 0.20 | 0.21 | 0.21 | 0.19 | 0.18 | 0.18 |
| | 30 | 0.18 | 0.19 | 0.20 | 0.20 | 0.20 | 0.20 | 0.20 | 0.19 | 0.19 | 0.20 | 0.20 |

**Table S1.** Values of the norm ratio between Gaussian and empirical composites computed for 2 m temperature anomaly.

## Geopotential

| | | τ [days] | | | | | | | | | |
|---|---|---|---|---|---|---|---|---|---|---|---|
| | | 0 | 3 | 6 | 9 | 12 | 15 | 18 | 21 | 24 | 27 | 30 |
| T [days] | 1 | 0.25 | 0.26 | 0.28 | 0.30 | 0.31 | 0.33 | 0.36 | 0.34 | 0.29 | 0.26 | 0.27 |
| | 3 | 0.24 | 0.23 | 0.26 | 0.28 | 0.29 | 0.31 | 0.34 | 0.31 | 0.27 | 0.25 | 0.25 |
| | 7 | 0.21 | 0.23 | 0.27 | 0.30 | 0.33 | 0.38 | 0.37 | 0.32 | 0.30 | 0.30 | 0.25 |
| | 14 | 0.21 | 0.24 | 0.30 | 0.34 | 0.34 | 0.33 | 0.33 | 0.34 | 0.32 | 0.28 | 0.28 |
| | 30 | 0.22 | 0.25 | 0.29 | 0.32 | 0.33 | 0.33 | 0.34 | 0.32 | 0.30 | 0.31 | 0.31 |

**Table S2.** Values of the norm ratio between Gaussian and empirical composites computed for 500 hPa geopotential height anomaly

## Soil moisture

| | | τ [days] | | | | | | | | | |
|---|---|---|---|---|---|---|---|---|---|---|---|
| | | 0 | 3 | 6 | 9 | 12 | 15 | 18 | 21 | 24 | 27 | 30 |
| T [days] | 1 | 0.10 | 0.11 | 0.12 | 0.12 | 0.13 | 0.13 | 0.13 | 0.13 | 0.14 | 0.14 | 0.14 |
| | 3 | 0.14 | 0.12 | 0.12 | 0.13 | 0.12 | 0.12 | 0.12 | 0.12 | 0.12 | 0.12 | 0.12 |
| | 7 | 0.22 | 0.20 | 0.19 | 0.18 | 0.18 | 0.17 | 0.16 | 0.16 | 0.15 | 0.15 | 0.14 |
| | 14 | 0.30 | 0.28 | 0.27 | 0.26 | 0.25 | 0.24 | 0.23 | 0.22 | 0.21 | 0.20 | 0.19 |
| | 30 | 0.40 | 0.38 | 0.36 | 0.35 | 0.33 | 0.31 | 0.30 | 0.28 | 0.27 | 0.25 | 0.24 |

**Table S3.** Values of the norm ratio between Gaussian and empirical composites computed for soil moisture anomaly.

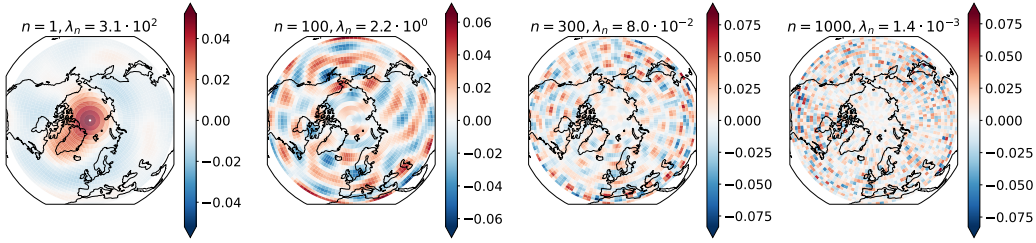

**Figure S6.** Four EOFs $e_n$. The higher the $n$ the smaller the spatial scales of the characteristic features represented. On top of each plot we report the value of $\lambda_n$.

To do so we move to the basis of Empirical Orthogonal Functions (EOFs) (Hannachi et al., 2007), which diagonalizes the covariance matrix $\Sigma_{XX}$. We call its eigenvalues $\lambda_1 \geq \lambda_2 \geq \cdots \geq \lambda_n \geq \cdots \geq \lambda_d$, and the corresponding eigenvectors $e_n$. Here we will use as an example the prediction of $T = 14$ day heatwaves at delay time $\tau = 0$, performed on the ERA5 reanalysis dataset. In this context there are $d = 2816$ degrees of freedom, corresponding to the pixels of the 500 hPa geopotential height anomaly field.

The first important point is that as $n$ increases, the variance $\lambda_n$ explained by $e_n$ decreases, and so does decrease the typical spatial size of the features that appear in $e_n$ (fig. S6). In particular, features with a typical size of the order of the synoptic scale are represented around $n = 100$. Moreover, almost two thirds of the EOFs ($n > 1000$) explain less than 0.05 % of the variance and are extremely noisy.

Second, the Gaussian composite map is proportional to the correlation map $\Sigma_{XA}$ (eq. (15)), and when we write it in the EOF basis,

$$C_{\mathcal{G}} \propto \Sigma_{XA} = \sum_{n=1}^{d} c_n e_n, \tag{S6}$$

it is dominated by EOFs at low values of $n$ (see black lines in fig. S7), and thus it appears spatially smooth.

Third, in the EOF representation the non-regularized ($\epsilon = 0$) projection map is written as

$$M_0 = \sum_{n=1}^{d} M_n^0 e_n \propto \Sigma_{XX}^{-1} \Sigma_{XA} \propto \sum_{n=1}^{d} \frac{c_n}{\lambda_n} e_n, \tag{S6}$$

and $\lambda_n$ goes to zero much faster than $c_n$ (dashed dark blue lines in fig. S7), resulting in $M_0$ being dominated by large $n$, high spatial frequency modes. This is what makes the non regularized pattern utterly non-interpretable.

If we perform $L_2$ regularization, the aim of the regularization coefficient is to prevent the contribution of these high frequency modes to explode, making them proportional to their values in the composite map:

$$M_\epsilon \propto (\Sigma_{XX} + \epsilon \mathbb{I})^{-1} \Sigma_{XA} \propto \sum_{n=1}^{d} \frac{c_n}{\lambda_n + \epsilon} e_n \approx \sum_{n=1}^{n_\epsilon - 1} M_n^0 e_n + \frac{1}{\epsilon} \sum_{n=n_\epsilon}^{d} c_n e_n, \tag{S6}$$

where $\lambda_{n_\epsilon} \approx \epsilon$. It is then clear that as $\epsilon$ increases $n_\epsilon \to 1$, and the projection map smoothly converges to the composite map (left panel of fig. S7).

On the other hand, when we perform $H_2$ regularization, as we do in this work, we regularize with matrix $W$, which doesn't share the same eigenvectors of $\Sigma_{XX}$. We can

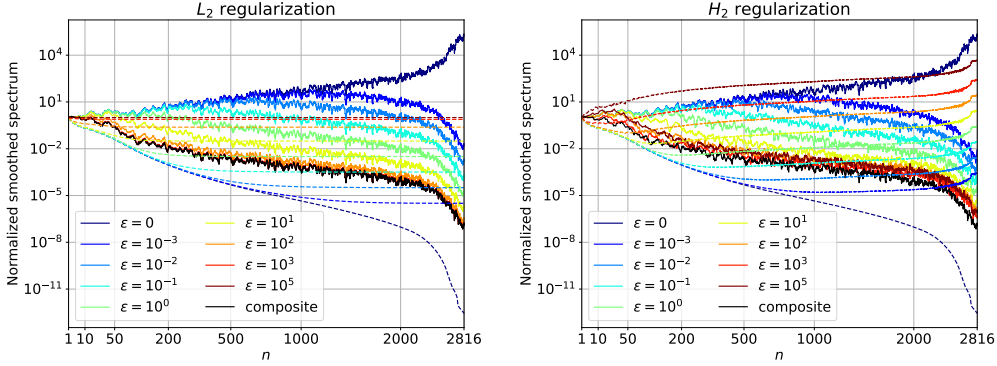

**Figure S7.** EOF spectra $|M_n^\epsilon|$ of the projection pattern at different values of the regularization coefficient $\epsilon$ (solid lines) when penalizing the $L_2$ norm of the pattern (left) or its spatial gradient (right). All spectra are normalized so that the term at $n = 1$ has unitary values. To ease the visualization, the spectra have been smoothed with a running average. On the left panel the dashed lines represent the spectra of the regularized covariance matrix: $\lambda_n + \epsilon$. On the right panel they represent the diagonal part of the gradient regularized covariance matrix in the original EOF basis, i.e. $\lambda_n + \epsilon e_n^\top W e_n$. The black line is the spectrum $c_n$ of the Gaussian composite map. For high $n$ (i.e. EOFs with small spatial scales), the values of $\lambda_n$ decay faster than those of $c_n$, which makes the non-regularized pattern extremely noisy. For $L_2$ regularization, increasing values of $\epsilon$ progressively reduce the contribution of EOFs at high $n$, and the projection pattern converges to the composite. On the other hand, $H_2$ regularization directly penalizes the spatial gradient of the projection pattern, so the small scales are first suppressed and then brought back to achieve the spatially uniform pattern. These spectra are presented for $T = 14$ and $\tau = 0$ on the ERA5 dataset, where $\epsilon_{\text{best}} = 10$ (light green lines).

in any case write $W$ in the EOF basis as

$$W = \sum_{mn} W_{nm} e_n e_m^\top. \quad \text{(S6)}$$

If we compute the terms $W_{mn}$, we notice that $W_{nn} \gg \max_{m \neq n} |W_{nm}|$. We can then say that the $W$ is almost diagonal and thus

$$M_\epsilon \propto (\Sigma_{XX} + \epsilon W)^{-1} \Sigma_{XA} \approx \sum_{n=1}^d \frac{c_n}{\lambda_n + \epsilon W_{nn}} e_n. \quad \text{(S6)}$$

This lets us apply a similar reasoning to the one explained above for $L_2$ regularization, where $W_{nn}$ is the norm of the spatial gradient of EOF $e_n$, which, considering the spatial structure of the EOFs (fig. S6), clearly increases with $n$. For this reason, when we increase $\epsilon$, we remove the high spatial frequencies faster than we would with $L_2$ regularization (right panel of fig. S7). On the other hand, for very high regularization, the approximation of $W$ being diagonal falls apart and the high frequencies are brought back to achieve a spatially uniform pattern, similarly to the Fourier representation of a square wave. So there is no asymptotic convergence to the composite map (brown curve). However, for intermediate values of $\epsilon$ (yellow, orange and red curves), the projection pattern is smoothed in a similar way as with $L_2$ regularization and thus it may look similar to the composite map.



\end{document}